\documentclass[twocolumn,aps,prb,showpacs,amsmath,superscriptaddress,longbibliography,notitlepage]{revtex4-1}
\usepackage{amssymb}
\usepackage{mathrsfs}
\usepackage{graphicx}
\usepackage{float}
\usepackage{array}
\usepackage{epstopdf}

\usepackage[normalem]{ulem}
\usepackage{color}
\usepackage{bm}
\usepackage[left=23mm,right=13mm,top=35mm,columnsep=15pt]{geometry}
\usepackage{subcaption}

\def\red{\textcolor{black}}
\def\GJ{\textcolor{black}}

\usepackage[pdftex, pdftitle={Article}, pdfauthor={Author}]{hyperref} % For hyperlinks in the PDF
\begin{document}

%\title{The winding number of Majorana Star as a topological invariant in a non-Hermitian multiband system}
\title{Topological characterization of non-Hermitian multiband systems using Majorana's Stellar Representation}

\author{Wei Xin Teo}
%    \email[Correspondence email address: ]{email@institution.com}% Your name
    \affiliation{National University of Singapore, Department of Physics}
\author{Linhu Li}
    \email{phylli@nus.edu.sg}% Your name
    \affiliation{National University of Singapore, Department of Physics}
\author{Xizheng Zhang}
%    \email[Correspondence email address: ]{email@institution.com}% Your name
    \affiliation{National University of Singapore, Department of Physics}
\author{Jiangbin Gong}
    \email{phygj@nus.edu.sg}% Your name
    \affiliation{National University of Singapore, Department of Physics}

\date{\today} % Leave empty to omit a date

\begin{abstract}
\GJ{For topological characterization of non-Hermitian multiband systems, Majorana's stellar representation (MSR) is applied \GJ{to}  1D multiband models consisting of asymmetric nearest-neighbor hopping and imaginary on-site potentials.  The number of edge states isolated from the continuous bulk bands in the complex energy plane is successfully linked with a topological invariant constructed from MSR. Specifically,  the number of isolated edge states can be obtained from a winding number defined for the Majorana stars, which also allows for a geometric visualization of the topology related to the isolated edge modes.  A remarkable success of our approach is that our winding number characterization remains valid even in the presence of exceptional points of the continuous bulk bands, where the Hamiltonian becomes non-diagonalizable and hence conventional topological invariants such as the Zak phase and the Chern number cannot be properly defined. Furthermore,
 cases with the so-called non-Hermitian skin effect are also studied, showing that the bulk-boundary correspondence between our defined winding numbers and isolated edge states can be restored}. {Of particular interest is a four-band example with an odd number of isolated edge states, where the Zak phase approach necessarily fails upon removing the skin effect, but our MSR-based characterization works equally well.  For these reasons, our study is expected to be widely useful in topological studies of non-Hermitian multiband systems, regardless of the skin effect or the presence of the exceptional points in non-Hermitian systems. }
%Although there is a rapid growth of the research in the non-Hermitian systems, less attention focus on the non-Hermitian multiband systems. Therefore, an idea that studies a Hermitian multiband system through the Majorana Star Representation is extended to the non-Hermitian system in this article. We propose a non-Hermitian 1D $n$-band chain model, which consists of asymmetric nearest-neighbor hopping and on-site potentials, and define the winding number of the Majorana stars as a new topological invariant. By presenting several examples, including the situations with the presence of the skin effect, we elaborate that the winding number can characterize the number of isolated edge states under open boundary conditions. Moreover, we have also shown that under the presence of the exceptional points, the winding number is a more potent topological invariant than the Zak phase because the Zak phase is ill-defined in this case.
\end{abstract}

\maketitle
%\tableofcontents
%\newpage

\section{Introduction} \label{sec:intro}

Non-Hermitian Hamiltonians \cite{PhysRevLett.80.5243,bender2007making} \GJ{are now widely recognized to be physically relevant} as effective Hamiltonians in many physical systems,
such as open quantum systems \cite{rotter2009non} with finite life time introduced by electron-electron or electron-phonon interactions \cite{yoshida2018non,yamamoto2019theory}, and photonic systems with gain and loss \cite{ruter2010observation,longhi2018parity,ozawa2019topological}.
%{Non-Hermitian Hamiltonians have gained a physically relevant space from the so-called effective Hamiltonians, where the non-Hermitian part serves different purposes.
%Such a rapidly growing field mainly consists of open quantum systems \cite{rotter2009non}, where the finite life time can be introduced by electron-electron or electron-phonon interactions \cite{yoshida2018non,yamamoto2019theory}, and photonic systems with gain and loss \cite{ruter2010observation,longhi2018parity,ozawa2019topological}.}
In particular, non-Hermitian topological phases have been one of the most intriguing research subjects during the past few years, because they possess many exotic topological phenomena beyond Hermitian systems.
In non-Hermitian systems, exceptional degeneracies emerge when two or more energy levels coalesce into one, becoming identical not only in eigenenergies, but also eigenstates
%two or more eigenstates can coalesce into one, resulting in exceptional degeneracies 
\cite{berry2004physics,jin2009solutions,longhi2010pt,heiss2012physics,lee2016anomalous,xu2016topological,hassan2017dynamically,hu2017exceptional,shen2018topological,wang2019arbitrary,ghatak2019new,miri2019exceptional,zhang2020non,yuce2020non,jin2019hybrid}.
Such degeneracies can form various manifolds with distinct topological structures in the Brillouin zone of systems beyond one-dimension
\cite{xu2017weyl,carlstrom2018exceptional,zhou2019exceptional,moors2019disorder,wang2019non,yang2019non,carlstrom2019knotted,okugawa2019topological,luo2018nodal,lee2018tidal}.
 The celebrated ten-fold symmetry classification has also been extended into non-Hermitian systems, and is much enriched due to the extra nonspatial symmetries of non-Hermitian matrices \cite{gong2018topological,liu2019topological,kawabata2019symmetry,zhou2019periodic,li2019geometric,liu2019topological2,lieu2018topological,kawabata2019topological,wu2019inversion,yoshida2019symmetry,yoshida2019exceptional,kawabata2019classification}.
%and an incomplete Hilbert basis.
%with rich geometric structures \cite{xu2017weyl,carlstrom2018exceptional,moors2019disorder,wang2019non,yang2019non,carlstrom2019knotted,yoshida2019symmetry,yoshida2019exceptional,zhou2019exceptional,okugawa2019topological,luo2018nodal}.
It has also been shown that the non-Hermitian skin effect (NHSE) \cite{yao2018edge}, reflected by enormous accumulation of eigenmodes at system boundaries, can modify the topological bulk-boundary correspondence \cite{Xiong_2018,PhysRevLett.121.026808,yao2018edge,lee2019anatomy,li2019geometric,PhysRevLett.123.246801,borgnia2020nonH,zhang2019correspondence,yoshida2019mirror,lee2019unraveling},
and lead to other novel phenomena when interplayed with different physical effects, e.g. non-Hermitian quasicrystal \cite{longhi2019topological,jiang2019interplay,zeng2020topological}, hybrid skin-topological modes and topology-controled non-reciprocal pumping \cite{lee2019hybrid,li2019topology}, emergence of a real space Fermi surface \cite{mu2019emergent}, and singularities in Berry curvature \cite{lee2019unraveling}.
%a new type of hybrid skin-topological modes when interplayed with topological localization in a transverse direction \cite{lee2019hybrid,li2019topology}.
%It has also be shown that the bulk-boundary correspondence is violated in the presence of non-Hermitian skin effect (NHSE) \blue{[cite]}, and can be restored in a properly defined generalized Brillouin zone \blue{[cite]}.
To understand these interesting behaviors, many efforts have been made to develop new tools for characterization and geometric visualization \GJ{of} non-Hermitian topology, such as the generalized Brillouin zone \cite{yao2018edge,yokomizo2019non,yang2019auxiliary}, energy vorticity \cite{shen2018topological} and \GJ{the associated} winding numbers \cite{zhang2019correspondence,okuma2019topological}, the singularity ring in pseudospin space \cite{li2019geometric}, and a graphic approach of eigenstates on torus \GJ{as certain parameter space}  \cite{yang2019visualizing}.

To date, topology of eigenstates in non-Hermitian systems can already be geometrically visualized through various methods \cite{yin2018geometrical,jiang2018topological,li2019geometric,yang2019visualizing,jiang2019topological}.
%To date, geometric features of non-Hermitian eigenstate topology have be intensively studied and well understood with various methods \cite{yin2018geometrical,jiang2018topological,li2019geometric,yang2019visualizing,jiang2019topological},
However, many of them have focused on two-band systems only. \GJ{It is possibly challenging but of great interest to study non-Hermitian multiband systems}.
In Hermitian systems, multiband topology can be analyzed with the Majorana's stellar representation (MSR) \cite{majorana_1932,bloch_rabi_1945,biedenharn_dam_1965,hannay_1998,bruno_2012}, which maps a $n$-band eigenstate to $n-1$ Majorana Stars (MSs) on a Bloch sphere, each representing a different spin-1/2 state \cite{bruno_2012,liu_fu_2014,yang_guo_fu_chen_2015,PhysRevA.94.022123}. The distributions and motions of these MSs on a Bloch sphere are collectively connected to the Berry phase of the original $n$-band eigenstates, thus providing an intuitive way towards understanding \GJ{the underlying} topological properties.
Inspired by \GJ{such progresses in Hermitian multiband systems}, in this work we extend the MSR approach to non-Hermitian multiband systems by defining MSs for only the right (or left) eigenstates.
With explicit analysis of two-band systems and several intriguing examples of multiband systems, we find a one-to-one correspondence between the number of edge states isolated from continuous bands and a winding number defined \GJ{from} these MSs. \GJ{Remarkably}, the winding number we propose is well-defined even when different energy bands coalesce at one or more exceptional points (EPs), whereas conventional topological invariants such as the Zak phase and the Chern number are generally ill-defined in such cases.  \GJ{Furthermore}, our method can be directly extended to systems with NHSE \GJ{via a known procedure, namely, by considering the so-called non-Bloch Hamiltonian obtained by a complex deformation of the quasi-momentum of the studied system}. {Of particular importance, in a 4-band model with NHSE and possessing an odd number of isolated edge states, our defined winding number continues to work because it can equally predict the number of edge states in this subtle case, whereas the Zak phase necessarily fails in this situation.
Putting all these results together, it can be concluded that our defined winding number provides the most potent topological invariant to date in characterizing non-Hermitian topological phases in multiband systems}, {where both continuous bands and isolated edge states behave very differently from Hermitian systems.}
%as edge states can exist and be isolated from continuous bands even when these bands share EPs and cannot be separated from each other.

The rest of this paper is organized as \GJ{follows}. In Sec.~\ref{sec:model} we introduce our multiband models with \GJ{non-Hermiticity} induced by non-reciprocal hoppings and imaginary on-site potentials. In Sec. \ref{sec:MSR} we briefly review the MSR of high-(pseudo)spin states, and define a winding number for the MSs of each band. Sec.~\ref{sec:nonherm} contains the main results of this work, where we illustrate the bulk-boundary correspondence between isolated edge states and our defined winding numbers in several different scenarios with/without EPs and/or NHSE. A brief summary and discussion are given in Sec.~\ref{sec:conclusion}.

\section{Non-Hermitian 1D Multiband Chain Model}
\label{sec:model}
\begin{figure}[H]
    \centering
    \includegraphics[width=0.48\textwidth]{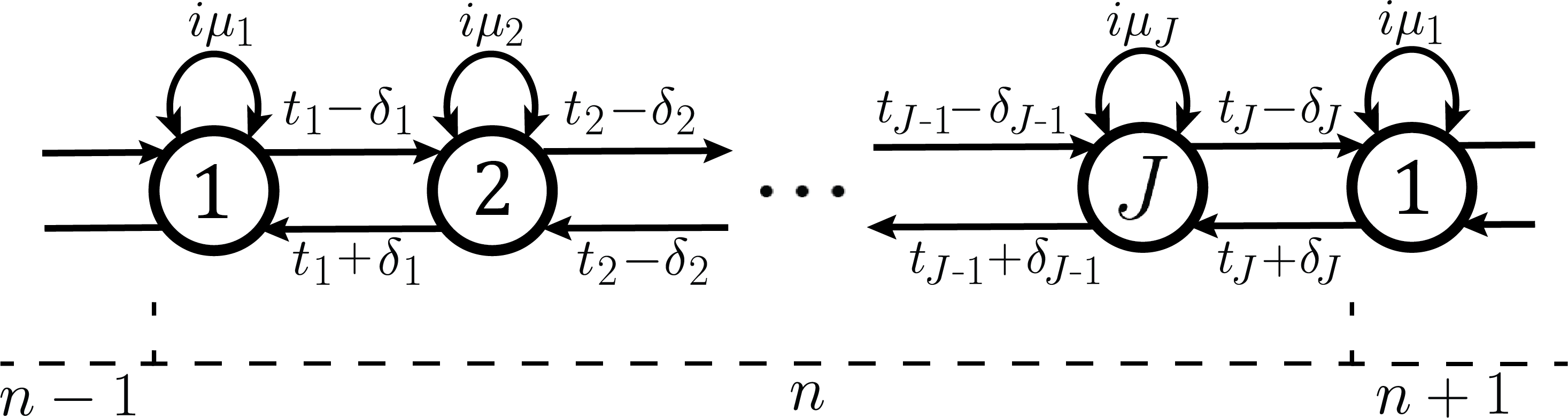}
    \caption{A simple illustration of the non-Hermitian 1D $J$-band chain model. Circles indicate the lattice site and arrows indicate the hopping between the lattices. }
    \label{fig:lattice}
\end{figure}

We consider a 1D lattice model with $J$ lattice sites in a unit cell, as illustrated in Fig.~\ref{fig:lattice}. The corresponding tight-binding Hamiltonian is given by
\begin{equation}
\label{eqn:model}
\begin{split}
    H=&\sum_n^N\sum_{j}^J(i\mu_{j}\hat{c}_{j,n}^{\dagger}\hat{c}_{j,n}+ (t_{j}+\delta_{j})\hat{c}_{j,n}^{\dagger}\hat{c}_{j+1,n} \\
    &  + (t_{j}-\delta_{j})\hat{c}_{j+1,n}^{\dagger}\hat{c}_{j,n}),
\end{split}
\end{equation}
where $\hat{c}_{j,n}$ ($\hat{c}^\dagger_{j,n}$) is the annihilation (creation) operator of a particle at the $j$th lattice site in the $n$th unit cell, and
$\hat{c}_{J+1,n}\equiv\hat{c}_{1,n+1}$. \GJ{Fig.~\ref{fig:lattice} presents
a more specific configuration of the lattice.}  As indicated in Fig.~\ref{fig:lattice}, our model consists of $(N\times J)$ lattice sites, with non-reciprocal hopping $t_j\pm\delta_j$ (which may induce NHSE in this system), and on-site imaginary potential $\mu_j$ depicting particle gain and loss.
% {$t_j$ and $\delta_j$ describe the Hermitian and non-Hermitian hopping amplitudes between nearest neighbor sites, and $i\mu_j$ is an imaginary on-site potential corresponding to particle gain and loss on each lattice site.}

By performing the Fourier transformation, $\hat{c}_{j,n} = 1/\sqrt{N}\sum_k e^{ink}\hat{c}_{j,k}$ with $j\in\{1,2,...,J\}$ to the real space Hamiltonian (\ref{eqn:model}), we can obtain the Bloch Hamiltonian $H_k=\sum_k\psi_k^{\dagger}h(k)\psi_k$ with \GJ{$k\in[0, 2\pi]$ being the quasimomentum variable},  $\psi_k=(\hat{c}_{1,k},\hat{c}_{2,k},...,\hat{c}_{J,k})^T$ and
\begin{equation}
\label{eqn:tband}
\begin{split}
&h(k) = \\
&\scalebox{0.9}{$\quad
\begin{bmatrix}
i\mu_1 & t_1+\delta_1 & ... & 0 & (t_J-\delta_J)e^{-ik} \\
t_1-\delta_1 & i\mu_2  & ... & 0 & 0 \\
\vdots & \vdots  & \ddots & \vdots & \vdots \\
0 & 0 & ... & i\mu_{J-1} & t_{J-1}+\delta_{J-1} \\
(t_J+\delta_J) e^{ik} & 0 & ... & t_{J-1}-\delta_{J-1} & i\mu_J
\end{bmatrix}
\quad$}.
\end{split}
\end{equation}

Conventionally, topological properties of 1D Hermitian systems can be characterized by the Zak phase \cite{zak1989berry}, which is the Berry phase \cite{berry1984quantal} \GJ{associated with each band as the quasi-momentum $k$ adiabatically runs over one cycle in the Brillouin zone}.
In particular, when the system consists of only two bands, the Zak phase can be obtained from the solid angle on a Bloch sphere of an eigenstate varying throughout the Brillouin zone (BZ), thus providing an intuitive geometric picture of the topology of 1D systems.
In non-Hermitian systems, a pair \GJ{of}  left and right eigenstates satisfy the biorthogonal normalization condition \cite{brody2013biorthogonal,PhysRevLett.121.026808}, and the non-Hermitian \GJ{(always real)} Zak phase can be defined as \cite{zhang_wang_gong_2019_2,zhang_wang_gong_2019}
\begin{equation}
\label{eqn:zaknh}
    \gamma^{(m)} = -{\rm Im}\oint_{BZ} \langle\psi_{L,m}(k)|\partial_k|\psi_{R,m}(k)\rangle dk,
\end{equation}
with $\psi_{L,m}(k)$ $\psi_{R,m}(k)$ the left (right) Bloch eigenstate and $m$ denoting the band index.
In a discrete lattice system, by considering discrete wavenumber of $k_n = 2\pi n/N_k$ and $N_k=N$ as a large integer, the Zak phase can be obtained numerically with
\begin{equation}
\label{eqn:zaknum}
    \gamma^{(m)} = -{\rm Im}\sum_n^{N_k}\log[\langle\psi_{L,m}(k_n)|\psi_{R,m}(k_{n+1})\rangle].
\end{equation}
%where {$n$ labeling discrete quasi-momenta.} %and the summation sums over the Brillouin Zone (BZ).
However, even for a two-band non-Hermitian system, this Zak phase cannot be directly mapped onto a Bloch sphere for visualization, because the definition of left and right eigenstates leads to complex winding angles of the pseduspin vector (see Appendix \ref{sec:2band}). Furthermore, as we illustrate in later discussions, non-Hermitian band structures can host topological edge states even in the presence of exceptional points. In this scenario, the Hamiltonian is not diagonalizable and the Zak phase becomes ill-defined. \GJ{For these two reasons}, a more \GJ{versatile} topological invariant is required to topologically characterize non-Hermitian multiband systems.

\section{The Winding Number of the Majorana Stars} \label{sec:MSR}
MSR is conventionally used to represent a pure quantum high-spin state with multiple spin-1/2 states \cite{majorana_1932,bloch_rabi_1945,biedenharn_dam_1965,hannay_1998,bruno_2012}, and has \GJ{later} been extended to describe Hermitian multiband topological systems \cite{bruno_2012,liu_fu_2014,yang_guo_fu_chen_2015,PhysRevA.94.022123}.
{For a spin-$L$ state $|\Phi\rangle$ which has 2$L$+1 components, it can be expressed in terms of 2$L$ spin-1/2 states according to the Schwinger boson representation theory \cite{biedenharn_dam_1965}:
\begin{equation}
|\Phi\rangle = \frac{1}{2N_L} \prod_{l=1}^{2L} (\cos(\frac{\theta_l}{2}) a^{\dagger}_{\uparrow} + \sin(\frac{\theta_l}{2}) e^{i\phi_l} a^{\dagger}_{\downarrow})|{0}\rangle.
\end{equation}
%In MSR, the spin-$J$ state can be decomposed to a symmetrized state $2J$ spin-1/2 systems, and hence $2J$ points on a Bloch Sphere which we call them Majorana Stars.
%This idea has been extended to Hermitian multiband topological systems \cite{bruno_2012,liu_fu_2014,yang_guo_fu_chen_2015,PhysRevA.94.022123}. %(cite
%We would like to further extend it to the non-Hermitian system.
%In fact, regardless of the hermiticity, the eigenstate of the n-band system, which is a column vector with $n$ complex elements, can be mapped to a spin-$J$ state, with $J=(n-1)/2$.
%Hence, we can apply MSR in the non-Hermitian system to decompose our eigenstates as well.
Hence, a spin-$L$ state can be decomposed to 2$L$ spin-1/2 states by finding the roots of the following MSR equation \cite{yang_guo_fu_chen_2015}:
\begin{equation}
\label{eqn:msr2}
    \sum_{l=0}^{2L}\frac{(-1)^l C_{2L-l+1}}{\sqrt{(2L-l)!l!}}x^{2L-l}=0,
\end{equation}
where $C_{\alpha}$ denotes the \GJ{wavefunction} components of a spin-$L$ state with $\alpha\in\{1,2,...,2L+1\}$, and $x_l$ being the found MS solutions.}

{In non-Hermitian J-band systems, the $m$th right eigenstate of the system has $J$ components in which each component can be understood as $C_{m,l}$, i.e. $|\psi_{R,m}(k)\rangle=(C_{m,1},C_{m,2},...,C_{m,J})^T$.
Therefore, the $m$-th right eigenstate is mapped to the spin-$L$ state by considering $2L+1=J$.
Further, the decomposition to MSRs can also be done for the right eigenstates by using the MSR equation Eq.~(\ref{eqn:msr2}), upon changing $2L$ to $J-1$ in the equation.}

{Since the decomposition is done for a given band $m$ and given wavenumber $k$, we denote the roots of the equation to be $x_{m,l}(k)= \tan\frac{\theta_{m,l}(k)}{2}e^{i\phi_{m,l}(k)}$ with $l\in\{1,2,...,J-1\}$, each representing a MS
%If we consider $\theta_l$ and $\phi_l$ as the altitude angle and the azimuth angle respectively, the roots of the MSR equation can be plotted
 on a Bloch sphere with the spherical coordinates $(1,\theta_{m,l}(k),\phi_{m,l}(k))$.
%Notice that the MSR equation is a polynomial with $n-1$ degree, we have $n-1$ roots and hence $n-1$ points on the Bloch sphere, which are the Majorana Stars.
%This is how we decompose a right eigenstate of a $n$-band system and visualize it in a Bloch sphere.
%Therefore, Under periodic boundary condition (PBC),
%we can apply this method to decompose all the $\ket{\psi_{R,m}(k)}$ for all $k\in [0,2\pi]$ in a given band $m$ and they will trace out curves in the Bloch sphere.
%These curves are related to the topology of the system.
A full MSR can be obtained by tracing each MS of $x_{m,l}(k)$ on the Bloch sphere with $k$ varying throughout the Brillouin zone.}

\red{In Hermitian systems, it has been rigorously proven that the geometrical phases of MSs, which consist of the solid angles formed by each stars and the correlation between two stars, are closely related to the Zak phases of the system \cite{liu_fu_2014,bruno_2012}.}
The existence of edge states was also known to be associated with a nontrivial winding of the system in a certain 2D plane. \cite{mong2011edge}. %By using this idea, we define a new topology invariant, called the winding number of the MSs as follow:
 Here we \GJ{define} a winding number for the MSs as the total winding of their \GJ{azimuthal} angles. That is,
\begin{equation}
\label{eqn:windinggen}
    \nu_m = -\frac{1}{2\pi} \sum_{l=1}^{J-1}\oint \partial_k\phi_{m,l}dk,
\end{equation}
for a given band $m$.
In two-band systems, we have proven in Appendix \ref{sec:2band} that this winding number is equivalent to the nontrivial winding of the Zak phase, which can be defined for both left and right eigenstates.
\red{However, we would like to emphasize that this winding number does not have a direct relation to the geometrical phases (which are related to solid angles instead).}
\GJ{Still, our hope is that the winding of MSs defined for right or left eigenstates only can also be used to characterize the existence of topological edge states in multiband systems.  This is confirmed by our extensive} numerical results for the 1D lattice model of Eq. (\ref{eqn:model}), as shown in the following sections.  Of particular importance is that in non-Hermitian systems, an isolated edge state may correspond to the coalescence of multiple eigenstates.  \GJ{We find that the winding number thus defined summed over all energy bands satisfy the following relation}:
\begin{equation}
\sum_m\nu_m=\sum_r D_r,
\end{equation}
with $D_r$ being the number of eigenstates \GJ{under the open boundary condition (OBC)} that \GJ{coalesce} into the $r$th isolated edge state, dubbed as a $D_r$-fold coalescent edge state hereafter. \GJ{Note however}, in a finite-size system under OBC, a $D_r$-fold coalescent edge state will be de-coalescent into totally $D_r$ edge states with slightly different eigenenergies and spatial distributions, with the total number of such edge states still directly given by $\sum_m\nu_m$.

Before moving on to the next section, we stress that we have only applied the MSR to right eigenstates in the above discussions.  \GJ{We have checked that using left eigenstates will yield the same conclusions.}

%We will show that this winding number will help us characterize the number of edge states of the system under OBC.
%Notice that we have only applied the MSR to the right eigenstate in the above discussions.
%However, since it can be shown that the Zak phase, which is evaluated from both the left and right eigenstates, is connected with either the winding number of the Majorana stars that is evaluated from the right or the left eigenstates in a 2-band system, it is our freedom to choose the left or the right eigenstates for the calculation (see Appendix\ref{sec:2band}).
%Therefore, we will only use the Majorana stars of the right eigenstates in the rest of the work.

\section{The Bulk-Boundary Correspondence} \label{sec:nonherm}

In a Hermitian 1D system, there is the bulk-boundary correspondence between the Zak phase evaluated under PBC and the number of edge states of the system under OBC \cite{zak1989berry,karoly2016short,PhysRevB.95.035421,chen2019zak}.
In this section, we will show that the winding number of the MSs, as defined above, can help us characterize the number of isolated edge states in non-Hermitian multiband systems,
even when EPs \GJ{are} present in the band structure and the Zak phase is ill-defined.
%Moreover, in the case of the presence of exceptional points, it behaves better than the Zak phase.
 \GJ{For} the rest of this paper, \GJ{we consider} the model of Eq. (\ref{eqn:model}) with $J=3,4$ as \GJ{representative} examples, with more demonstrations with larger $J$, i.e. $J=5$, are shown in Appendix \ref{sec:5band}.
 \begin{figure}[H]
        \begin{subfigure}[b]{0.24\textwidth}
                \includegraphics[width=\linewidth]{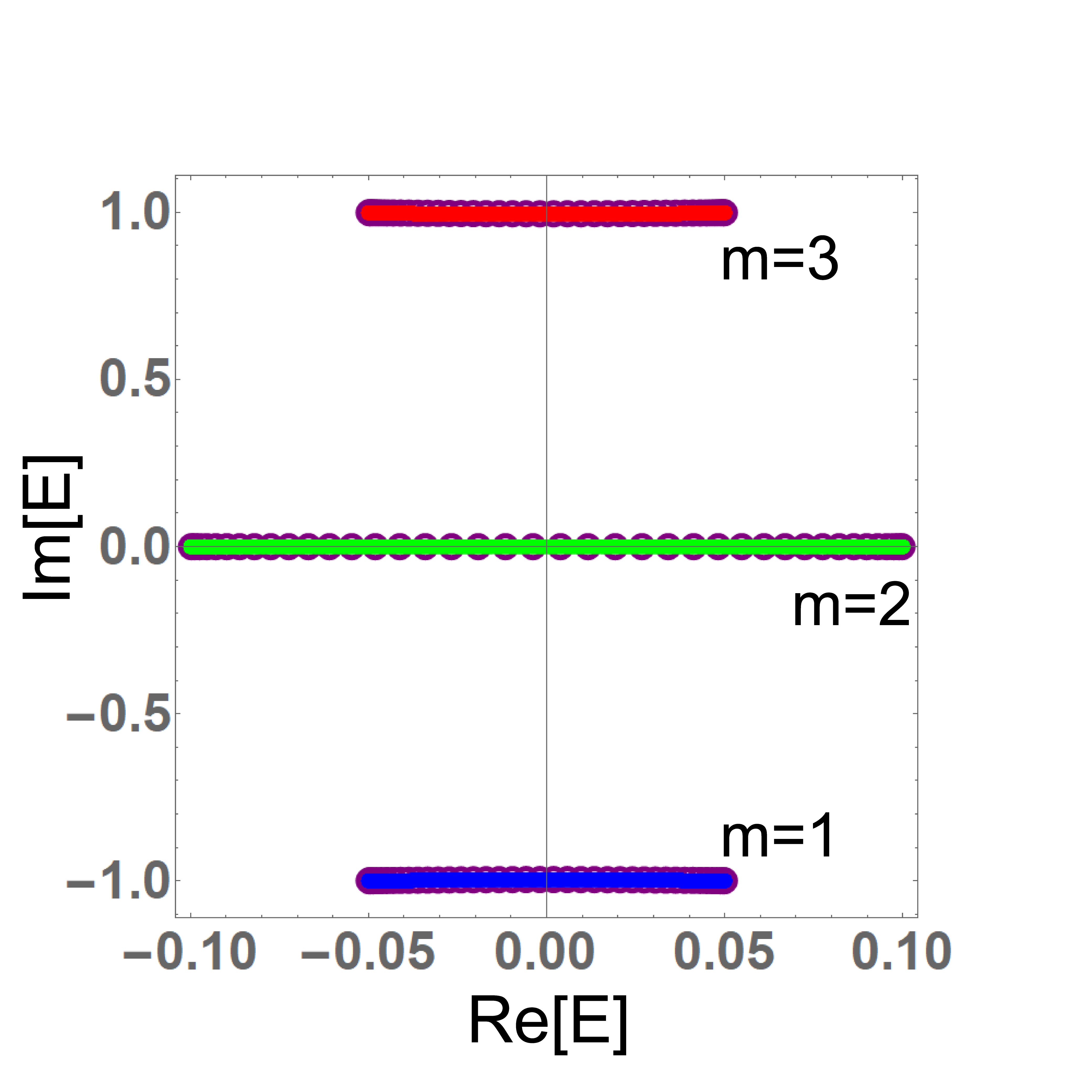}
                \caption{Trivial Case}
                \label{fig:energytrivial}
        \end{subfigure}%
        \begin{subfigure}[b]{0.24\textwidth}
                \includegraphics[width=\linewidth]{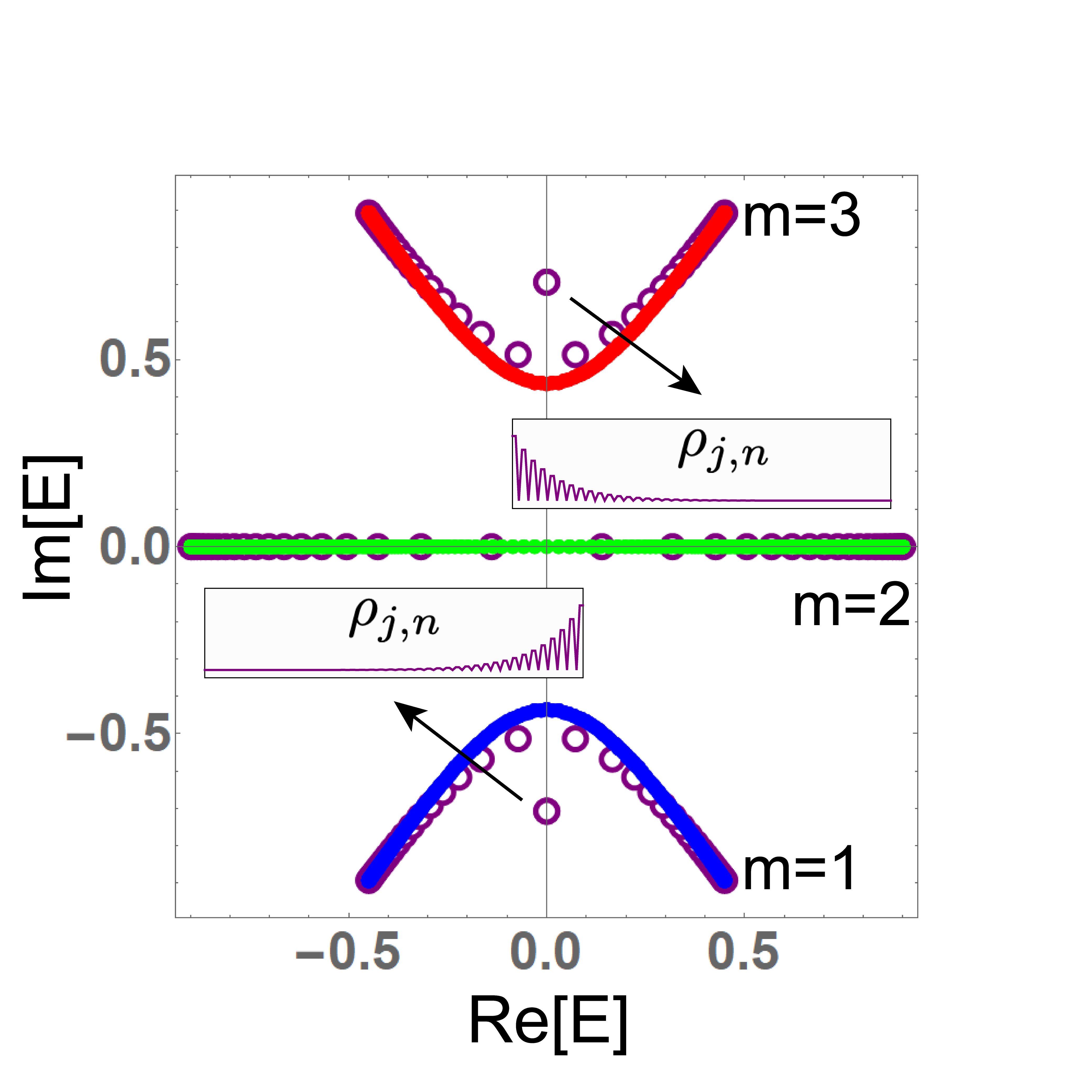}
                \caption{Non-trivial Case}
                \label{fig:energynontrivial}
        \end{subfigure}%
        \caption{(a) and (b) respectively show the energy spectra of the topologically trivial case ($t_3=0.1$) and non-trivial case ($t_3=0.9$). The purple circles indicate the energy spectrum under OBC, and the red, blue, and green curves indicate the three energy bands of the system under PBC. Parameters for both cases: $N=120$, $t_1=t_2=\sqrt{2}/2$, $\mu_1=\sqrt{2}$, $\mu_2=0$, $\mu_3=-\sqrt{2}$, $\delta_j=0$ for $j\in\{1,2,3\}$. Insets show the real space distributions $\rho_{j,n}=|\psi_{j,n}|^2$ of the two-fold coalescent edge states, with $\psi_{j,n}$ being the amplitude of their wavefunctions at lattice site $(j,n)$.}
        \label{fig:energy}
\end{figure}

\subsection{Reciprocal system with separable bands}
\label{sec:nse}
We first consider a reciprocal system with $\delta_j=0$ for $j\in\{1,2,3\}$, and non-Hermiticity is introduced solely from the imaginary on-site potential.
In Fig.~\ref{fig:energy}, we illustrate the PBC and OBC spectra for two typical cases with and without edge states isolated from the three separable continuous bands.

\begin{figure}[H]
    \begin{subfigure}[b]{0.48\textwidth}
        \includegraphics[width=\linewidth]{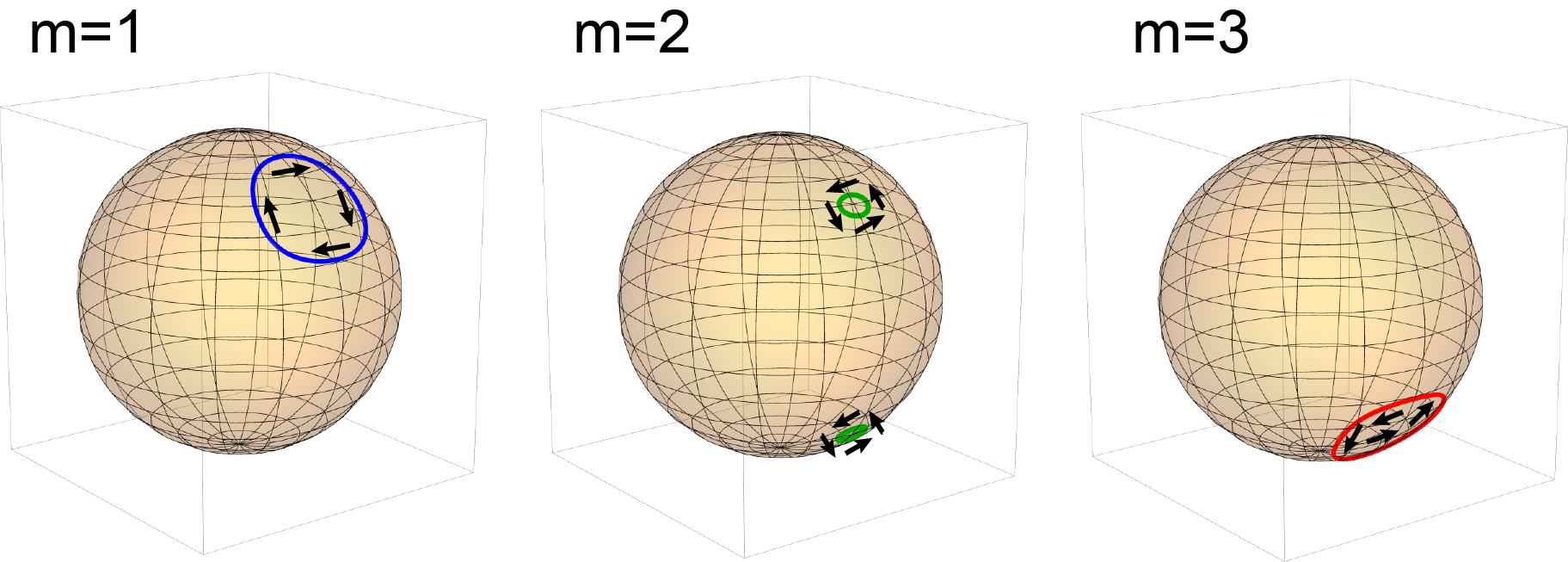}
        \caption{Trivial Case}
        \label{fig:b1t}
    \end{subfigure}%
    \vskip\baselineskip
    \begin{subfigure}[b]{0.48\textwidth}
        \includegraphics[width=\linewidth]{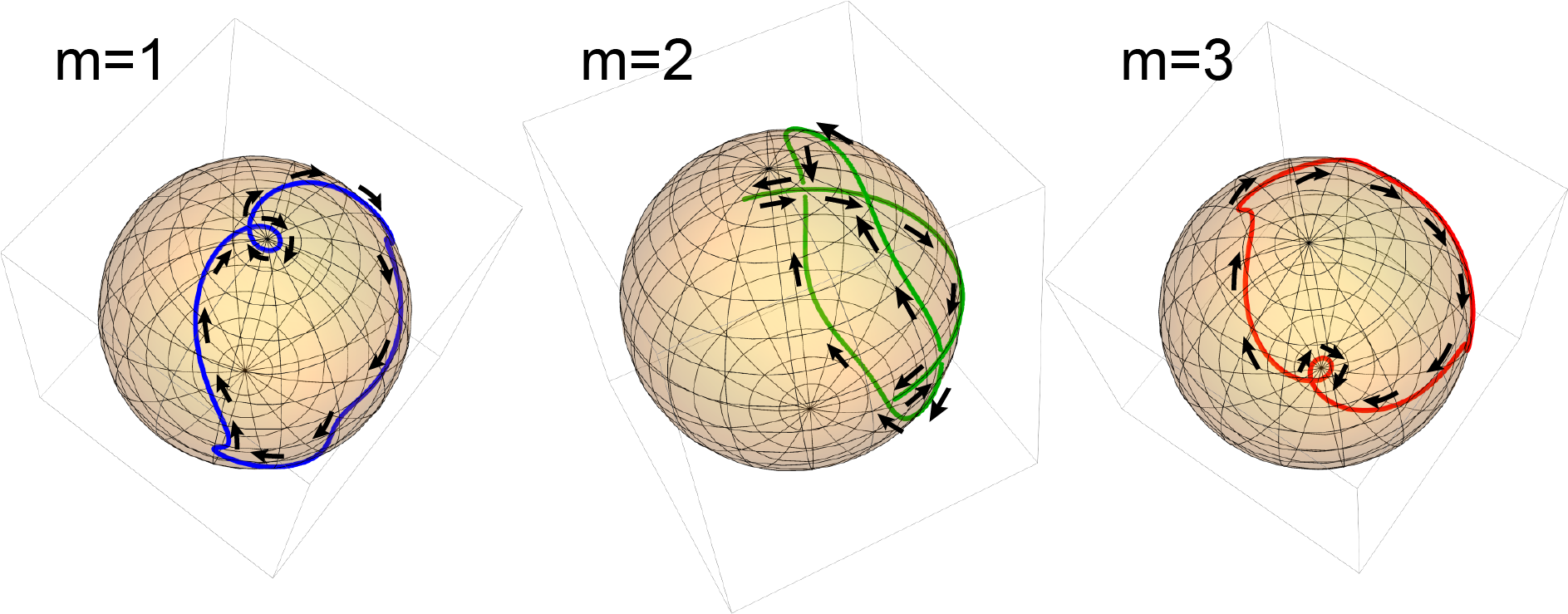}
        \caption{Non-trivial Case}
        \label{fig:b1n}
    \end{subfigure}%
    \caption{Trajectories of MSs for (a) the trivial case and (b) the nontrivial case of Fig. \ref{fig:energy}. \red{The arrows indicate the orientation of the MSs when $k$ runs over BZ.} In (a), each trajectory does not enclose the z-axis. In (b), the trajectories of MSs for the first and third bands enclose the z-axis, but those of the second band does not. {Note that in most cases there is only one closed loop on the sphere, which is formed by the trajectories of two MSs of a given band connecting to each other.}}
    \label{fig:msrtn}
\end{figure}

In both cases, the PBC spectrum follows well with the OBC spectrum, suggesting the absence of NHSE. Thus bulk-boundary correspondence between the PBC system and isolated edge states under OBCs is expected to hold.
%However, in the non-trivial case, there are 4 isolating energy outside the energy bands in the PBC spectrum, which indicates the presence of edge states in this case.
Indeed, for the case of Fig. \ref{fig:energy}(a) without any isolated edge state, it is found that the trajectories of MSs for each band do not enclose the z-axis [Fig. \ref{fig:msrtn}(a)], meaning that the winding number takes $\nu_m=0$ for $m=1,2,3$. On the other hand, MSs for two of the three bands in Fig. \ref{fig:energy}(b) wind around the z-axis, indicating a nontrivial topology and corresponding edge states of the system. Note that in most cases, the trajectories of a pair of MSs for a single band exchange with each other as $k$ varies from $0$ to $2\pi$, and go back to themselves at $k=4\pi$. In such cases, the winding number $\nu_m$ defined as summation over all MSs for the $m$th band is equivalent to the total winding of one of the MSs with $k$ varying from $0$ to $4\pi$.

\begin{figure}[H]
        \begin{subfigure}[b]{0.48\textwidth}
                \includegraphics[width=\linewidth]{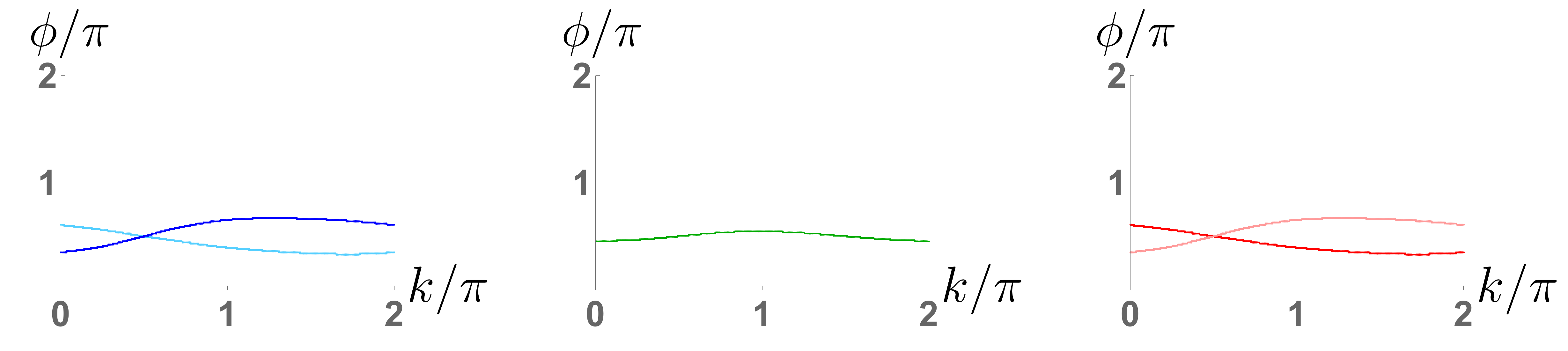}
                \caption{Trivial Case}
                \label{fig:phitrivial}
        \end{subfigure}%
        \vskip\baselineskip
        \begin{subfigure}[b]{0.48\textwidth}
                \includegraphics[width=\linewidth]{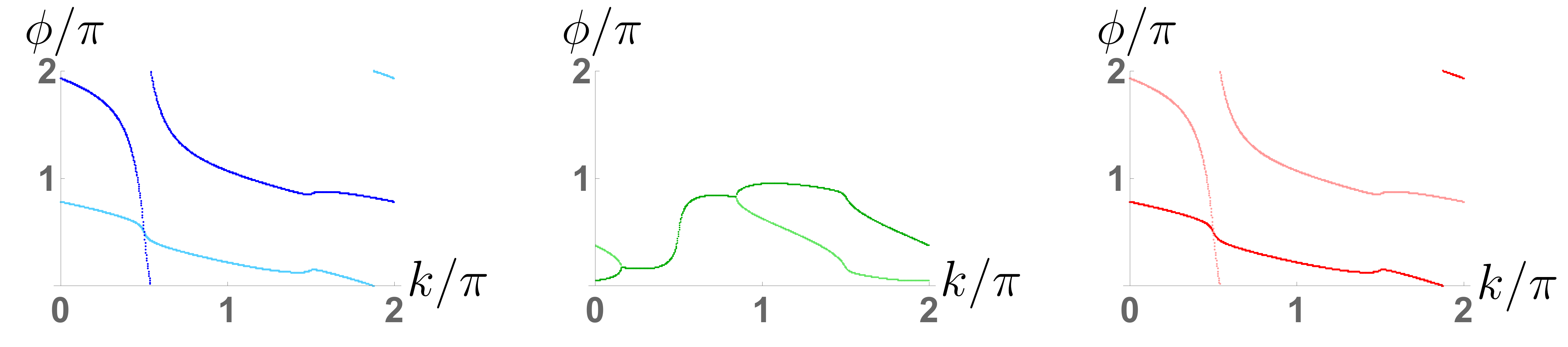}
                \caption{Non-trivial Case}
                \label{fig:phinontrivial}
        \end{subfigure}%
        \caption{(a) and (b) show the variation of angle $\phi$ of the MSs in the trivial and non-trivial cases respectively. In the trivial case, all three bands does not cross over $2\pi$, hence their winding numbers are 0 according to the definition (\ref{eqn:windinggen}). In contrast, the first band (blue) and the third band (red) in the non-trivial case cross over $2\pi$ twice, indicating that they have winding numbers $\nu_m=2$, but the winding number of the second band (green) is 0. In each panel, lighter and darker colors indicate the two MSs of each band. {In most cases, each MS does not go back to itself, but connects to the other MS when $k$ varies from $0$ to $2\pi$. The quantized winding number is thus given by the summed windings of all MSs for each band.}}
        \label{fig:phint}
\end{figure}

In Fig.~\ref{fig:phint}, we further illustrate the \GJ{azimuthal} angle $\phi$ for each MS as a function of  $k$. We can see that for band 1 (blue) and band 3 (red) of the topologically nontrivial case, $\phi$ changes by $4\pi$ as $k$ goes through the BZ twice, corresponding to winding numbers $\nu_{1,3}=2$. On the other hand, the second band in the same case has a winding number of $\nu_2=0$, which is the same for any band in the topologically trivial case in Fig. \ref{fig:energy}(a).  The total winding number of the nontrivial case is hence given by
\begin{eqnarray}
\sum_m \nu_m=4,
\end{eqnarray}
which agrees with the fact that we obtain two two-fold coalescent edge states isolated from the continuous bands in Fig.~\ref{fig:energy}(b).

Finally, \GJ{to understand if our proposed topological characterization at least covers the traditional Zak phase based approach}, we inspect the associated Zak phases by use of Eq. (\ref{eqn:zaknum}) for a comparison. Numerically, we obtain $\gamma^{(1,2,3)}=(0.0035,-0.0035,0)\pi$ for the topologically trivial case in Fig. \ref{fig:energy}(a), and $(0.1671, 3.8329,0)\pi$ for the topologically nontrivial case in Fig. \ref{fig:energy}(b).
Note that the Zak phases are not quantized for each \GJ{individual} band here, because the system we have chosen does not possess chiral symmetry nor inversion symmetry that protects a quantized Zak phase \red{and the edge states.}
Nevertheless, it has been shown that in Hermitian cases, the number of edge states can be related to the summation of Zak phases of all energy bands \cite{PhysRevB.95.035421,chen2019zak}. Consistent with this, we respectively obtain $\sum_m \gamma^{(m)}=0$ and $4\pi$ for the two cases here, which are in agreement with our defined winding number and the number of two-fold coalescent edge states. \GJ{This being the case, for the examples here our approach does not outperform the Zak phase approach yet}.

\subsection{Inseparable bands with EPs}
\label{sec:EP}
Since the energy spectrum of a non-Hermitian system is complex in general,
EPs can exist when different energy bands at the same quasi-momentum $k$ coincide on the complex energy plane. In the presence of EPs, the eigenstates of these bands coalesce into one, leading to an incomplete Hilbert space and a non-diagonalizable Hamiltonian.
\GJ{Such situations are of more interest to us, because the Zak phase cannot be properly defined for each individual band. The central question is then the following: in such cases with EPs, can our defined winding numbers be used to predict the number of edge modes isolated from continuous bands on the complex energy plane?}

\begin{figure}[H]
    \centering
    \includegraphics[width=0.3\textwidth]{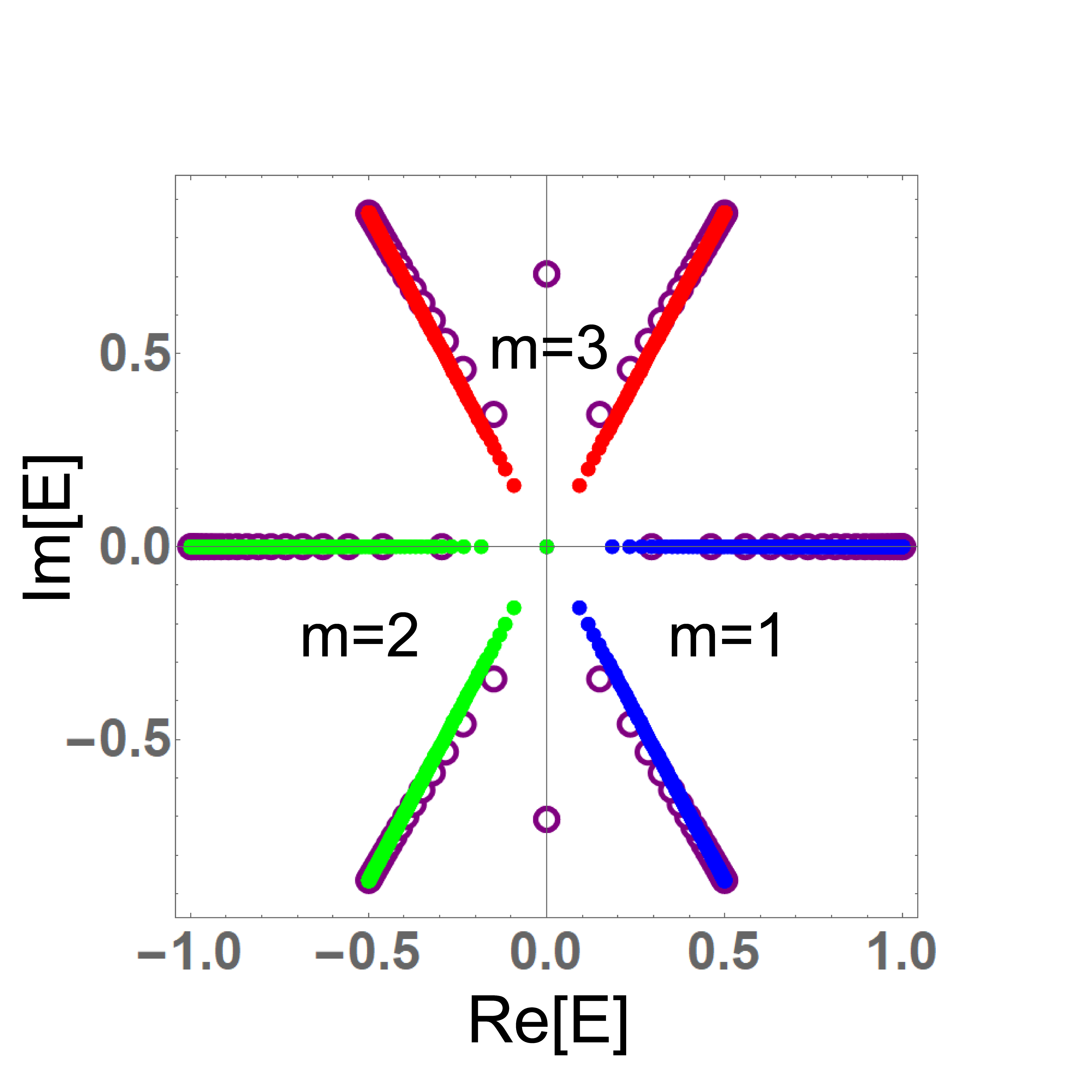}
    \caption{Purple circles show the energy spectrum under OBC, and the red, green and blue curves show the energy spectrum of the 3 bands under PBC. All three bands coincide with each other at energy 0, causing an EP in the system. \red{ The energy gap between the EP (energy 0) and the consecutive energy is large, because the energy changes abruptly near the EP, and hence the numerical calculation cannot fully capture the continuous spectrum near the EP.} Parameters are $N=120$, $t_1=t_2=\sqrt{2}/2$, $t_3=1$, $\mu_1=\sqrt{2}$, $\mu_2=0$, $\mu_3=-\sqrt{2}$ and $\delta_j=0$.}
    \label{fig:EP3}
\end{figure}

\begin{figure}[H]
    \centering
    \includegraphics[width=0.48\textwidth]{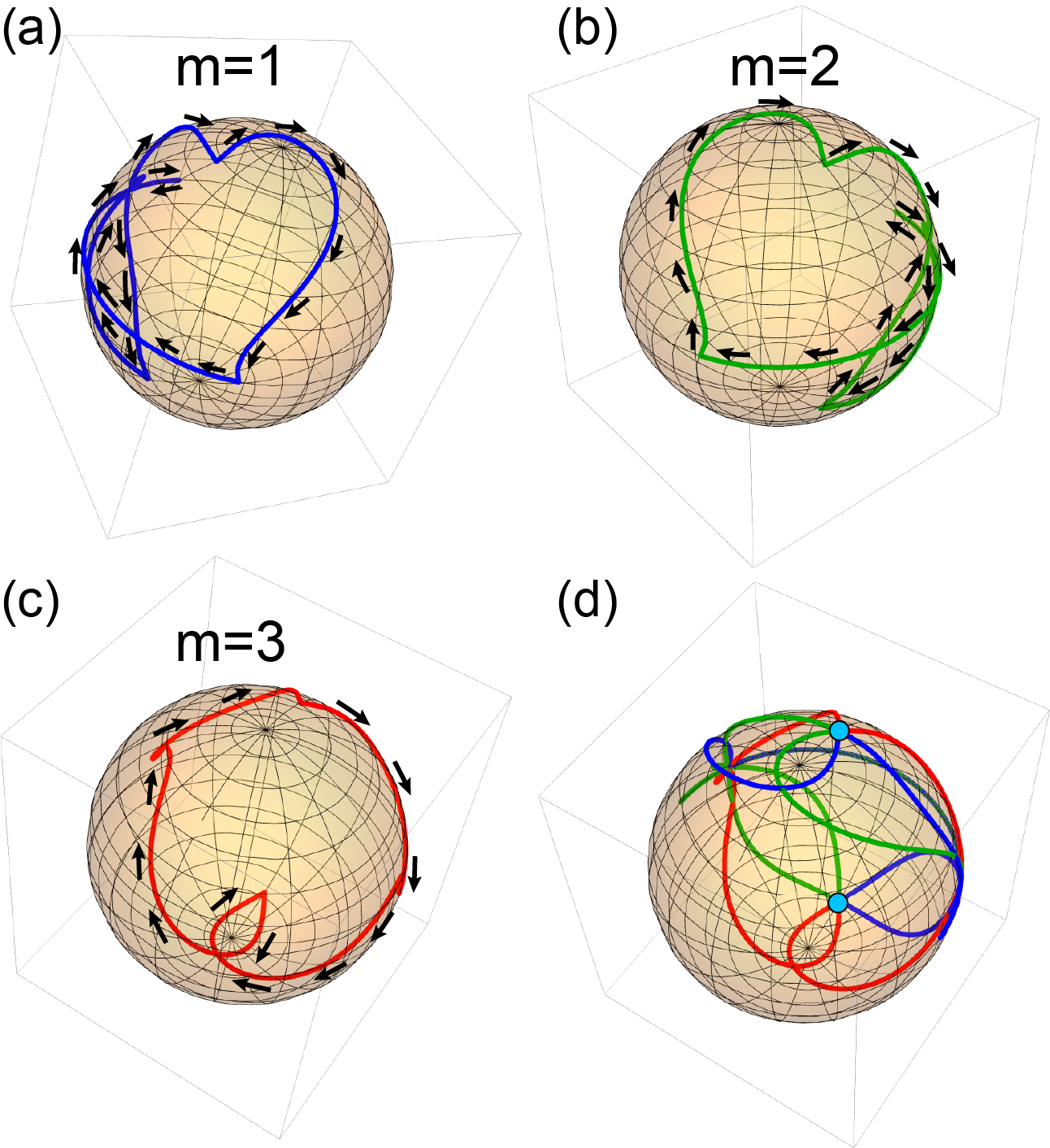}
    \caption{(a-c) MSR for each of the three bands, corresponding to $\nu_{1,2,3}=(1,1,2)$ respectively. (d) the MSR of all energy bands, where two EPs are marked with light blue dots. \red{The arrows indicate the orientation of the MSs when $k$ runs over BZ.}}
    \label{fig:EPmsr}
\end{figure}

In Fig.~\ref{fig:EP3}, the three energy bands coalesce at zero energy when $k=\pm\pi/2$, and two two-fold coalescent edge states exist, well separated from the continuous bands along the imaginary axis.
To see the topological origin of these coalesecent edge states, we plot the three bands by three different colors in Fig.~\ref{fig:EP3}, and then inspect the trajectories of their associated MSs of each band in
Fig.~\ref{fig:EPmsr}. It is found that the first two bands give winding numbers $\nu_m=1$, and the third band gives $\nu_m=2$.
Collectively, the summation of the winding number agrees with the summed total number of  all coalescence edge states.
By contrast, the calculated Zak phase summed over all energy bands for the situation here with EPs does not give a quantized value. \GJ{Clearly then, our topological characterization prevails but the Zak phase approach breaks down.}
{As a side note, it is not always straightforward to directly observe the winding from the trajectories of MSs on the Bloch sphere, especially when the system has many bands (see Appendix  \ref{sec:5bandep}).
Therefore, a numerical calculation through Eq. (\ref{eqn:windinggen}) should always be done to obtain the winding number.}

To further \GJ{illustrate the power of our method,}%To illustrate this idea further,
we provide a phase diagram by evaluating the summed winding number of MSs of the system with varying parameters $t_1=t_2$ and $t_3$,
as shown in Fig.~\ref{fig:phase} with different winding numbers represented by different colors.
{The region where one or more EPs presents is enclosed by solid black curves, \red{and covered by a translucent white layer on top,} which are obtained from [See Appendix \ref{sec:phaseboundary}]
\begin{eqnarray}
2t_1^2+t_3^2-2&=&0,\nonumber\\
\sqrt{\frac{2t_1^2+t_3^2-2}{3}}^3\frac{1}{t_1^2t_3}&=&1,\label{eq:phaseboundary}
\end{eqnarray}}
corresponding to the ellipse curve and the two almost straight lines in Fig.~\ref{fig:phase}, respectively. Across these lines, the number of EPs of the continuous bands changes between zero and nonzero, representing an EP phase transition.
In the absence of EP, the summed winding number agrees with the summed Zak phase (divided by $\pi$), as we have discussed in Sec.~\ref{sec:nse}.
When EPs present, the Zak phase is ill-defined but we can still use the summed winding number as a topology invariant to characterize the number of isolated edge states.
Note that there is a clear transition line (yellow) inside the region with the presence of EPs. Such a region with EPs is usually considered as a critical region lying between different topological phases. Interestingly, our results indicate that within such a critical region with EPs, there is a further boundary separating different topological phases. This transition is unique in non-Hermitian systems, as edge states may be isolated from inseparable continuous bands only when they possess complex energies.
%Therefore, the winding number of Majorana star behaves more powerful than the Zak phase in this scenario.

\red{Before we proceed to the next section, we would like to strengthen that there is no gap closing along the phase transition line (yellow line). As we have briefly mentioned in Sec.~\ref{sec:nse}, the Zak phases of the bands in the system are not quantized individually, suggesting that the edge states are not protected by a band gap but only have topological origin characterized by the winding number proposed here.
In fact,  even in Hermitian systems it was known that topological transitions and their associated changes in edge states could occur without gap closing  \cite{ezawa_tanaka_nagaosa_2013}.
Therefore, it is not surprising that the edge states here can emerge without the topological protection by the band gaps.}

\subsection{The presence of NHSE}
\label{sec:nhse}
Although NHSE is absent in our previous examples, it plays a crucial role in many non-Hermitian systems \cite{yao2018edge,lee2019anatomy,li2019geometric,PhysRevLett.123.246801,zhang2019correspondence} and can greatly affect our understanding of their topological properties.
With the presence of NHSE, bulk-boundary correspondence breaks down, and the energy spectrum under PBCs and OBCs behaves completely different \cite{yao2018edge,lee2019anatomy,lee2019unraveling}. %(cite
Therefore, we cannot directly apply the MSR method to the PBCs system and calculate the winding number to characterize isolated edge states under OBCs in this case.

To recover the bulk-boundary correspondence, we need to consider the so-called non-Bloch Hamiltonian $\Tilde{h}(k)=h(k+i\kappa)$ such that for a certain $\kappa$, the energy spectrum of $\Tilde{h}(k)$ does not form loops and reproduce OBC spectrum in the complex plane \cite{yao2018edge,lee2019anatomy,li2019geometric,zhang2019correspondence,lee2019unraveling}.
Therefore, the procedure to deal with a non-Hermitian system typically consists of the following steps in general:

\begin{figure}[H]
    \centering
    \includegraphics[width=0.5\textwidth]{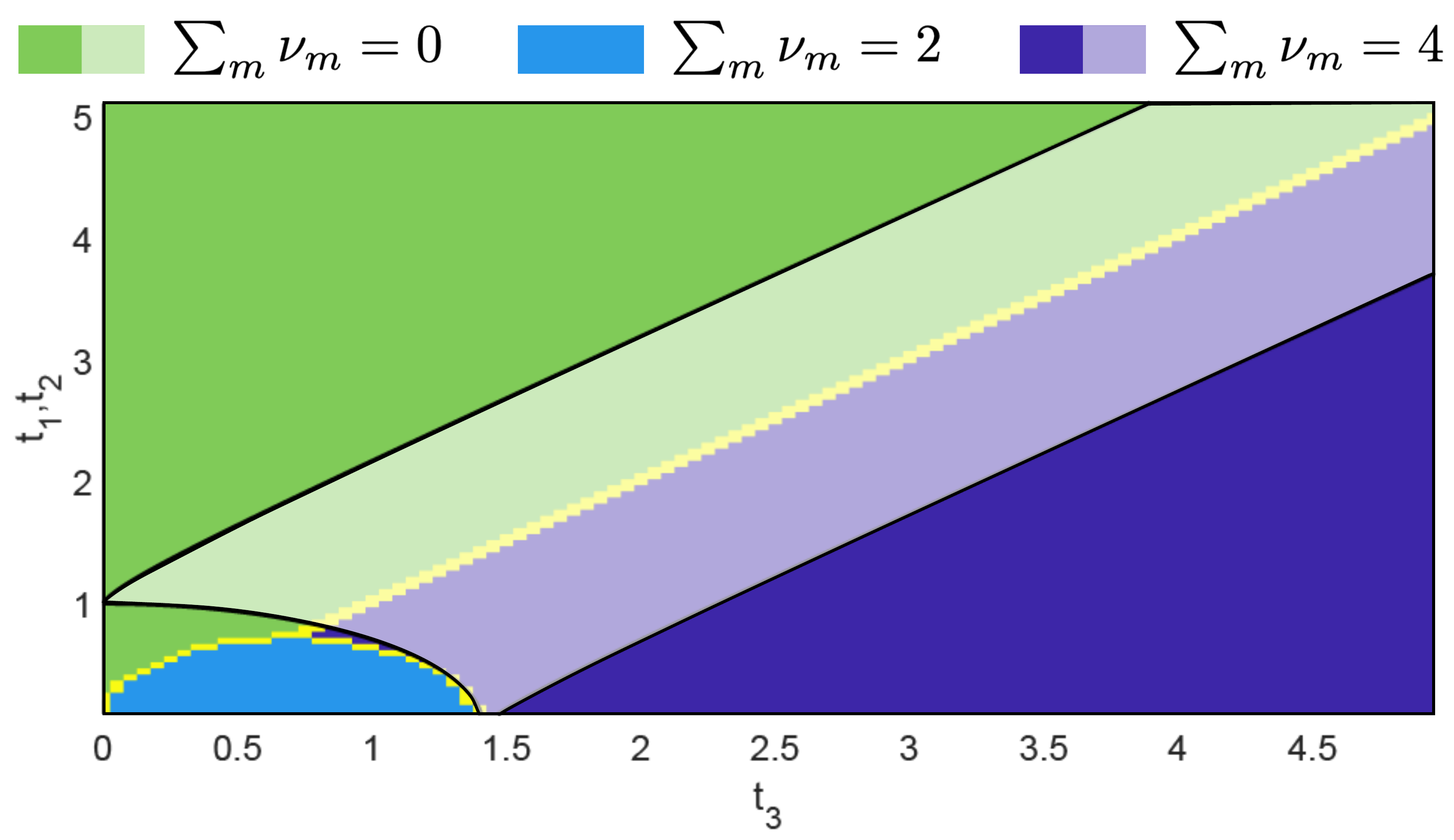}
    \caption{The phase diagram obtained by varying the parameters $t_1=t_2$ and $t_3$. The other parameters remain unchanged, as in the previous example in Fig.~\ref{fig:EP3}. {The winding numbers in different regions are labeled in the figure. Topological transitions occur when parameters vary across the yellow lines separating regions with different winding numbers. \red{EP phase transitions are given by the black solid lines, and the region enclosed is covered by a translucent white layer, where the system possesses one or more EPs of continuous bands in this region.}}}
    \label{fig:phase}
\end{figure}

\begin{enumerate}
    \item[(i)] Obtain the Bloch Hamiltonian of the system.
    \item[(ii)] \red{Calculate the energy spectrum under both PBC and OBC.}
    \item[(iii)] \red{Check the PBC energy spectrum. If it forms loops, it means that the conventional bulk-boundary correspondence breaks down, and one needs to calculate the GBZ described by a complex deformation of the quasimomentum $k\rightarrow k+i\kappa$ \cite{yao2018edge,yokomizo2019non,zhang2019correspondence,yang2019auxiliary}. With GBZ, the non-Bloch Hamiltonian $H(k+i\kappa)$, which recovers the OBC spectrum, can be then obtained. If the PBC energy spectrum does not form any loop, it means it is free from NHSE, and we can directly proceed to the next step.} 
    \item[(iv)] \red{From the non-Bloch Hamiltonian obtained, calculate the corresponding eigenstates over GBZ.}
    \item[(v)] Use the MSR equation (\ref{eqn:msr2}) to decompose the eigenstates, and then calculate the azimuthal angle of $\phi_{m,l}$.
    \item[(vi)] Calculate the winding number of MSs, which can then be used to indicate the number of isolated edge states of the system.
\end{enumerate}	

\begin{figure}[H]
        \begin{subfigure}[b]{0.16\textwidth}
                \includegraphics[width=\linewidth]{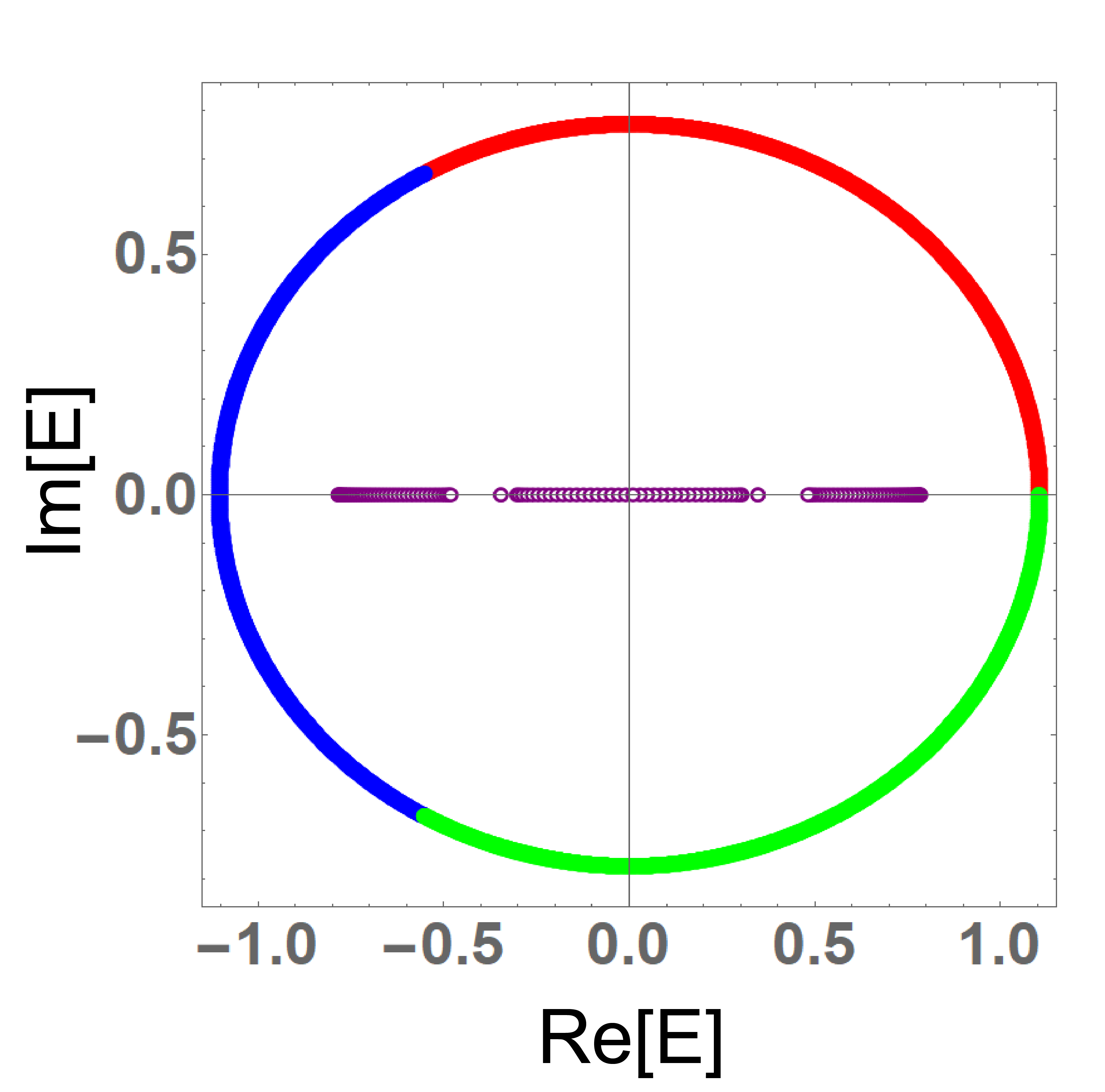}
                \caption{}
                \label{fig:SEenergy}
        \end{subfigure}%
        \begin{subfigure}[b]{0.16\textwidth}
                \includegraphics[width=\linewidth]{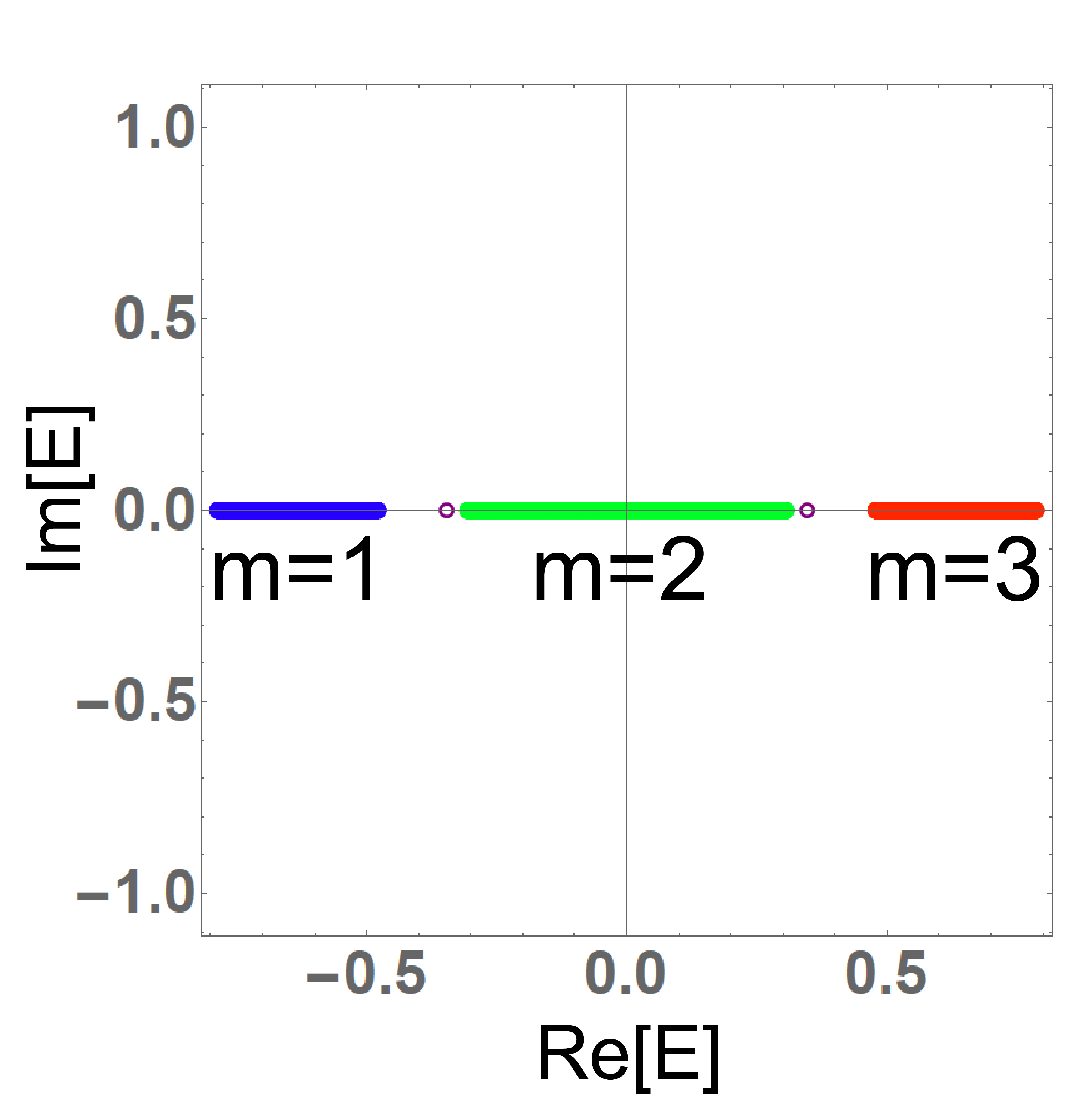}
                \caption{}
                \label{fig:SEenergyNB}
        \end{subfigure}%
        \begin{subfigure}[b]{0.16\textwidth}
                \includegraphics[width=\linewidth]{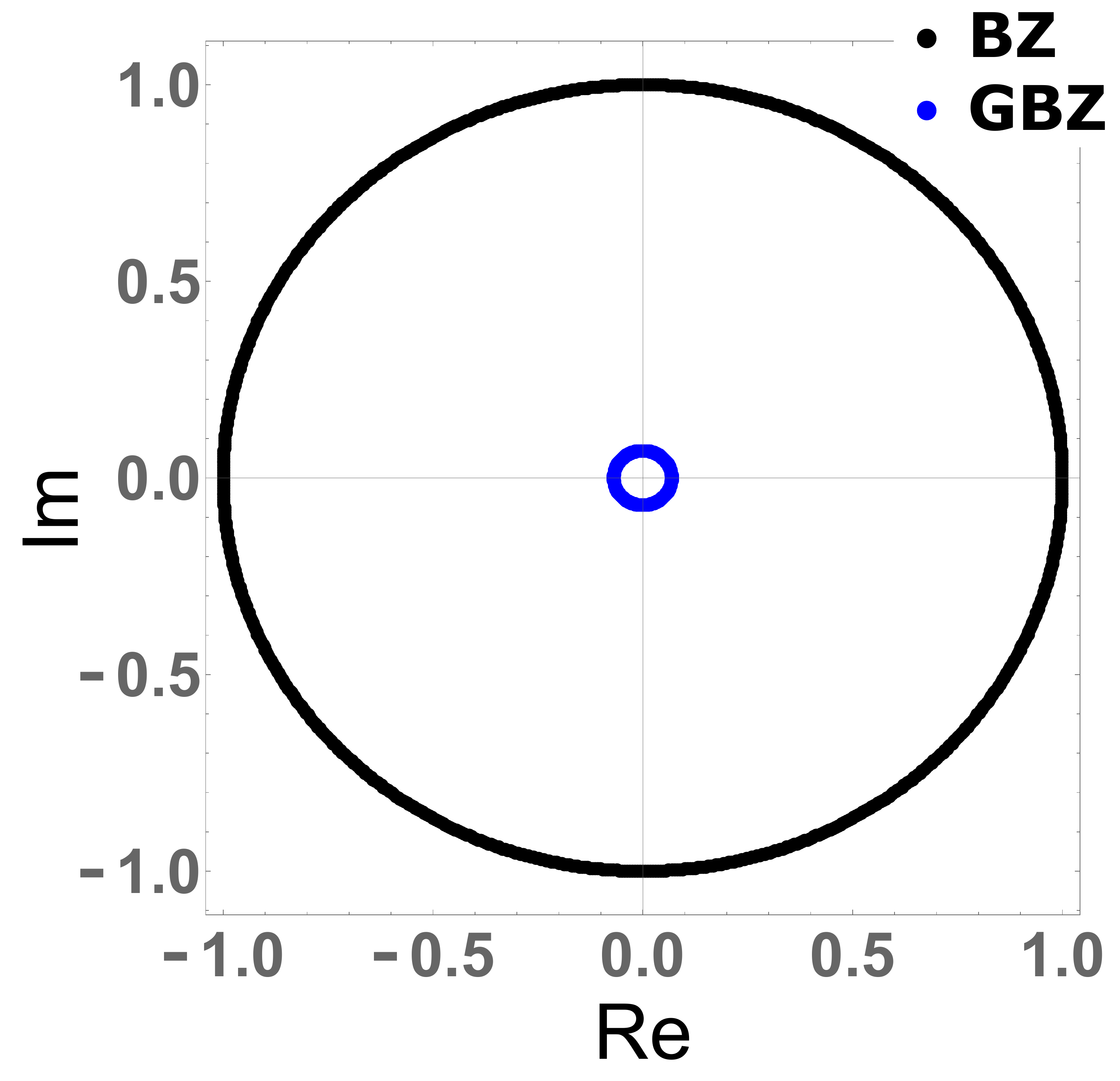}
                \caption{}
                \label{fig:gbz3}
        \end{subfigure}%
        \caption{\red{(a) and (b) show the energy spectra of both PBC and OBC for the Bloch Hamiltonian and non-Bloch Hamiltonian, respectively. Purple circles indicate the energy under OBC, and the red, green and blue curves show the energy under PBC. (c) shows both the BZ and the GBZ. In (a), energy spectrum under PBC forms a loop, showing the presence of the skin effect.  With GBZ, the $\kappa$ can be evaluated to be $2.67$, and the energy spectrum of the non-Bloch Hamiltonian under PBC recovers the OBC spectrum.} Other parameters for both cases are $N=120$, $t_1=t_2=0.4$, $t_3=1.2$, $\delta_1=\delta_2=0.2$, $\delta_3=1.1$ and $\mu_j=0$.}
        \label{fig:see}
\end{figure}

\begin{figure}[H]
    \centering
    \begin{subfigure}[b]{0.475\textwidth}
        \centering
        \includegraphics[width=\textwidth]{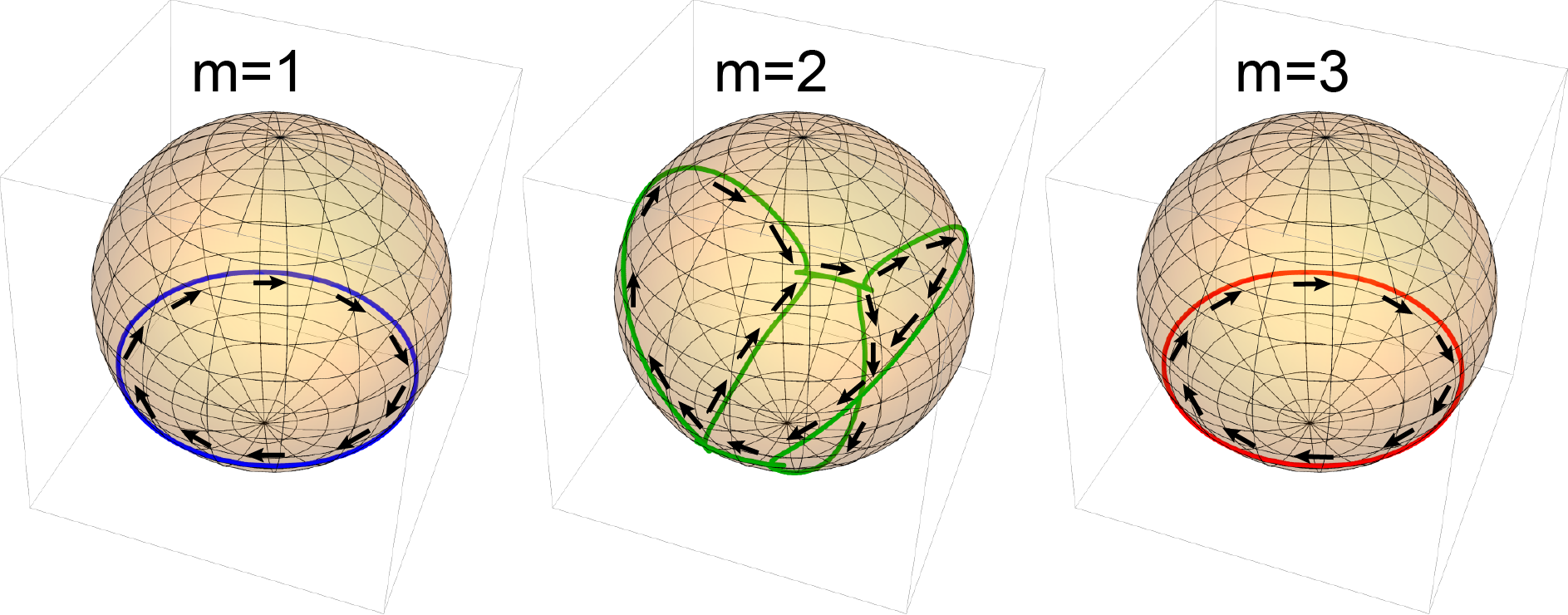}
        \caption{{\small MSR}}
        \label{fig:semsr1}
    \end{subfigure}
    \vskip\baselineskip
    \begin{subfigure}[b]{0.475\textwidth}
        \centering
        \includegraphics[width=\textwidth]{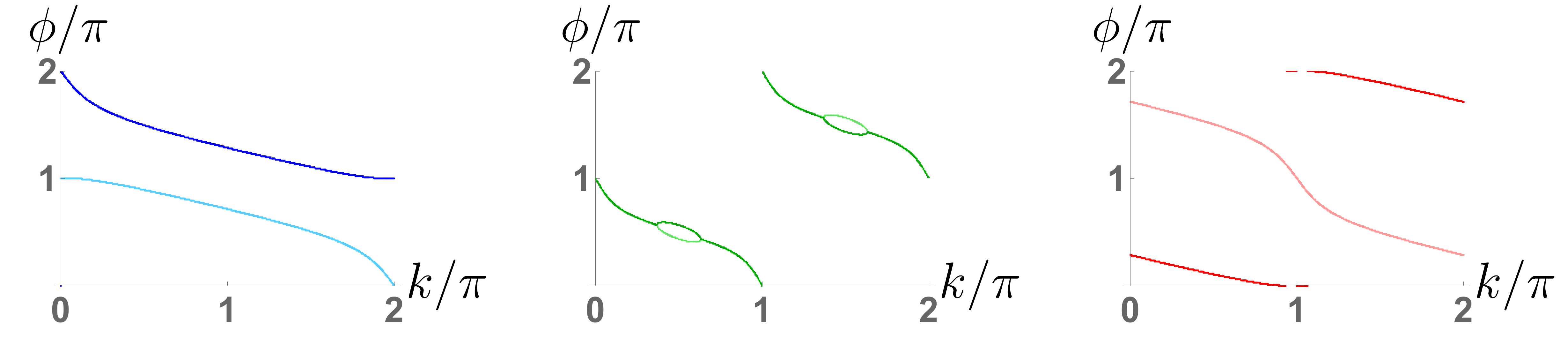}
        \caption{{\small Variation of $\phi$}}
        \label{fig:sephi}
    \end{subfigure}
    \caption{\small The three images in (a) and (b) are corresponding to the three bands (red, blue, and green in Fig.\ref{fig:SEenergyNB} respectively). \red{The arrows indicate the orientation of the MSs when $k$ runs over GBZ.} The two colors in each plot here indicate the two MSs for a given band. Their winding number are 1, 2, 1 respectively.}
    \label{fig:SEMSR}
\end{figure}

In Fig.~\ref{fig:SEenergy}, we illustrate an example with the presence of both NHSE and edge states isolated from the continuous OBC band. \red{We can evaluate the GBZ, as shown in Fig.~\ref{fig:gbz3}, and hence obtain the non-Bloch Hamiltonian at $\kappa\sim 2.67$ [Fig.~\ref{fig:SEenergyNB}], and bulk-boundary correspondence is expected to be restored.}
%reproduces the OBC spectrum (except for isolated edge states)
%We can numerically find the $\kappa = 2.67$ and calculate the energy spectrum of the corresponding non-Bloch Hamiltonian, which is shown in Fig.\ref{fig:SEenergyNB}.
%In this case, the energy spectrum under PBC well matches the one under OBC, so the bulk-boundary correspondence is recovered.
We then calculate the MSR for the non-Bloch Hamiltonian, and the results are shown in Fig.~\ref{fig:SEMSR}.
Both the summed winding number and the summed Zak phase are 4, in agreement with the fact that there are  two two-fold coalescent edge states in this case.
Therefore, the bulk-boundary correspondence is fully recovered and our characterization with the winding number of the MSs remains valid for the non-Bloch Hamiltonian.

%\begin{figure}[H]
%    \centering
%    \includegraphics[width=0.3\textwidth]{gbz3.pdf}
%    \caption{The Brillouin Zone (BZ) and the generalized Brillouin Zone (GBZ) of  }
%    \label{fig:gbz3}
%\end{figure}

\begin{figure}[H]
        \begin{subfigure}[b]{0.16\textwidth}
                \includegraphics[width=\linewidth]{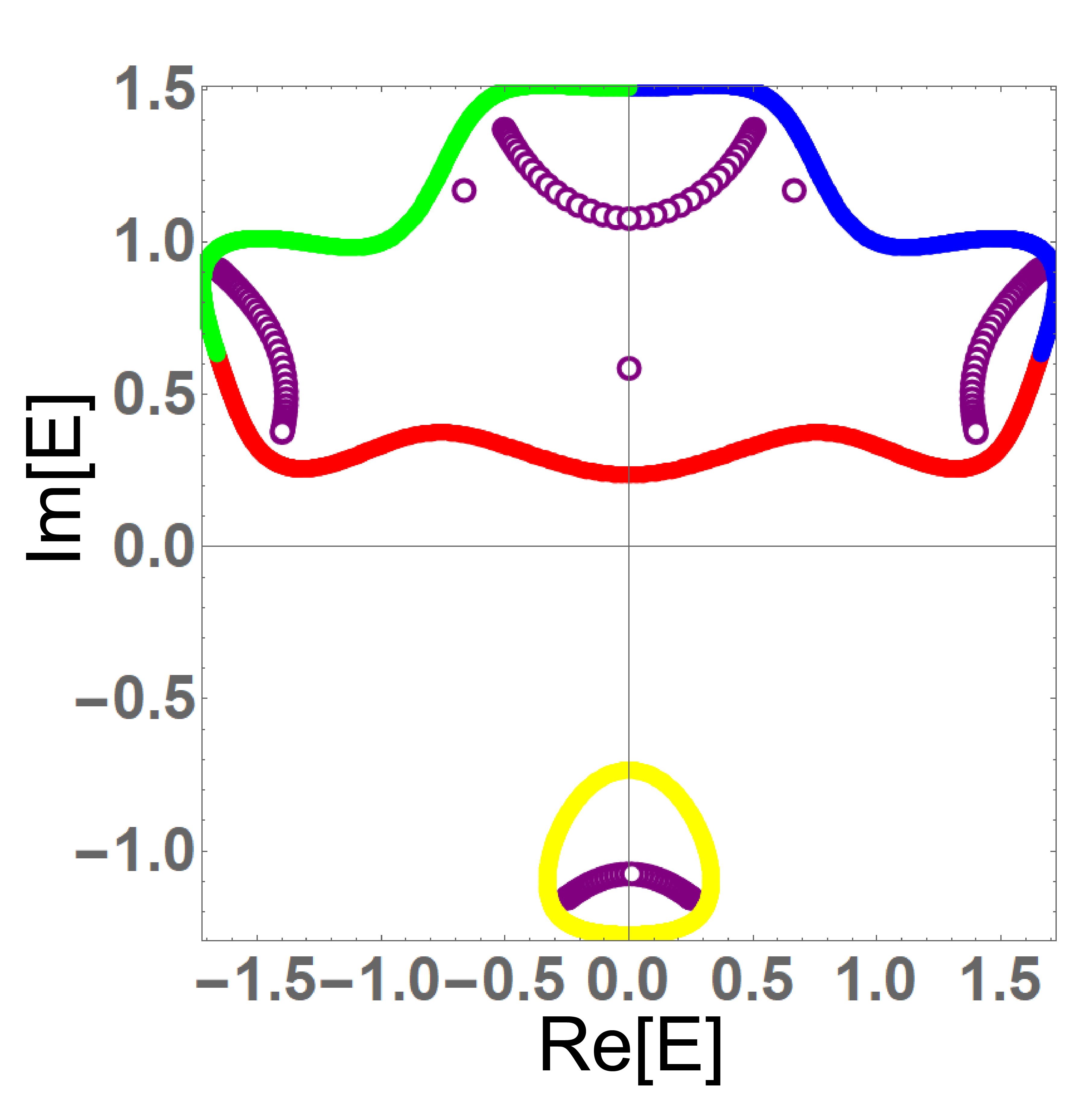}
                \caption{}
                \label{fig:4bandbloch}
        \end{subfigure}%
        \begin{subfigure}[b]{0.16\textwidth}
                \includegraphics[width=\linewidth]{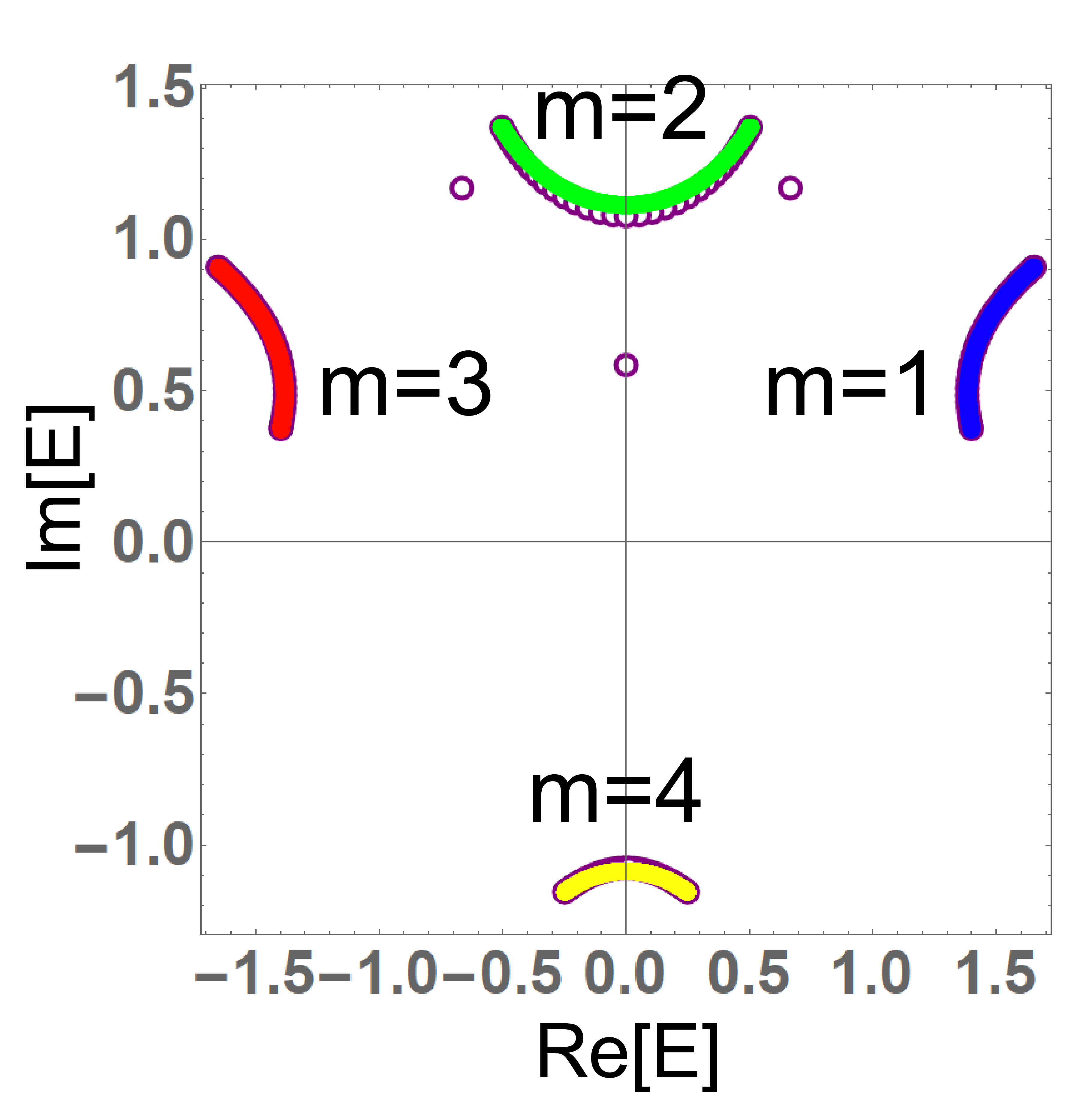}
                \caption{}
                \label{fig:4bandnonbloch}
        \end{subfigure}%
        \begin{subfigure}[b]{0.16\textwidth}
                \includegraphics[width=\linewidth]{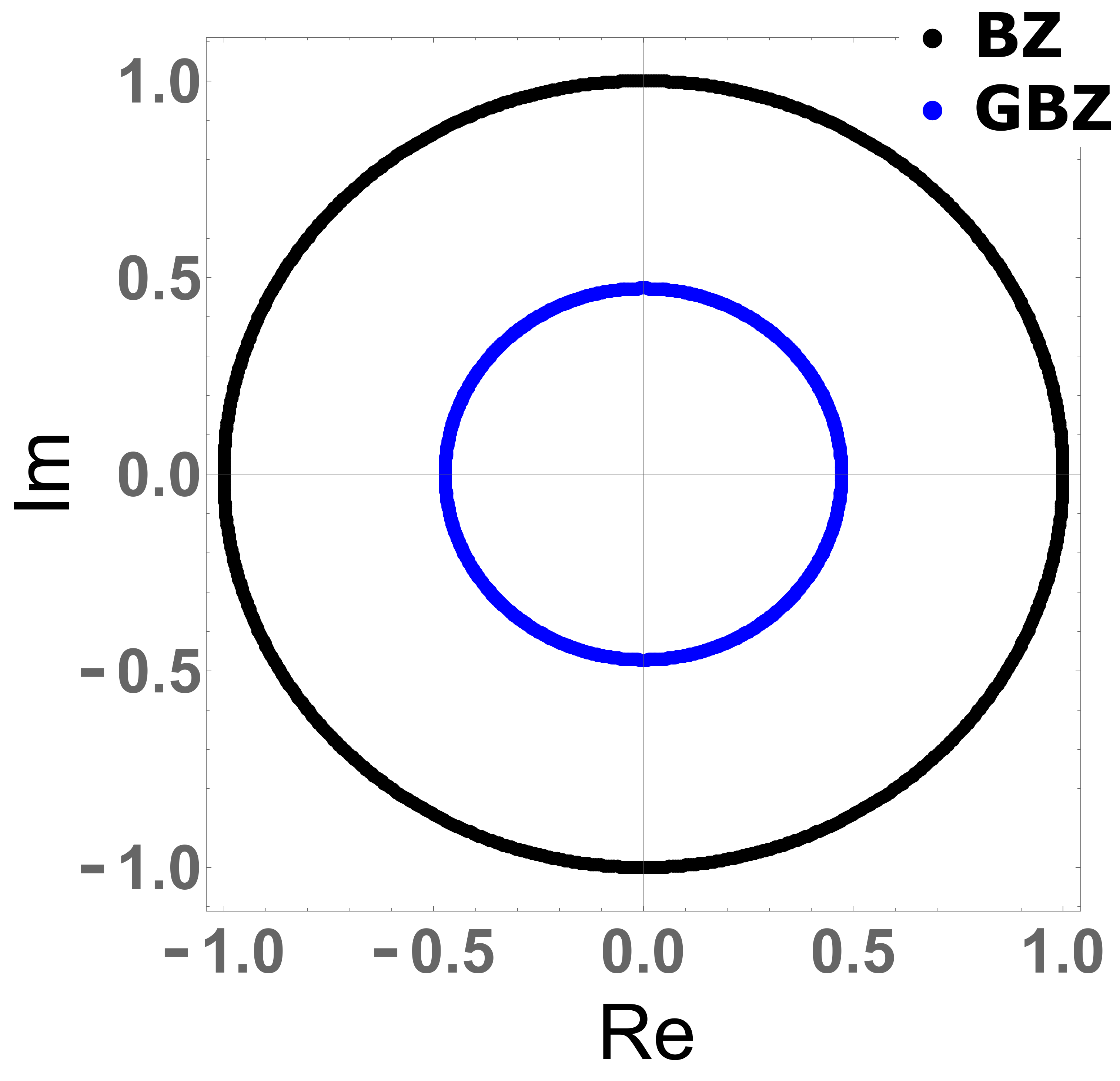}
                \caption{}
                \label{fig:gbz4}
        \end{subfigure}%
        \caption{(a) and (b) show the energy spectra of the Bloch Hamiltonian and non-Bloch Hamiltonian respectively. \red{(c) shows both the BZ and the GBZ. The $\kappa$ of the non-Bloch Hamiltonian is found to be 0.75.} Parameters are $t_1=1$, $t_2=0.8$, $t_3=1$, $t_4=1.2$, $\delta_1=0.2$, $\delta_2=1.7$, $\delta_3=0.2$, $\delta_4=0.2$, $\mu_1=0.5$, $\mu_2=-0.3$, $\mu_3=0.4$, $\mu_4=0.9$.}
        \label{fig:4band}
\end{figure}

{\subsection{4-band system with 3 isolated edge states}}
\label{sec:3es}
Besides the cases with EPs, let us examine a case where the system has an odd number of isolated edge states, \GJ{a situation that can never be connected with the Zak phase. In particular},
we consider below a 4-band case from our model with $J=4$, and in certain parameter regime, this case yields three non-coalescent edge states isolated from the continuous bands, as shown in Fig.~\ref{fig:4bandbloch}. NHSE also presents in this case as the PBC Bloch bands form some closed loops and are distinguished from the OBC bands. \red{Following the procedure in Sec.~\ref{sec:nhse}, we can obtain $\kappa\sim0.75$ for the non-Bloch Hamiltonian to recover the OBC spectrum.  This is illustrated in Fig.~\ref{fig:4bandnonbloch}.}

\begin{figure}[H]
    \centering
    \includegraphics[width=0.48\textwidth]{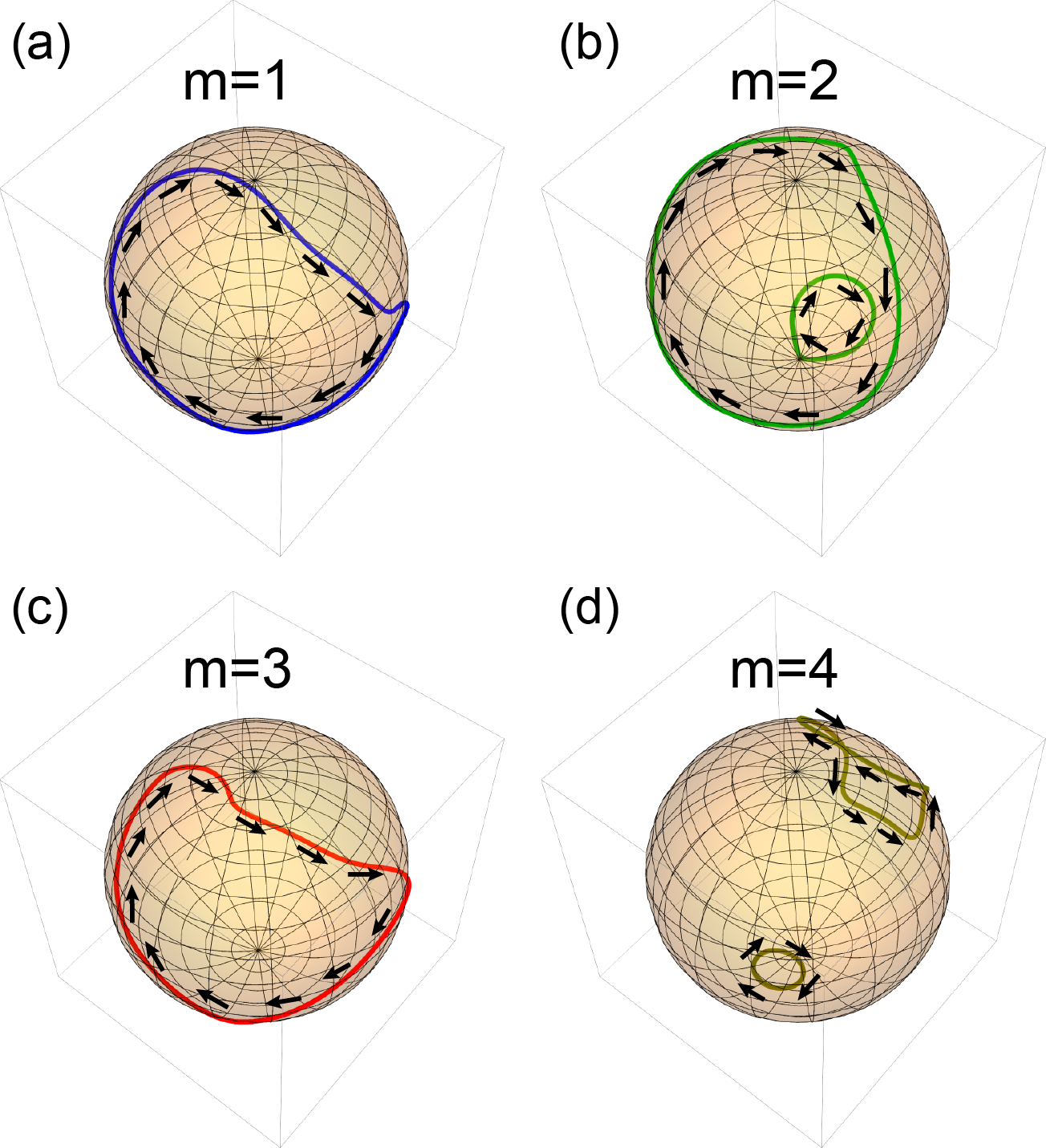}

    \caption{The MSR of the 4 bands. (a-c) correspond to the 3 band on top in Fig.\ref{fig:4bandnonbloch}, and the bottom right correspond to the 1 band at the bottom in Fig.\ref{fig:4bandnonbloch}. \red{The arrows indicate the orientation of the MSs when $k$ runs over GBZ.} Their winding numbers are 1,1,1,0 respectively.}
    \label{fig:3ESMSR}
\end{figure}

By applying Eq.~(\ref{eqn:msr2}), we obtain the corresponding MSRs, and the obtained results are shown in Fig.~\ref{fig:3ESMSR}. It is seen that the winding number is $\nu_m=1$ for $m=1,2,3$, and $0$ for $m=4$. This observation is also verified by our direct numerical approach to $\nu_m$. Thus the summation of the winding numbers reflects the total number of isolated edge states in this case.

Remarkably, the Zak phases summed over all energy bands \GJ{of non-Bloch Hamiltonian} may quantize to an odd multiple of $\pi$ only when the system goes around an exceptional degeneracy \GJ{as we scan $k$ from 0 to $2\pi$}, the precise situation where the system must have NHSE \cite{mailybaev2005geometric,li2019geometric}. However, \GJ{by construction} the NHSE has been ``removed" from the non-Bloch Hamiltonian through the complex deformation $k\rightarrow k+i\kappa$. That is, the summed Zak phase can only take an even multiple of $\pi$ for a non-Bloch Hamiltonian, hence cannot reflect the number of isolated edge states in this case. Indeed, numerically, we obtain $\gamma^{(1,2,3,4)}=(1.0128,1.9608,1.0128,0.0136)\pi$, which sum to a quantized value of $4\pi$, but our system has only three edge states!  \GJ{We have thus demonstrated again that our topological characterization is superior to the Zak phase approach.}

%The MSR method can be used in the system with more bands. Here, we provide an example of a non-Hermitian 4-band system.
%We consider the parameters: $t_1=1$, $t_2=0.8$, $t_3=1$, $t_4=1.2$, $\delta_1=0.2$, $\delta_2=1.7$, $\delta_3=0.2$, $\delta_4=0.2$, $\mu_1=0.5$, $\mu_2=-0.3$, $\mu_3=0.4$, $\mu_4=0.9$, and hence obtain the energy spectrum as shown in Fig.\ref{fig:4bandbloch}.
%We notice that there is the presence of the skin effect, so we numerically find the $\kappa=0.75$ and obtain the energy spectra for the non-Bloch Hamiltonian which is shown in Fig.\ref{fig:4bandnonbloch}.
%We can count that there are only 3 energy under OBC that are isolated from the energy spectrum under PBC, indicating that we have 3 isolated non-coalescent edge states.
%We can decompose the eigenstate by using the MSR equation (\ref{eqn:msr2}) and the results are shown in Fig.\ref{fig:3ESMSR}.

\section{Conclusions} \label{sec:conclusion}
\GJ{In summary, we have used the MSRs to decompose the eigenstates of a non-Hermitian multiband system such that we can visualize them on the Bloch sphere.
We have proposed a winding number of the MSs as a new topological invariant. Our defined topological invariant successfully characterizes the number of isolated edge states of non-Hermitian multiband systems under OBC.
As a marked result, our topological characterization is effective  even in the presence of EPs. Indeed, the winding numbers of the MSs are more generally applicable than the Zak phase, with the latter being ill-defined as the Hamiltonian becomes non-diagonalizable.
Via our proposed topological invariant, we are able to predict the change in the number of isolated edge states and hence identify topological transitions in a parameter region where EPs are present and the band are inseparable.
We have further applied our method to examples with non-Hermitian skin effects, and again verified that the bulk-boundary correspondence between the isolated edge states and the topological invariant we define can be restored by considering a non-Bloch Hamiltonian, obtained by a complex shifting of the quasi-momentum.
 Along this line, we have also discussed an example with an odd number of isolated edge states, where the Zak phase necessarily fails to predict the number of isolated edge states, whereas our topological invariant again agrees with the number of edge states.
 Such extraordinarily wide applicability of our approach  may trigger further studies and bring us more insights into the interplay between exceptional degeneracies and topological properties.
 }

{\it Note added.}
\GJ{During the final stage of our manuscript preparation, we became aware of a preprint \cite{xu2020ms}, which uses MSRs to study a three-band non-Hermitian Lieb lattice model, with a very different focus}.

\section*{Acknowledgements} \label{sec:acknowledgements}
J.G. acknowledges fund support by  the  Singapore  NRF  Grant  No. NRF-NRFI2017-04 (WBS No. 
R-144-000-378-281)  and  by  the  Singapore Ministry  of  Education  Academic  Research  Fund  
Tier-3 Grant No. MOE2017-T3-1-001 (WBS. No. R-144-000-425-592). X.Z. is supported by National Natural Science Foundation of China (Grant No. 11975166) before he joined NUS.
L.L. would like to thank Chao Yang for helpful discussions.
\appendix

\begin{center}
{\bf Appendix}
\end{center}

%\section{Linkage between the Zak phase and the winding number of Majorana stars formed by the right eigenstate and the left eigenstate}
\section{Non-Hermitian Zak phase and the winding number of the right or left eigenstates for two-band systems}
\label{sec:2band}
A two-band non-Hermitian system can be described by the Hamiltonian
\begin{eqnarray}
H=\sum_k h_x(k)\sigma_x+h_y(k)\sigma_y+h_z(k)\sigma_z+h_0(k)\mathbb{I},
\end{eqnarray}
with $\sigma_{x,y,z}$ the Pauli matrices acting on a pseudospin-1/2 space, and $\mathbb{I}$ the corresponding $2\times2$ identity matrix.
Its left and right eigenstates are given by
\begin{eqnarray}
|\psi^R_+\rangle=\left(\begin{matrix}
e^{-i\phi}\cos\frac{\theta}{2}  \\
\sin\frac{\theta}{2}
\end{matrix}\right),~~|\psi^R_-\rangle=\left(\begin{matrix}
e^{-i\phi}\sin\frac{\theta}{2} \\
-\cos\frac{\theta}{2}
\end{matrix}\right)
\end{eqnarray}
and
\begin{eqnarray}
\langle\psi^L_+|=\left(\begin{matrix}
e^{i\phi}\cos\frac{\theta}{2}  \\
\sin\frac{\theta}{2}
\end{matrix}\right)
^T,~~\langle\psi^L_-|=\left(\begin{matrix}
e^{i\phi}\sin\frac{\theta}{2}  \\
-\cos\frac{\theta}{2}
\end{matrix}\right)
^T,
\end{eqnarray}
with $\cos\theta=h_z/\sqrt{h_x^2+h_y^2+h_z^2}$ and $\cos\phi=h_x/\sqrt{h_x^2+h_y^2}$, and $\pm$ denoting the two bands. Note that we have omitted the dependence of $k$ for simplicity.
Without loss of generality, we consider the Zak phase of the lower band, which is given by
\begin{eqnarray}
\gamma^{(-)}={\rm Re}\oint_{BZ}\frac{\partial \phi}{\partial k}\sin^2\frac{\theta}{2}dk.
\end{eqnarray}
In Hermitian systems with a chiral symmetry $\sigma_z h(k)\sigma_z=-h(k)$, we have $\theta=0$, and the Zak phase is given by half of the winding angle of $\phi$ throughout Brillouin zone, corresponding to a quantized winding number. Degenerate topological edge states exist whenever the winding angle is nonzero.
In the absence of a chiral symmetry (and other symmetries that protect 1D topology),
The degeneracy of edge states will be lifted, and they are no longer protected by topology. Nevertheless, the existence of these edge states can still be characterized by the winding number of the phase angle $\phi$ \cite{mong2011edge}, which indicates a topological origin of them despite the lack of topological protection.

For non-Hermitian systems, both angle parameters take complex values in general, thus we rewrite them as $\phi=\phi_r+i\phi_i$ and $\theta=\theta_r+i\theta_i$, with $\phi_{r,i}$ and $\theta_{r,i}$ being real. Nevertheless, due to the periodicity of the Brillouin zone, imaginary phase factors $\theta_i$ and $\phi_i$ must go back to their original values when $k$ varies a period, and so does $\theta_r$ as it is originated from the altitude angle in the Hermitian limit.
Therefore the winding corresponding to the Zak phase is solely given by the winding of $\phi_r$.

\red{In the main text, we have applied MSR to decompose the right eigenstates to the MSs only, containing no information of the left eigenstates. }
In the simplest two-band picture, each band consists of only one MS, directly given by the right eigenstates. Therefore we shall rewrite the right eigenstates with the orthogonal normalization condition, which yields
\begin{eqnarray}
|\psi'_+\rangle=\left(\begin{matrix}
e^{-i\phi'}\cos\frac{\theta'}{2}  \\
\sin\frac{\theta'}{2}
\end{matrix}\right),~~|\psi'_-\rangle=\left(\begin{matrix}
e^{-i\phi'}\sin\frac{\theta'}{2} \\
-\cos\frac{\theta'}{2}
\end{matrix}\right)
\end{eqnarray}
with $\phi'$ and $\theta'$ being real phase parameters. $|\psi'_\pm\rangle$ and $|\psi^R_\pm\rangle$ shall be equivalent up to an overall coefficient.
Taking the eigenstates of ``$-$'' band as an example, requiring $|\psi'_-\rangle=c_-|\psi^R_-\rangle$ with $c_-$ a coefficient, it is straightforward to obtain
\begin{eqnarray}
e^{-i\phi'}&=&e^{-i(\phi_r+\frac{\pi}{2})}e^{\phi_i}c_-(e^{-\frac{\theta_i}{2}}e^{i\frac{\theta_r}{2}}-e^{\frac{\theta_i}{2}}e^{-i\frac{\theta_r}{2}}),\nonumber\\
\cos \frac{\theta'}{2}&=&\frac{c_-}{2}(e^{\frac{-\theta_i}{2}}e^{\frac{i\theta_r}{2}}+e^{\frac{\theta_i}{2}}e^{\frac{-i\theta_r}{2}}).
\end{eqnarray}
We can see that for each individual point in the Brillouin zone, $\phi'$ is given by $\phi_r$ plus $\pi/2$ and some phases given by $c_-$ and $\theta_{i,r}$.
However, neither of these extra phases may yield a nonzero winding over a period, as $c_-$ dependents only on $\theta_{i,r}$, which must go back to themselves as discussed earlier. Therefore, with $k$ varying from $0$ to $2\pi$, $\phi'$ and $\phi_r$ must give the same winding number, reflecting the number of isolated edge states as demonstrated in the main text.
\red{Note that we always express the decomposed MSs to be $\tan\frac{\theta}{2}e^{i\phi}$ with real phases $\theta$ and $\phi$. Therefore, in the main text, we do not encounter the complex phases that we have discussed earlier in this section.}

%However, it is possible to consider complex phases for MSs. To have complex phases, an extended Bloch sphere is required and different geometrical properties of the phases on the extended Bloch sphere should be considered. For instance, people has studied a 2-band system with chiral symmetry, where only one circle of the Bloch sphere matters, and the circle is extended to a torus to characterize the complex phase \cite{yang2019visualizing}.

Finally, we note that though we have only considered the right eigenstates in above discussion, this analysis also applies to left eigenstates.

\section{Examples in 5-band systems}
\label{sec:5band}
MSR can also be applied to multiband systems with even more bands. 
Here, we provide two examples with separable bands and inseparable bands in a 5-band system that is free from the skin effect (i.e. $\delta_j=0$ for $j\in\{1,2,3,4,5\}$).

\subsection{Separable bands}
\begin{figure}[H]
    \centering
    \includegraphics[width=0.3\textwidth]{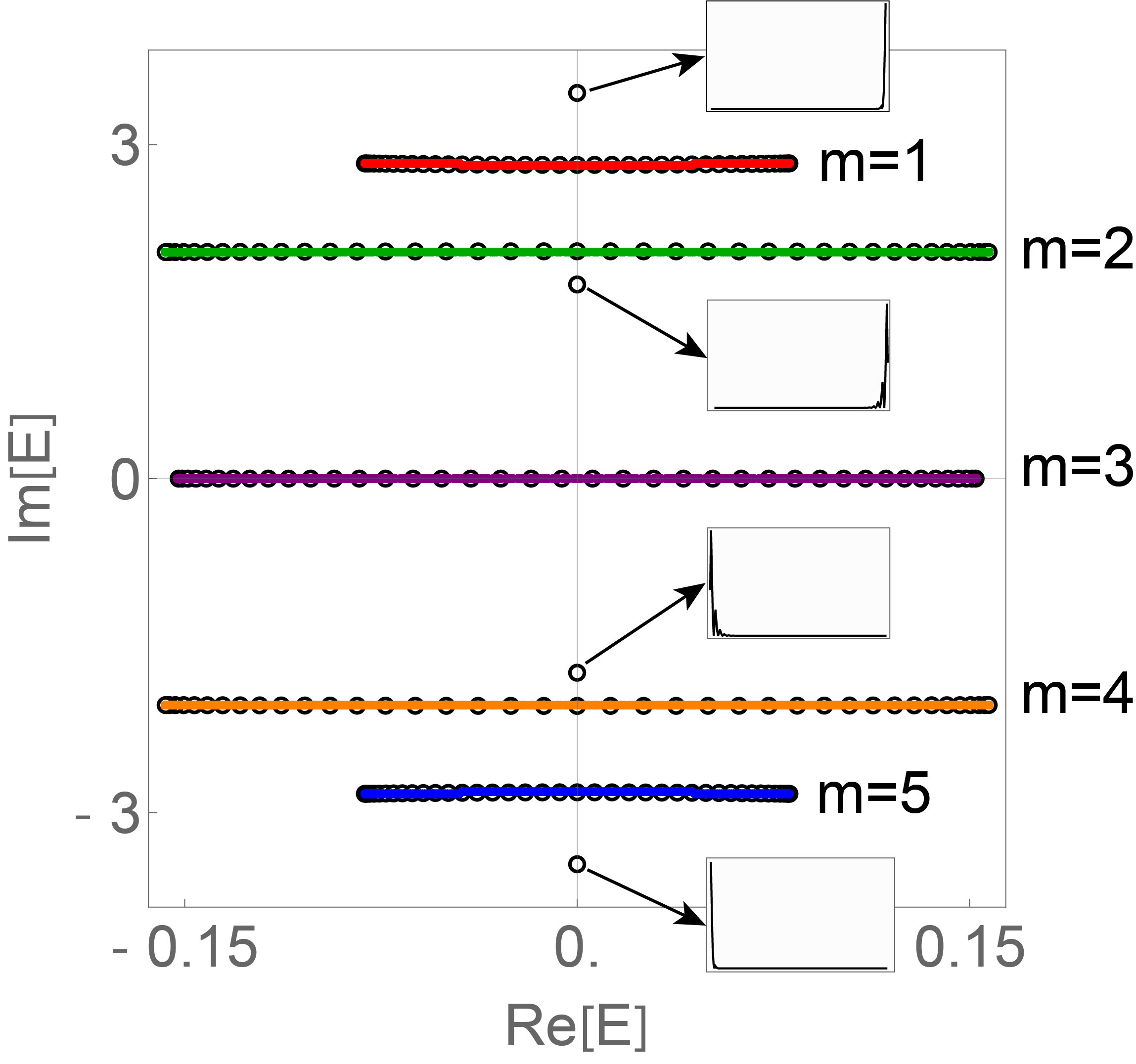}
    \caption{Black circles show the energy spectrum under OBC, and the red, green, blue, orange and purple curves show the energy spectrum of the 5 bands under PBC. Parameters are $N=200$, $t_1=t_4=2$, $t_2=t_3=\sqrt{6}/2$,  $t_5=0.9$, $\mu_1=-\mu_5=4$, $\mu_2=-\mu_4=2$, $\mu_3=0$ and $\delta_j=0$. Insets show the real space distribution of the edge states.}
    \label{fig:5bandnoepenergy}
\end{figure}

Considering the parameters as follow: $t_1=t_4=2$, $t_2=t_3=\sqrt{6}/2$, $t_5=1.7$, $\mu_1=-\mu_5=4$, $\mu_2=-\mu_4=2$ and $\mu_3=0$.
The energy spectrum is shown in Fig.~\ref{fig:5bandnoepenergy}.

\begin{figure}[H]
    \centering
    \includegraphics[width=0.3\textwidth]{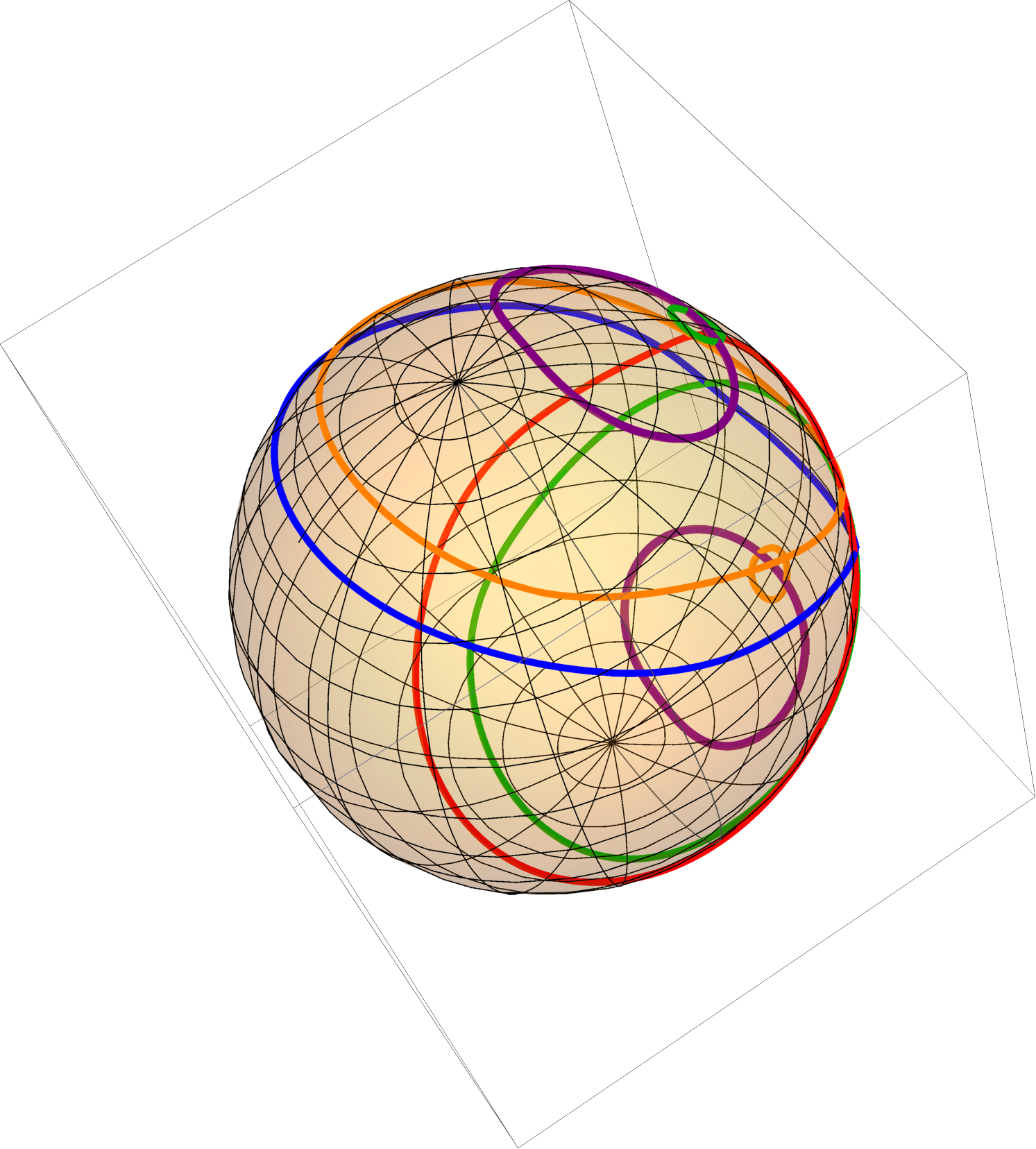}
    \caption{The MSR of all bands are combined in the same Bloch sphere, in which the colors are corresponding to the energy bands in Fig.\ref{fig:5bandnoepenergy}}
    \label{fig:5bandnoepmsr}
\end{figure}

Four non-coalescent edge states isolated from the continuous bands can be observed in the energy spectrum of Fig.~\ref{fig:5bandnoepenergy}.
By obtaining the MSR of this model and plotting the stars on the Bloch sphere, we can also observe that the winding number is $\nu_m=1$ for $m=1,2,4,5$ and $\nu_3=0$, as shown in Fig.~\ref{fig:5bandnoepmsr}.

\subsection{Inseparable bands with EPs}
\label{sec:5bandep}
\begin{figure}[H]
    \centering
    \includegraphics[width=0.3\textwidth]{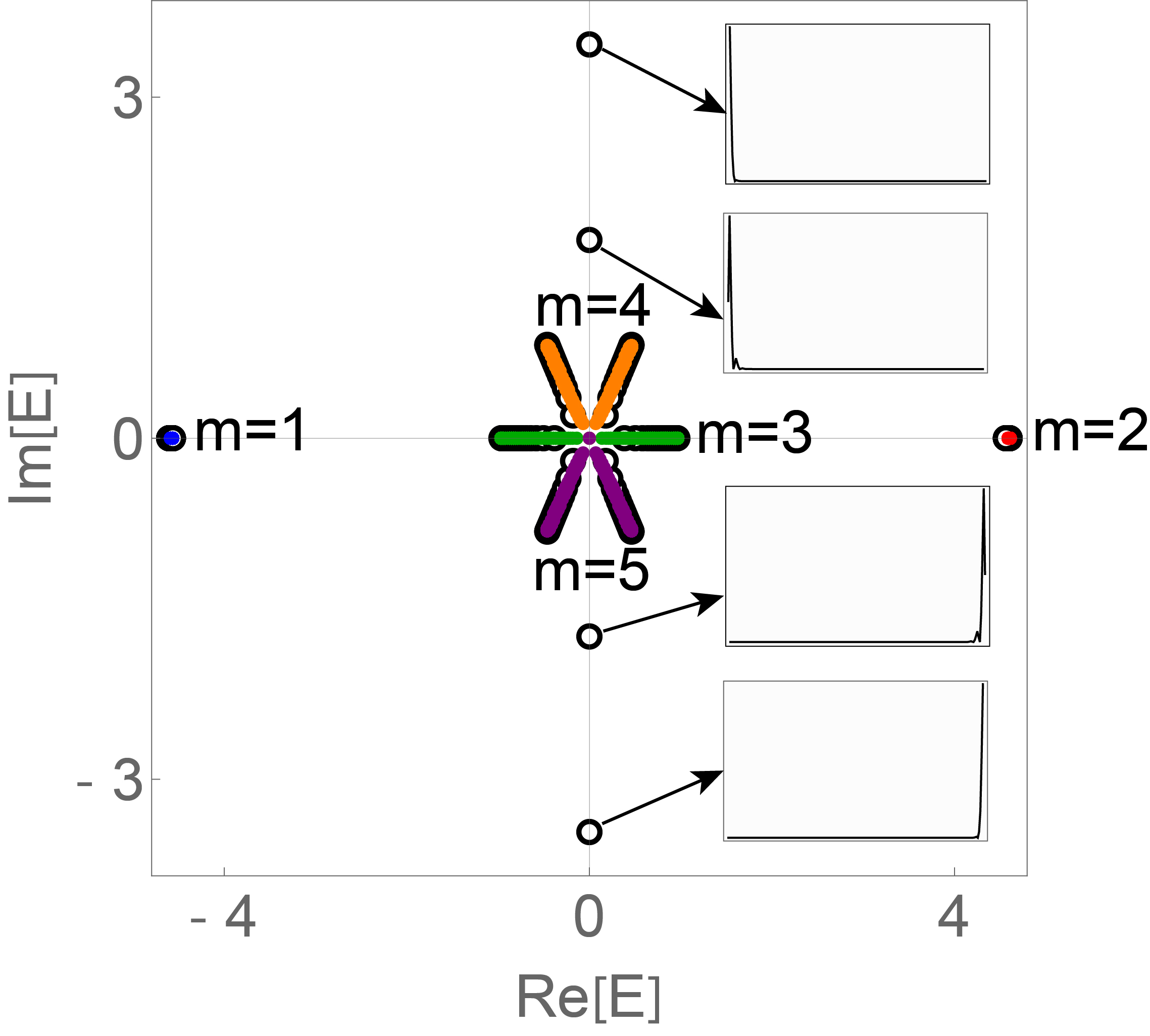}
    \caption{Black circles show the energy spectrum under OBC, and the red, green, blue, orange and purple curves show the energy spectrum of the 5 bands under PBC. Parameters are $N=200$, $t_1=t_4=2$, $t_2=t_3=\sqrt{6}/2$,  $t_5=6$, $\mu_1=-\mu_5=4$, $\mu_2=-\mu_4=2$, $\mu_3=0$ and $\delta_j=0$. . Insets show the real space distribution of the edge states.}
    \label{fig:5bandepenergy}
\end{figure}

Considering $t_5=6$ and the rest of the parameters remains the same as in the previous section.
The energy spectrum is shown in Fig.~\ref{fig:5bandepenergy}. There are also 4 isolating edge states in this case as shown in the figure. {Note that the two bands with $m=1,2$ (blue,red) are pinned at two points on the complex energy plane, corresponding to two flat bands throughout the BZ}.

\begin{figure}[H]
    \centering
    \includegraphics[width=0.3\textwidth]{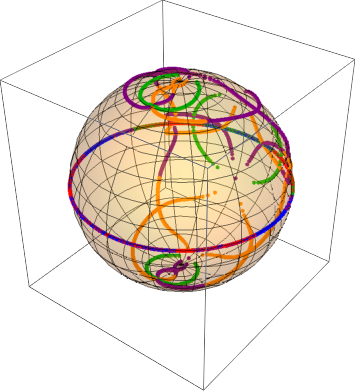}
    \caption{The MSR of all bands are combined in the same Bloch sphere, in which the colors are corresponding to the energy bands in Fig.\ref{fig:5bandepenergy}.}
    \label{fig:5bandepmsr}
\end{figure}

However, if we solve the MSR and plot it in the Bloch sphere, the stars are rather messy in this case, as shown in Fig.\ref{fig:5bandepmsr}.
It is hard to directly obtain the winding information from the plot in higher multiband systems with the presence of EPs. Nevertheless, the winding numbers can still be obtained numerically  through Eq. (\ref{eqn:windinggen}).
In this specific example, the winding numbers are $\nu_m=0$ for $m=1,2$, $\nu_m=1$ for $m=4,5$, and $\nu_3=2$.

\section{solutions of EP phase boundaries}
\label{sec:phaseboundary}
The eigenenergies of our system under PBC can be obtained from the eigenequation $\det[h(k)-E_k]=0$ of Eq. (\ref{eqn:tband}) in the main text, yielding
\begin{eqnarray}
E_k^3-(2t_1^2+t_3^2-2)E_k-2t_1^2t_3\cos k = 0\label{eq:det}
\end{eqnarray}
for the parameters we considered in Fig. \ref{fig:phase}.
In the presence of EPs, this equation can be written as $(E_k-E_1)^2(E_k-E_2)=0$. Compared with the coefficients of Eq. (\ref{eq:det}), we have
%\begin{eqnarray}
%E_2+2E_1&=&0,\\
%E_1^2+2E_1E_2&=&2-2t_1^2-t_3^2,\\
%E_1^2E_2&=&t_1^2t_3\cos k,
%\end{eqnarray}
\begin{eqnarray}
3E_1^2&=&2t_1^2+t_3^2-2,\\
E_1^3&=&-t_1^2t_3\cos k.
\end{eqnarray}
Therefore, to have a real solution of $k$ (and hence EPs in the BZ), $-E_1^3/t_1^2t_3$ must takes a real value between $-1$ and $1$, and we can obtain the following two conditions:
\begin{eqnarray}
2t_1^2+t_3^2-2\geqslant0
\end{eqnarray}
to give a real value of $E_1$, and
\begin{eqnarray}
|\cos k|=|\sqrt{\frac{2t_1^2+t_3^2-2}{3}}^3\frac{1}{t_1^2t_3}|\leqslant1.
\end{eqnarray}
Thus the EP phase boundaries in Fig. \ref{fig:phase} is obtained when the above inequalities takes the equal sign, i.e. Eqs. (\ref{eq:phaseboundary}) in the main text.

%\bibliographystyle{apsrev4-1}
%\bibliography{reference}

\begin{thebibliography}{86}%
\makeatletter
\providecommand \@ifxundefined [1]{%
 \@ifx{#1\undefined}
}%
\providecommand \@ifnum [1]{%
 \ifnum #1\expandafter \@firstoftwo
 \else \expandafter \@secondoftwo
 \fi
}%
\providecommand \@ifx [1]{%
 \ifx #1\expandafter \@firstoftwo
 \else \expandafter \@secondoftwo
 \fi
}%
\providecommand \natexlab [1]{#1}%
\providecommand \enquote  [1]{``#1''}%
\providecommand \bibnamefont  [1]{#1}%
\providecommand \bibfnamefont [1]{#1}%
\providecommand \citenamefont [1]{#1}%
\providecommand \href@noop [0]{\@secondoftwo}%
\providecommand \href [0]{\begingroup \@sanitize@url \@href}%
\providecommand \@href[1]{\@@startlink{#1}\@@href}%
\providecommand \@@href[1]{\endgroup#1\@@endlink}%
\providecommand \@sanitize@url [0]{\catcode `\\12\catcode `\$12\catcode
  `\&12\catcode `\#12\catcode `\^12\catcode `\_12\catcode `\%12\relax}%
\providecommand \@@startlink[1]{}%
\providecommand \@@endlink[0]{}%
\providecommand \url  [0]{\begingroup\@sanitize@url \@url }%
\providecommand \@url [1]{\endgroup\@href {#1}{\urlprefix }}%
\providecommand \urlprefix  [0]{URL }%
\providecommand \Eprint [0]{\href }%
\providecommand \doibase [0]{http://dx.doi.org/}%
\providecommand \selectlanguage [0]{\@gobble}%
\providecommand \bibinfo  [0]{\@secondoftwo}%
\providecommand \bibfield  [0]{\@secondoftwo}%
\providecommand \translation [1]{[#1]}%
\providecommand \BibitemOpen [0]{}%
\providecommand \bibitemStop [0]{}%
\providecommand \bibitemNoStop [0]{.\EOS\space}%
\providecommand \EOS [0]{\spacefactor3000\relax}%
\providecommand \BibitemShut  [1]{\csname bibitem#1\endcsname}%
\let\auto@bib@innerbib\@empty
%</preamble>
\bibitem [{\citenamefont {Bender}\ and\ \citenamefont
  {Boettcher}(1998)}]{PhysRevLett.80.5243}%
  \BibitemOpen
  \bibfield  {author} {\bibinfo {author} {\bibfnamefont {Carl~M.}\ \bibnamefont
  {Bender}}\ and\ \bibinfo {author} {\bibfnamefont {Stefan}\ \bibnamefont
  {Boettcher}},\ }\bibfield  {title} {\enquote {\bibinfo {title} {Real spectra
  in non-hermitian hamiltonians having $\mathcal{PT}$ symmetry},}\ }\href
  {\doibase 10.1103/PhysRevLett.80.5243} {\bibfield  {journal} {\bibinfo
  {journal} {Phys. Rev. Lett.}\ }\textbf {\bibinfo {volume} {80}},\ \bibinfo
  {pages} {5243--5246} (\bibinfo {year} {1998})}\BibitemShut {NoStop}%
\bibitem [{\citenamefont {Bender}(2007)}]{bender2007making}%
  \BibitemOpen
  \bibfield  {author} {\bibinfo {author} {\bibfnamefont {Carl~M}\ \bibnamefont
  {Bender}},\ }\bibfield  {title} {\enquote {\bibinfo {title} {Making sense of
  non-hermitian hamiltonians},}\ }\href@noop {} {\bibfield  {journal} {\bibinfo
   {journal} {Reports on Progress in Physics}\ }\textbf {\bibinfo {volume}
  {70}},\ \bibinfo {pages} {947} (\bibinfo {year} {2007})}\BibitemShut
  {NoStop}%
\bibitem [{\citenamefont {Rotter}(2009)}]{rotter2009non}%
  \BibitemOpen
  \bibfield  {author} {\bibinfo {author} {\bibfnamefont {Ingrid}\ \bibnamefont
  {Rotter}},\ }\bibfield  {title} {\enquote {\bibinfo {title} {A non-hermitian
  hamilton operator and the physics of open quantum systems},}\ }\href@noop {}
  {\bibfield  {journal} {\bibinfo  {journal} {Journal of Physics A:
  Mathematical and Theoretical}\ }\textbf {\bibinfo {volume} {42}},\ \bibinfo
  {pages} {153001} (\bibinfo {year} {2009})}\BibitemShut {NoStop}%
\bibitem [{\citenamefont {Yoshida}\ \emph {et~al.}(2018)\citenamefont
  {Yoshida}, \citenamefont {Peters},\ and\ \citenamefont
  {Kawakami}}]{yoshida2018non}%
  \BibitemOpen
  \bibfield  {author} {\bibinfo {author} {\bibfnamefont {Tsuneya}\ \bibnamefont
  {Yoshida}}, \bibinfo {author} {\bibfnamefont {Robert}\ \bibnamefont
  {Peters}}, \ and\ \bibinfo {author} {\bibfnamefont {Norio}\ \bibnamefont
  {Kawakami}},\ }\bibfield  {title} {\enquote {\bibinfo {title} {Non-hermitian
  perspective of the band structure in heavy-fermion systems},}\ }\href@noop {}
  {\bibfield  {journal} {\bibinfo  {journal} {Physical Review B}\ }\textbf
  {\bibinfo {volume} {98}},\ \bibinfo {pages} {035141} (\bibinfo {year}
  {2018})}\BibitemShut {NoStop}%
\bibitem [{\citenamefont {Yamamoto}\ \emph {et~al.}(2019)\citenamefont
  {Yamamoto}, \citenamefont {Nakagawa}, \citenamefont {Adachi}, \citenamefont
  {Takasan}, \citenamefont {Ueda},\ and\ \citenamefont
  {Kawakami}}]{yamamoto2019theory}%
  \BibitemOpen
  \bibfield  {author} {\bibinfo {author} {\bibfnamefont {Kazuki}\ \bibnamefont
  {Yamamoto}}, \bibinfo {author} {\bibfnamefont {Masaya}\ \bibnamefont
  {Nakagawa}}, \bibinfo {author} {\bibfnamefont {Kyosuke}\ \bibnamefont
  {Adachi}}, \bibinfo {author} {\bibfnamefont {Kazuaki}\ \bibnamefont
  {Takasan}}, \bibinfo {author} {\bibfnamefont {Masahito}\ \bibnamefont
  {Ueda}}, \ and\ \bibinfo {author} {\bibfnamefont {Norio}\ \bibnamefont
  {Kawakami}},\ }\bibfield  {title} {\enquote {\bibinfo {title} {Theory of
  non-hermitian fermionic superfluidity with a complex-valued interaction},}\
  }\href@noop {} {\bibfield  {journal} {\bibinfo  {journal} {Physical review
  letters}\ }\textbf {\bibinfo {volume} {123}},\ \bibinfo {pages} {123601}
  (\bibinfo {year} {2019})}\BibitemShut {NoStop}%
\bibitem [{\citenamefont {R{\"u}ter}\ \emph {et~al.}(2010)\citenamefont
  {R{\"u}ter}, \citenamefont {Makris}, \citenamefont {El-Ganainy},
  \citenamefont {Christodoulides}, \citenamefont {Segev},\ and\ \citenamefont
  {Kip}}]{ruter2010observation}%
  \BibitemOpen
  \bibfield  {author} {\bibinfo {author} {\bibfnamefont {Christian~E}\
  \bibnamefont {R{\"u}ter}}, \bibinfo {author} {\bibfnamefont {Konstantinos~G}\
  \bibnamefont {Makris}}, \bibinfo {author} {\bibfnamefont {Ramy}\ \bibnamefont
  {El-Ganainy}}, \bibinfo {author} {\bibfnamefont {Demetrios~N}\ \bibnamefont
  {Christodoulides}}, \bibinfo {author} {\bibfnamefont {Mordechai}\
  \bibnamefont {Segev}}, \ and\ \bibinfo {author} {\bibfnamefont {Detlef}\
  \bibnamefont {Kip}},\ }\bibfield  {title} {\enquote {\bibinfo {title}
  {Observation of parity-time symmetry in optics},}\ }\href@noop {} {\bibfield
  {journal} {\bibinfo  {journal} {Nature physics}\ }\textbf {\bibinfo {volume}
  {6}},\ \bibinfo {pages} {192--195} (\bibinfo {year} {2010})}\BibitemShut
  {NoStop}%
\bibitem [{\citenamefont {Longhi}(2018)}]{longhi2018parity}%
  \BibitemOpen
  \bibfield  {author} {\bibinfo {author} {\bibfnamefont {Stefano}\ \bibnamefont
  {Longhi}},\ }\bibfield  {title} {\enquote {\bibinfo {title} {Parity-time
  symmetry meets photonics: A new twist in non-hermitian optics},}\ }\href@noop
  {} {\bibfield  {journal} {\bibinfo  {journal} {EPL (Europhysics Letters)}\
  }\textbf {\bibinfo {volume} {120}},\ \bibinfo {pages} {64001} (\bibinfo
  {year} {2018})}\BibitemShut {NoStop}%
\bibitem [{\citenamefont {Ozawa}\ \emph {et~al.}(2019)\citenamefont {Ozawa},
  \citenamefont {Price}, \citenamefont {Amo}, \citenamefont {Goldman},
  \citenamefont {Hafezi}, \citenamefont {Lu}, \citenamefont {Rechtsman},
  \citenamefont {Schuster}, \citenamefont {Simon}, \citenamefont {Zilberberg}
  \emph {et~al.}}]{ozawa2019topological}%
  \BibitemOpen
  \bibfield  {author} {\bibinfo {author} {\bibfnamefont {Tomoki}\ \bibnamefont
  {Ozawa}}, \bibinfo {author} {\bibfnamefont {Hannah~M}\ \bibnamefont {Price}},
  \bibinfo {author} {\bibfnamefont {Alberto}\ \bibnamefont {Amo}}, \bibinfo
  {author} {\bibfnamefont {Nathan}\ \bibnamefont {Goldman}}, \bibinfo {author}
  {\bibfnamefont {Mohammad}\ \bibnamefont {Hafezi}}, \bibinfo {author}
  {\bibfnamefont {Ling}\ \bibnamefont {Lu}}, \bibinfo {author} {\bibfnamefont
  {Mikael~C}\ \bibnamefont {Rechtsman}}, \bibinfo {author} {\bibfnamefont
  {David}\ \bibnamefont {Schuster}}, \bibinfo {author} {\bibfnamefont
  {Jonathan}\ \bibnamefont {Simon}}, \bibinfo {author} {\bibfnamefont {Oded}\
  \bibnamefont {Zilberberg}},  \emph {et~al.},\ }\bibfield  {title} {\enquote
  {\bibinfo {title} {Topological photonics},}\ }\href@noop {} {\bibfield
  {journal} {\bibinfo  {journal} {Reviews of Modern Physics}\ }\textbf
  {\bibinfo {volume} {91}},\ \bibinfo {pages} {015006} (\bibinfo {year}
  {2019})}\BibitemShut {NoStop}%
\bibitem [{\citenamefont {Berry}(2004)}]{berry2004physics}%
  \BibitemOpen
  \bibfield  {author} {\bibinfo {author} {\bibfnamefont {Michael~V}\
  \bibnamefont {Berry}},\ }\bibfield  {title} {\enquote {\bibinfo {title}
  {Physics of nonhermitian degeneracies},}\ }\href@noop {} {\bibfield
  {journal} {\bibinfo  {journal} {Czechoslovak journal of physics}\ }\textbf
  {\bibinfo {volume} {54}},\ \bibinfo {pages} {1039--1047} (\bibinfo {year}
  {2004})}\BibitemShut {NoStop}%
\bibitem [{\citenamefont {Jin}\ and\ \citenamefont
  {Song}(2009)}]{jin2009solutions}%
  \BibitemOpen
  \bibfield  {author} {\bibinfo {author} {\bibfnamefont {L}~\bibnamefont
  {Jin}}\ and\ \bibinfo {author} {\bibfnamefont {Z}~\bibnamefont {Song}},\
  }\bibfield  {title} {\enquote {\bibinfo {title} {Solutions of p t-symmetric
  tight-binding chain and its equivalent hermitian counterpart},}\ }\href@noop
  {} {\bibfield  {journal} {\bibinfo  {journal} {Physical Review A}\ }\textbf
  {\bibinfo {volume} {80}},\ \bibinfo {pages} {052107} (\bibinfo {year}
  {2009})}\BibitemShut {NoStop}%
\bibitem [{\citenamefont {Longhi}(2010)}]{longhi2010pt}%
  \BibitemOpen
  \bibfield  {author} {\bibinfo {author} {\bibfnamefont {Stefano}\ \bibnamefont
  {Longhi}},\ }\bibfield  {title} {\enquote {\bibinfo {title} {Pt-symmetric
  laser absorber},}\ }\href@noop {} {\bibfield  {journal} {\bibinfo  {journal}
  {Physical Review A}\ }\textbf {\bibinfo {volume} {82}},\ \bibinfo {pages}
  {031801} (\bibinfo {year} {2010})}\BibitemShut {NoStop}%
\bibitem [{\citenamefont {Heiss}(2012)}]{heiss2012physics}%
  \BibitemOpen
  \bibfield  {author} {\bibinfo {author} {\bibfnamefont {WD}~\bibnamefont
  {Heiss}},\ }\bibfield  {title} {\enquote {\bibinfo {title} {The physics of
  exceptional points},}\ }\href@noop {} {\bibfield  {journal} {\bibinfo
  {journal} {Journal of Physics A: Mathematical and Theoretical}\ }\textbf
  {\bibinfo {volume} {45}},\ \bibinfo {pages} {444016} (\bibinfo {year}
  {2012})}\BibitemShut {NoStop}%
\bibitem [{\citenamefont {Lee}(2016)}]{lee2016anomalous}%
  \BibitemOpen
  \bibfield  {author} {\bibinfo {author} {\bibfnamefont {Tony~E}\ \bibnamefont
  {Lee}},\ }\bibfield  {title} {\enquote {\bibinfo {title} {Anomalous edge
  state in a non-hermitian lattice},}\ }\href@noop {} {\bibfield  {journal}
  {\bibinfo  {journal} {Physical review letters}\ }\textbf {\bibinfo {volume}
  {116}},\ \bibinfo {pages} {133903} (\bibinfo {year} {2016})}\BibitemShut
  {NoStop}%
\bibitem [{\citenamefont {Xu}\ \emph {et~al.}(2016)\citenamefont {Xu},
  \citenamefont {Mason}, \citenamefont {Jiang},\ and\ \citenamefont
  {Harris}}]{xu2016topological}%
  \BibitemOpen
  \bibfield  {author} {\bibinfo {author} {\bibfnamefont {Haitan}\ \bibnamefont
  {Xu}}, \bibinfo {author} {\bibfnamefont {David}\ \bibnamefont {Mason}},
  \bibinfo {author} {\bibfnamefont {Luyao}\ \bibnamefont {Jiang}}, \ and\
  \bibinfo {author} {\bibfnamefont {JGE}\ \bibnamefont {Harris}},\ }\bibfield
  {title} {\enquote {\bibinfo {title} {Topological energy transfer in an
  optomechanical system with exceptional points},}\ }\href@noop {} {\bibfield
  {journal} {\bibinfo  {journal} {Nature}\ }\textbf {\bibinfo {volume} {537}},\
  \bibinfo {pages} {80} (\bibinfo {year} {2016})}\BibitemShut {NoStop}%
\bibitem [{\citenamefont {Hassan}\ \emph {et~al.}(2017)\citenamefont {Hassan},
  \citenamefont {Zhen}, \citenamefont {Solja{\v{c}}i{\'c}}, \citenamefont
  {Khajavikhan},\ and\ \citenamefont
  {Christodoulides}}]{hassan2017dynamically}%
  \BibitemOpen
  \bibfield  {author} {\bibinfo {author} {\bibfnamefont {Absar~U}\ \bibnamefont
  {Hassan}}, \bibinfo {author} {\bibfnamefont {Bo}~\bibnamefont {Zhen}},
  \bibinfo {author} {\bibfnamefont {Marin}\ \bibnamefont {Solja{\v{c}}i{\'c}}},
  \bibinfo {author} {\bibfnamefont {Mercedeh}\ \bibnamefont {Khajavikhan}}, \
  and\ \bibinfo {author} {\bibfnamefont {Demetrios~N}\ \bibnamefont
  {Christodoulides}},\ }\bibfield  {title} {\enquote {\bibinfo {title}
  {Dynamically encircling exceptional points: exact evolution and polarization
  state conversion},}\ }\href@noop {} {\bibfield  {journal} {\bibinfo
  {journal} {Physical review letters}\ }\textbf {\bibinfo {volume} {118}},\
  \bibinfo {pages} {093002} (\bibinfo {year} {2017})}\BibitemShut {NoStop}%
\bibitem [{\citenamefont {Hu}\ \emph {et~al.}(2017)\citenamefont {Hu},
  \citenamefont {Wang}, \citenamefont {Shum},\ and\ \citenamefont
  {Chong}}]{hu2017exceptional}%
  \BibitemOpen
  \bibfield  {author} {\bibinfo {author} {\bibfnamefont {Wenchao}\ \bibnamefont
  {Hu}}, \bibinfo {author} {\bibfnamefont {Hailong}\ \bibnamefont {Wang}},
  \bibinfo {author} {\bibfnamefont {Perry~Ping}\ \bibnamefont {Shum}}, \ and\
  \bibinfo {author} {\bibfnamefont {Yi~Dong}\ \bibnamefont {Chong}},\
  }\bibfield  {title} {\enquote {\bibinfo {title} {Exceptional points in a
  non-hermitian topological pump},}\ }\href@noop {} {\bibfield  {journal}
  {\bibinfo  {journal} {Physical Review B}\ }\textbf {\bibinfo {volume} {95}},\
  \bibinfo {pages} {184306} (\bibinfo {year} {2017})}\BibitemShut {NoStop}%
\bibitem [{\citenamefont {Shen}\ \emph {et~al.}(2018)\citenamefont {Shen},
  \citenamefont {Zhen},\ and\ \citenamefont {Fu}}]{shen2018topological}%
  \BibitemOpen
  \bibfield  {author} {\bibinfo {author} {\bibfnamefont {Huitao}\ \bibnamefont
  {Shen}}, \bibinfo {author} {\bibfnamefont {Bo}~\bibnamefont {Zhen}}, \ and\
  \bibinfo {author} {\bibfnamefont {Liang}\ \bibnamefont {Fu}},\ }\bibfield
  {title} {\enquote {\bibinfo {title} {Topological band theory for
  non-hermitian hamiltonians},}\ }\href@noop {} {\bibfield  {journal} {\bibinfo
   {journal} {Physical review letters}\ }\textbf {\bibinfo {volume} {120}},\
  \bibinfo {pages} {146402} (\bibinfo {year} {2018})}\BibitemShut {NoStop}%
\bibitem [{\citenamefont {Wang}\ \emph
  {et~al.}(2019{\natexlab{a}})\citenamefont {Wang}, \citenamefont {Hou},
  \citenamefont {Lu}, \citenamefont {Chen}, \citenamefont {Zhang},\ and\
  \citenamefont {Chan}}]{wang2019arbitrary}%
  \BibitemOpen
  \bibfield  {author} {\bibinfo {author} {\bibfnamefont {Shubo}\ \bibnamefont
  {Wang}}, \bibinfo {author} {\bibfnamefont {Bo}~\bibnamefont {Hou}}, \bibinfo
  {author} {\bibfnamefont {Weixin}\ \bibnamefont {Lu}}, \bibinfo {author}
  {\bibfnamefont {Yuntian}\ \bibnamefont {Chen}}, \bibinfo {author}
  {\bibfnamefont {ZQ}~\bibnamefont {Zhang}}, \ and\ \bibinfo {author}
  {\bibfnamefont {CT}~\bibnamefont {Chan}},\ }\bibfield  {title} {\enquote
  {\bibinfo {title} {Arbitrary order exceptional point induced by photonic
  spin--orbit interaction in coupled resonators},}\ }\href@noop {} {\bibfield
  {journal} {\bibinfo  {journal} {Nature communications}\ }\textbf {\bibinfo
  {volume} {10}},\ \bibinfo {pages} {1--9} (\bibinfo {year}
  {2019}{\natexlab{a}})}\BibitemShut {NoStop}%
\bibitem [{\citenamefont {Ghatak}\ and\ \citenamefont
  {Das}(2019)}]{ghatak2019new}%
  \BibitemOpen
  \bibfield  {author} {\bibinfo {author} {\bibfnamefont {Ananya}\ \bibnamefont
  {Ghatak}}\ and\ \bibinfo {author} {\bibfnamefont {Tanmoy}\ \bibnamefont
  {Das}},\ }\bibfield  {title} {\enquote {\bibinfo {title} {New topological
  invariants in non-hermitian systems},}\ }\href@noop {} {\bibfield  {journal}
  {\bibinfo  {journal} {Journal of Physics: Condensed Matter}\ }\textbf
  {\bibinfo {volume} {31}},\ \bibinfo {pages} {263001} (\bibinfo {year}
  {2019})}\BibitemShut {NoStop}%
\bibitem [{\citenamefont {Miri}\ and\ \citenamefont
  {Al{\`u}}(2019)}]{miri2019exceptional}%
  \BibitemOpen
  \bibfield  {author} {\bibinfo {author} {\bibfnamefont {Mohammad-Ali}\
  \bibnamefont {Miri}}\ and\ \bibinfo {author} {\bibfnamefont {Andrea}\
  \bibnamefont {Al{\`u}}},\ }\bibfield  {title} {\enquote {\bibinfo {title}
  {Exceptional points in optics and photonics},}\ }\href@noop {} {\bibfield
  {journal} {\bibinfo  {journal} {Science}\ }\textbf {\bibinfo {volume}
  {363}},\ \bibinfo {pages} {eaar7709} (\bibinfo {year} {2019})}\BibitemShut
  {NoStop}%
\bibitem [{\citenamefont {Zhang}\ and\ \citenamefont
  {Gong}(2020)}]{zhang2020non}%
  \BibitemOpen
  \bibfield  {author} {\bibinfo {author} {\bibfnamefont {Xizheng}\ \bibnamefont
  {Zhang}}\ and\ \bibinfo {author} {\bibfnamefont {Jiangbin}\ \bibnamefont
  {Gong}},\ }\bibfield  {title} {\enquote {\bibinfo {title} {Non-hermitian
  floquet topological phases: Exceptional points, coalescent edge modes, and
  the skin effect},}\ }\href@noop {} {\bibfield  {journal} {\bibinfo  {journal}
  {Physical Review B}\ }\textbf {\bibinfo {volume} {101}},\ \bibinfo {pages}
  {045415} (\bibinfo {year} {2020})}\BibitemShut {NoStop}%
\bibitem [{\citenamefont {Yuce}(2020)}]{yuce2020non}%
  \BibitemOpen
  \bibfield  {author} {\bibinfo {author} {\bibfnamefont {C}~\bibnamefont
  {Yuce}},\ }\bibfield  {title} {\enquote {\bibinfo {title} {Non-hermitian
  anomalous skin effect},}\ }\href@noop {} {\bibfield  {journal} {\bibinfo
  {journal} {Physics Letters A}\ }\textbf {\bibinfo {volume} {384}},\ \bibinfo
  {pages} {126094} (\bibinfo {year} {2020})}\BibitemShut {NoStop}%
\bibitem [{\citenamefont {Jin}\ \emph {et~al.}(2019)\citenamefont {Jin},
  \citenamefont {Wu}, \citenamefont {Wei},\ and\ \citenamefont
  {Song}}]{jin2019hybrid}%
  \BibitemOpen
  \bibfield  {author} {\bibinfo {author} {\bibfnamefont {L}~\bibnamefont
  {Jin}}, \bibinfo {author} {\bibfnamefont {HC}~\bibnamefont {Wu}}, \bibinfo
  {author} {\bibfnamefont {Bo-Bo}\ \bibnamefont {Wei}}, \ and\ \bibinfo
  {author} {\bibfnamefont {Z}~\bibnamefont {Song}},\ }\bibfield  {title}
  {\enquote {\bibinfo {title} {Hybrid exceptional point created from type iii
  dirac point},}\ }\href@noop {} {\bibfield  {journal} {\bibinfo  {journal}
  {arXiv preprint arXiv:1908.10512}\ } (\bibinfo {year} {2019})}\BibitemShut
  {NoStop}%
\bibitem [{\citenamefont {Xu}\ \emph {et~al.}(2017)\citenamefont {Xu},
  \citenamefont {Wang},\ and\ \citenamefont {Duan}}]{xu2017weyl}%
  \BibitemOpen
  \bibfield  {author} {\bibinfo {author} {\bibfnamefont {Yong}\ \bibnamefont
  {Xu}}, \bibinfo {author} {\bibfnamefont {Sheng-Tao}\ \bibnamefont {Wang}}, \
  and\ \bibinfo {author} {\bibfnamefont {L-M}\ \bibnamefont {Duan}},\
  }\bibfield  {title} {\enquote {\bibinfo {title} {Weyl exceptional rings in a
  three-dimensional dissipative cold atomic gas},}\ }\href@noop {} {\bibfield
  {journal} {\bibinfo  {journal} {Physical review letters}\ }\textbf {\bibinfo
  {volume} {118}},\ \bibinfo {pages} {045701} (\bibinfo {year}
  {2017})}\BibitemShut {NoStop}%
\bibitem [{\citenamefont {Carlstr{\"o}m}\ and\ \citenamefont
  {Bergholtz}(2018)}]{carlstrom2018exceptional}%
  \BibitemOpen
  \bibfield  {author} {\bibinfo {author} {\bibfnamefont {Johan}\ \bibnamefont
  {Carlstr{\"o}m}}\ and\ \bibinfo {author} {\bibfnamefont {Emil~J}\
  \bibnamefont {Bergholtz}},\ }\bibfield  {title} {\enquote {\bibinfo {title}
  {Exceptional links and twisted fermi ribbons in non-hermitian systems},}\
  }\href@noop {} {\bibfield  {journal} {\bibinfo  {journal} {Physical Review
  A}\ }\textbf {\bibinfo {volume} {98}},\ \bibinfo {pages} {042114} (\bibinfo
  {year} {2018})}\BibitemShut {NoStop}%
\bibitem [{\citenamefont {Zhou}\ \emph {et~al.}(2019)\citenamefont {Zhou},
  \citenamefont {Lee}, \citenamefont {Liu},\ and\ \citenamefont
  {Zhen}}]{zhou2019exceptional}%
  \BibitemOpen
  \bibfield  {author} {\bibinfo {author} {\bibfnamefont {Hengyun}\ \bibnamefont
  {Zhou}}, \bibinfo {author} {\bibfnamefont {Jong~Yeon}\ \bibnamefont {Lee}},
  \bibinfo {author} {\bibfnamefont {Shang}\ \bibnamefont {Liu}}, \ and\
  \bibinfo {author} {\bibfnamefont {Bo}~\bibnamefont {Zhen}},\ }\bibfield
  {title} {\enquote {\bibinfo {title} {Exceptional surfaces in pt-symmetric
  non-hermitian photonic systems},}\ }\href@noop {} {\bibfield  {journal}
  {\bibinfo  {journal} {Optica}\ }\textbf {\bibinfo {volume} {6}},\ \bibinfo
  {pages} {190--193} (\bibinfo {year} {2019})}\BibitemShut {NoStop}%
\bibitem [{\citenamefont {Moors}\ \emph {et~al.}(2019)\citenamefont {Moors},
  \citenamefont {Zyuzin}, \citenamefont {Zyuzin}, \citenamefont {Tiwari},\ and\
  \citenamefont {Schmidt}}]{moors2019disorder}%
  \BibitemOpen
  \bibfield  {author} {\bibinfo {author} {\bibfnamefont {Kristof}\ \bibnamefont
  {Moors}}, \bibinfo {author} {\bibfnamefont {Alexander~A}\ \bibnamefont
  {Zyuzin}}, \bibinfo {author} {\bibfnamefont {Alexander~Yu}\ \bibnamefont
  {Zyuzin}}, \bibinfo {author} {\bibfnamefont {Rakesh~P}\ \bibnamefont
  {Tiwari}}, \ and\ \bibinfo {author} {\bibfnamefont {Thomas~L}\ \bibnamefont
  {Schmidt}},\ }\bibfield  {title} {\enquote {\bibinfo {title} {Disorder-driven
  exceptional lines and fermi ribbons in tilted nodal-line semimetals},}\
  }\href@noop {} {\bibfield  {journal} {\bibinfo  {journal} {Physical Review
  B}\ }\textbf {\bibinfo {volume} {99}},\ \bibinfo {pages} {041116} (\bibinfo
  {year} {2019})}\BibitemShut {NoStop}%
\bibitem [{\citenamefont {Wang}\ \emph
  {et~al.}(2019{\natexlab{b}})\citenamefont {Wang}, \citenamefont {Ruan},\ and\
  \citenamefont {Zhang}}]{wang2019non}%
  \BibitemOpen
  \bibfield  {author} {\bibinfo {author} {\bibfnamefont {Huaiqiang}\
  \bibnamefont {Wang}}, \bibinfo {author} {\bibfnamefont {Jiawei}\ \bibnamefont
  {Ruan}}, \ and\ \bibinfo {author} {\bibfnamefont {Haijun}\ \bibnamefont
  {Zhang}},\ }\bibfield  {title} {\enquote {\bibinfo {title} {Non-hermitian
  nodal-line semimetals with an anomalous bulk-boundary correspondence},}\
  }\href@noop {} {\bibfield  {journal} {\bibinfo  {journal} {Physical Review
  B}\ }\textbf {\bibinfo {volume} {99}},\ \bibinfo {pages} {075130} (\bibinfo
  {year} {2019}{\natexlab{b}})}\BibitemShut {NoStop}%
\bibitem [{\citenamefont {Yang}\ and\ \citenamefont {Hu}(2019)}]{yang2019non}%
  \BibitemOpen
  \bibfield  {author} {\bibinfo {author} {\bibfnamefont {Zhesen}\ \bibnamefont
  {Yang}}\ and\ \bibinfo {author} {\bibfnamefont {Jiangping}\ \bibnamefont
  {Hu}},\ }\bibfield  {title} {\enquote {\bibinfo {title} {Non-hermitian
  hopf-link exceptional line semimetals},}\ }\href@noop {} {\bibfield
  {journal} {\bibinfo  {journal} {Physical Review B}\ }\textbf {\bibinfo
  {volume} {99}},\ \bibinfo {pages} {081102} (\bibinfo {year}
  {2019})}\BibitemShut {NoStop}%
\bibitem [{\citenamefont {Carlstr{\"o}m}\ \emph {et~al.}(2019)\citenamefont
  {Carlstr{\"o}m}, \citenamefont {St{\aa}lhammar}, \citenamefont {Budich},\
  and\ \citenamefont {Bergholtz}}]{carlstrom2019knotted}%
  \BibitemOpen
  \bibfield  {author} {\bibinfo {author} {\bibfnamefont {Johan}\ \bibnamefont
  {Carlstr{\"o}m}}, \bibinfo {author} {\bibfnamefont {Marcus}\ \bibnamefont
  {St{\aa}lhammar}}, \bibinfo {author} {\bibfnamefont {Jan~Carl}\ \bibnamefont
  {Budich}}, \ and\ \bibinfo {author} {\bibfnamefont {Emil~J}\ \bibnamefont
  {Bergholtz}},\ }\bibfield  {title} {\enquote {\bibinfo {title} {Knotted
  non-hermitian metals},}\ }\href@noop {} {\bibfield  {journal} {\bibinfo
  {journal} {Physical Review B}\ }\textbf {\bibinfo {volume} {99}},\ \bibinfo
  {pages} {161115} (\bibinfo {year} {2019})}\BibitemShut {NoStop}%
\bibitem [{\citenamefont {Okugawa}\ and\ \citenamefont
  {Yokoyama}(2019)}]{okugawa2019topological}%
  \BibitemOpen
  \bibfield  {author} {\bibinfo {author} {\bibfnamefont {Ryo}\ \bibnamefont
  {Okugawa}}\ and\ \bibinfo {author} {\bibfnamefont {Takehito}\ \bibnamefont
  {Yokoyama}},\ }\bibfield  {title} {\enquote {\bibinfo {title} {Topological
  exceptional surfaces in non-hermitian systems with parity-time and
  parity-particle-hole symmetries},}\ }\href@noop {} {\bibfield  {journal}
  {\bibinfo  {journal} {Physical Review B}\ }\textbf {\bibinfo {volume} {99}},\
  \bibinfo {pages} {041202} (\bibinfo {year} {2019})}\BibitemShut {NoStop}%
\bibitem [{\citenamefont {Luo}\ \emph {et~al.}()\citenamefont {Luo},
  \citenamefont {Feng}, \citenamefont {Zhao},\ and\ \citenamefont
  {Yu}}]{luo2018nodal}%
  \BibitemOpen
  \bibfield  {author} {\bibinfo {author} {\bibfnamefont {Kaifa}\ \bibnamefont
  {Luo}}, \bibinfo {author} {\bibfnamefont {Jiajin}\ \bibnamefont {Feng}},
  \bibinfo {author} {\bibfnamefont {Y.~X.}\ \bibnamefont {Zhao}}, \ and\
  \bibinfo {author} {\bibfnamefont {Rui}\ \bibnamefont {Yu}},\ }\bibfield
  {title} {\enquote {\bibinfo {title} {Nodal manifolds bounded by exceptional
  points on non-hermitian honeycomb lattices and electrical-circuit
  realizations},}\ }\href@noop {} {\ }\Eprint
  {http://arxiv.org/abs/1810.09231v1} {1810.09231v1} \BibitemShut {NoStop}%
\bibitem [{\citenamefont {Lee}\ \emph {et~al.}({\natexlab{a}})\citenamefont
  {Lee}, \citenamefont {Li}, \citenamefont {Liu}, \citenamefont {Tai},
  \citenamefont {Thomale},\ and\ \citenamefont {Zhang}}]{lee2018tidal}%
  \BibitemOpen
  \bibfield  {author} {\bibinfo {author} {\bibfnamefont {Ching~Hua}\
  \bibnamefont {Lee}}, \bibinfo {author} {\bibfnamefont {Guangjie}\
  \bibnamefont {Li}}, \bibinfo {author} {\bibfnamefont {Yuhan}\ \bibnamefont
  {Liu}}, \bibinfo {author} {\bibfnamefont {Tommy}\ \bibnamefont {Tai}},
  \bibinfo {author} {\bibfnamefont {Ronny}\ \bibnamefont {Thomale}}, \ and\
  \bibinfo {author} {\bibfnamefont {Xiao}\ \bibnamefont {Zhang}},\ }\bibfield
  {title} {\enquote {\bibinfo {title} {Tidal surface states as fingerprints of
  non-hermitian nodal knot metals},}\ }\href@noop {} {\  ({\natexlab{a}})},\
  \Eprint {http://arxiv.org/abs/1812.02011v1} {1812.02011v1} \BibitemShut
  {NoStop}%
\bibitem [{\citenamefont {Gong}\ \emph {et~al.}(2018)\citenamefont {Gong},
  \citenamefont {Ashida}, \citenamefont {Kawabata}, \citenamefont {Takasan},
  \citenamefont {Higashikawa},\ and\ \citenamefont
  {Ueda}}]{gong2018topological}%
  \BibitemOpen
  \bibfield  {author} {\bibinfo {author} {\bibfnamefont {Zongping}\
  \bibnamefont {Gong}}, \bibinfo {author} {\bibfnamefont {Yuto}\ \bibnamefont
  {Ashida}}, \bibinfo {author} {\bibfnamefont {Kohei}\ \bibnamefont
  {Kawabata}}, \bibinfo {author} {\bibfnamefont {Kazuaki}\ \bibnamefont
  {Takasan}}, \bibinfo {author} {\bibfnamefont {Sho}\ \bibnamefont
  {Higashikawa}}, \ and\ \bibinfo {author} {\bibfnamefont {Masahito}\
  \bibnamefont {Ueda}},\ }\bibfield  {title} {\enquote {\bibinfo {title}
  {Topological phases of non-hermitian systems},}\ }\href@noop {} {\bibfield
  {journal} {\bibinfo  {journal} {Physical Review X}\ }\textbf {\bibinfo
  {volume} {8}},\ \bibinfo {pages} {031079} (\bibinfo {year}
  {2018})}\BibitemShut {NoStop}%
\bibitem [{\citenamefont {Liu}\ \emph {et~al.}(2019)\citenamefont {Liu},
  \citenamefont {Jiang},\ and\ \citenamefont {Chen}}]{liu2019topological}%
  \BibitemOpen
  \bibfield  {author} {\bibinfo {author} {\bibfnamefont {Chun-Hui}\
  \bibnamefont {Liu}}, \bibinfo {author} {\bibfnamefont {Hui}\ \bibnamefont
  {Jiang}}, \ and\ \bibinfo {author} {\bibfnamefont {Shu}\ \bibnamefont
  {Chen}},\ }\bibfield  {title} {\enquote {\bibinfo {title} {Topological
  classification of non-hermitian systems with reflection symmetry},}\
  }\href@noop {} {\bibfield  {journal} {\bibinfo  {journal} {Physical Review
  B}\ }\textbf {\bibinfo {volume} {99}},\ \bibinfo {pages} {125103} (\bibinfo
  {year} {2019})}\BibitemShut {NoStop}%
\bibitem [{\citenamefont {Kawabata}\ \emph
  {et~al.}(2019{\natexlab{a}})\citenamefont {Kawabata}, \citenamefont
  {Shiozaki}, \citenamefont {Ueda},\ and\ \citenamefont
  {Sato}}]{kawabata2019symmetry}%
  \BibitemOpen
  \bibfield  {author} {\bibinfo {author} {\bibfnamefont {Kohei}\ \bibnamefont
  {Kawabata}}, \bibinfo {author} {\bibfnamefont {Ken}\ \bibnamefont
  {Shiozaki}}, \bibinfo {author} {\bibfnamefont {Masahito}\ \bibnamefont
  {Ueda}}, \ and\ \bibinfo {author} {\bibfnamefont {Masatoshi}\ \bibnamefont
  {Sato}},\ }\bibfield  {title} {\enquote {\bibinfo {title} {Symmetry and
  topology in non-hermitian physics},}\ }\href@noop {} {\bibfield  {journal}
  {\bibinfo  {journal} {Physical Review X}\ }\textbf {\bibinfo {volume} {9}},\
  \bibinfo {pages} {041015} (\bibinfo {year} {2019}{\natexlab{a}})}\BibitemShut
  {NoStop}%
\bibitem [{\citenamefont {Zhou}\ and\ \citenamefont
  {Lee}(2019)}]{zhou2019periodic}%
  \BibitemOpen
  \bibfield  {author} {\bibinfo {author} {\bibfnamefont {Hengyun}\ \bibnamefont
  {Zhou}}\ and\ \bibinfo {author} {\bibfnamefont {Jong~Yeon}\ \bibnamefont
  {Lee}},\ }\bibfield  {title} {\enquote {\bibinfo {title} {Periodic table for
  topological bands with non-hermitian symmetries},}\ }\href@noop {} {\bibfield
   {journal} {\bibinfo  {journal} {Physical Review B}\ }\textbf {\bibinfo
  {volume} {99}},\ \bibinfo {pages} {235112} (\bibinfo {year}
  {2019})}\BibitemShut {NoStop}%
\bibitem [{\citenamefont {Li}\ \emph {et~al.}(2019)\citenamefont {Li},
  \citenamefont {Lee},\ and\ \citenamefont {Gong}}]{li2019geometric}%
  \BibitemOpen
  \bibfield  {author} {\bibinfo {author} {\bibfnamefont {Linhu}\ \bibnamefont
  {Li}}, \bibinfo {author} {\bibfnamefont {Ching~Hua}\ \bibnamefont {Lee}}, \
  and\ \bibinfo {author} {\bibfnamefont {Jiangbin}\ \bibnamefont {Gong}},\
  }\bibfield  {title} {\enquote {\bibinfo {title} {Geometric characterization
  of non-hermitian topological systems through the singularity ring in
  pseudospin vector space},}\ }\href {\doibase 10.1103/PhysRevB.100.075403}
  {\bibfield  {journal} {\bibinfo  {journal} {Phys. Rev. B}\ }\textbf {\bibinfo
  {volume} {100}},\ \bibinfo {pages} {075403} (\bibinfo {year}
  {2019})}\BibitemShut {NoStop}%
\bibitem [{\citenamefont {Liu}\ and\ \citenamefont
  {Chen}(2019)}]{liu2019topological2}%
  \BibitemOpen
  \bibfield  {author} {\bibinfo {author} {\bibfnamefont {Chun-Hui}\
  \bibnamefont {Liu}}\ and\ \bibinfo {author} {\bibfnamefont {Shu}\
  \bibnamefont {Chen}},\ }\bibfield  {title} {\enquote {\bibinfo {title}
  {Topological classification of defects in non-hermitian systems},}\
  }\href@noop {} {\bibfield  {journal} {\bibinfo  {journal} {Physical Review
  B}\ }\textbf {\bibinfo {volume} {100}},\ \bibinfo {pages} {144106} (\bibinfo
  {year} {2019})}\BibitemShut {NoStop}%
\bibitem [{\citenamefont {Lieu}(2018)}]{lieu2018topological}%
  \BibitemOpen
  \bibfield  {author} {\bibinfo {author} {\bibfnamefont {Simon}\ \bibnamefont
  {Lieu}},\ }\bibfield  {title} {\enquote {\bibinfo {title} {Topological
  symmetry classes for non-hermitian models and connections to the bosonic
  bogoliubov--de gennes equation},}\ }\href@noop {} {\bibfield  {journal}
  {\bibinfo  {journal} {Physical Review B}\ }\textbf {\bibinfo {volume} {98}},\
  \bibinfo {pages} {115135} (\bibinfo {year} {2018})}\BibitemShut {NoStop}%
\bibitem [{\citenamefont {Kawabata}\ \emph
  {et~al.}(2019{\natexlab{b}})\citenamefont {Kawabata}, \citenamefont
  {Higashikawa}, \citenamefont {Gong}, \citenamefont {Ashida},\ and\
  \citenamefont {Ueda}}]{kawabata2019topological}%
  \BibitemOpen
  \bibfield  {author} {\bibinfo {author} {\bibfnamefont {Kohei}\ \bibnamefont
  {Kawabata}}, \bibinfo {author} {\bibfnamefont {Sho}\ \bibnamefont
  {Higashikawa}}, \bibinfo {author} {\bibfnamefont {Zongping}\ \bibnamefont
  {Gong}}, \bibinfo {author} {\bibfnamefont {Yuto}\ \bibnamefont {Ashida}}, \
  and\ \bibinfo {author} {\bibfnamefont {Masahito}\ \bibnamefont {Ueda}},\
  }\bibfield  {title} {\enquote {\bibinfo {title} {Topological unification of
  time-reversal and particle-hole symmetries in non-hermitian physics},}\
  }\href@noop {} {\bibfield  {journal} {\bibinfo  {journal} {Nature
  communications}\ }\textbf {\bibinfo {volume} {10}},\ \bibinfo {pages} {1--7}
  (\bibinfo {year} {2019}{\natexlab{b}})}\BibitemShut {NoStop}%
\bibitem [{\citenamefont {Wu}\ \emph {et~al.}(2019)\citenamefont {Wu},
  \citenamefont {Jin},\ and\ \citenamefont {Song}}]{wu2019inversion}%
  \BibitemOpen
  \bibfield  {author} {\bibinfo {author} {\bibfnamefont {HC}~\bibnamefont
  {Wu}}, \bibinfo {author} {\bibfnamefont {L}~\bibnamefont {Jin}}, \ and\
  \bibinfo {author} {\bibfnamefont {Z}~\bibnamefont {Song}},\ }\bibfield
  {title} {\enquote {\bibinfo {title} {Inversion symmetric non-hermitian chern
  insulator},}\ }\href@noop {} {\bibfield  {journal} {\bibinfo  {journal}
  {Physical Review B}\ }\textbf {\bibinfo {volume} {100}},\ \bibinfo {pages}
  {155117} (\bibinfo {year} {2019})}\BibitemShut {NoStop}%
\bibitem [{\citenamefont {Yoshida}\ \emph {et~al.}(2019)\citenamefont
  {Yoshida}, \citenamefont {Peters}, \citenamefont {Kawakami},\ and\
  \citenamefont {Hatsugai}}]{yoshida2019symmetry}%
  \BibitemOpen
  \bibfield  {author} {\bibinfo {author} {\bibfnamefont {Tsuneya}\ \bibnamefont
  {Yoshida}}, \bibinfo {author} {\bibfnamefont {Robert}\ \bibnamefont
  {Peters}}, \bibinfo {author} {\bibfnamefont {Norio}\ \bibnamefont
  {Kawakami}}, \ and\ \bibinfo {author} {\bibfnamefont {Yasuhiro}\ \bibnamefont
  {Hatsugai}},\ }\bibfield  {title} {\enquote {\bibinfo {title}
  {Symmetry-protected exceptional rings in two-dimensional correlated systems
  with chiral symmetry},}\ }\href@noop {} {\bibfield  {journal} {\bibinfo
  {journal} {Physical Review B}\ }\textbf {\bibinfo {volume} {99}},\ \bibinfo
  {pages} {121101} (\bibinfo {year} {2019})}\BibitemShut {NoStop}%
\bibitem [{\citenamefont {Yoshida}\ and\ \citenamefont
  {Hatsugai}(2019)}]{yoshida2019exceptional}%
  \BibitemOpen
  \bibfield  {author} {\bibinfo {author} {\bibfnamefont {Tsuneya}\ \bibnamefont
  {Yoshida}}\ and\ \bibinfo {author} {\bibfnamefont {Yasuhiro}\ \bibnamefont
  {Hatsugai}},\ }\bibfield  {title} {\enquote {\bibinfo {title} {Exceptional
  rings protected by emergent symmetry for mechanical systems},}\ }\href@noop
  {} {\bibfield  {journal} {\bibinfo  {journal} {Physical Review B}\ }\textbf
  {\bibinfo {volume} {100}},\ \bibinfo {pages} {054109} (\bibinfo {year}
  {2019})}\BibitemShut {NoStop}%
\bibitem [{\citenamefont {Kawabata}\ \emph
  {et~al.}(2019{\natexlab{c}})\citenamefont {Kawabata}, \citenamefont
  {Bessho},\ and\ \citenamefont {Sato}}]{kawabata2019classification}%
  \BibitemOpen
  \bibfield  {author} {\bibinfo {author} {\bibfnamefont {Kohei}\ \bibnamefont
  {Kawabata}}, \bibinfo {author} {\bibfnamefont {Takumi}\ \bibnamefont
  {Bessho}}, \ and\ \bibinfo {author} {\bibfnamefont {Masatoshi}\ \bibnamefont
  {Sato}},\ }\bibfield  {title} {\enquote {\bibinfo {title} {Classification of
  exceptional points and non-hermitian topological semimetals},}\ }\href@noop
  {} {\bibfield  {journal} {\bibinfo  {journal} {Physical review letters}\
  }\textbf {\bibinfo {volume} {123}},\ \bibinfo {pages} {066405} (\bibinfo
  {year} {2019}{\natexlab{c}})}\BibitemShut {NoStop}%
\bibitem [{\citenamefont {Yao}\ and\ \citenamefont {Wang}(2018)}]{yao2018edge}%
  \BibitemOpen
  \bibfield  {author} {\bibinfo {author} {\bibfnamefont {Shunyu}\ \bibnamefont
  {Yao}}\ and\ \bibinfo {author} {\bibfnamefont {Zhong}\ \bibnamefont {Wang}},\
  }\bibfield  {title} {\enquote {\bibinfo {title} {Edge states and topological
  invariants of non-hermitian systems},}\ }\href@noop {} {\bibfield  {journal}
  {\bibinfo  {journal} {Physical review letters}\ }\textbf {\bibinfo {volume}
  {121}},\ \bibinfo {pages} {086803} (\bibinfo {year} {2018})}\BibitemShut
  {NoStop}%
\bibitem [{\citenamefont {Xiong}(2018)}]{Xiong_2018}%
  \BibitemOpen
  \bibfield  {author} {\bibinfo {author} {\bibfnamefont {Ye}~\bibnamefont
  {Xiong}},\ }\bibfield  {title} {\enquote {\bibinfo {title} {Why does bulk
  boundary correspondence fail in some non-hermitian topological models},}\
  }\href {\doibase 10.1088/2399-6528/aab64a} {\bibfield  {journal} {\bibinfo
  {journal} {Journal of Physics Communications}\ }\textbf {\bibinfo {volume}
  {2}},\ \bibinfo {pages} {035043} (\bibinfo {year} {2018})}\BibitemShut
  {NoStop}%
\bibitem [{\citenamefont {Kunst}\ \emph {et~al.}(2018)\citenamefont {Kunst},
  \citenamefont {Edvardsson}, \citenamefont {Budich},\ and\ \citenamefont
  {Bergholtz}}]{PhysRevLett.121.026808}%
  \BibitemOpen
  \bibfield  {author} {\bibinfo {author} {\bibfnamefont {Flore~K.}\
  \bibnamefont {Kunst}}, \bibinfo {author} {\bibfnamefont {Elisabet}\
  \bibnamefont {Edvardsson}}, \bibinfo {author} {\bibfnamefont {Jan~Carl}\
  \bibnamefont {Budich}}, \ and\ \bibinfo {author} {\bibfnamefont {Emil~J.}\
  \bibnamefont {Bergholtz}},\ }\bibfield  {title} {\enquote {\bibinfo {title}
  {Biorthogonal bulk-boundary correspondence in non-hermitian systems},}\
  }\href {\doibase 10.1103/PhysRevLett.121.026808} {\bibfield  {journal}
  {\bibinfo  {journal} {Phys. Rev. Lett.}\ }\textbf {\bibinfo {volume} {121}},\
  \bibinfo {pages} {026808} (\bibinfo {year} {2018})}\BibitemShut {NoStop}%
\bibitem [{\citenamefont {Lee}\ and\ \citenamefont
  {Thomale}(2019)}]{lee2019anatomy}%
  \BibitemOpen
  \bibfield  {author} {\bibinfo {author} {\bibfnamefont {Ching~Hua}\
  \bibnamefont {Lee}}\ and\ \bibinfo {author} {\bibfnamefont {Ronny}\
  \bibnamefont {Thomale}},\ }\bibfield  {title} {\enquote {\bibinfo {title}
  {Anatomy of skin modes and topology in non-hermitian systems},}\ }\href@noop
  {} {\bibfield  {journal} {\bibinfo  {journal} {Physical Review B}\ }\textbf
  {\bibinfo {volume} {99}},\ \bibinfo {pages} {201103} (\bibinfo {year}
  {2019})}\BibitemShut {NoStop}%
\bibitem [{\citenamefont {Song}\ \emph {et~al.}(2019)\citenamefont {Song},
  \citenamefont {Yao},\ and\ \citenamefont {Wang}}]{PhysRevLett.123.246801}%
  \BibitemOpen
  \bibfield  {author} {\bibinfo {author} {\bibfnamefont {Fei}\ \bibnamefont
  {Song}}, \bibinfo {author} {\bibfnamefont {Shunyu}\ \bibnamefont {Yao}}, \
  and\ \bibinfo {author} {\bibfnamefont {Zhong}\ \bibnamefont {Wang}},\
  }\bibfield  {title} {\enquote {\bibinfo {title} {Non-hermitian topological
  invariants in real space},}\ }\href {\doibase 10.1103/PhysRevLett.123.246801}
  {\bibfield  {journal} {\bibinfo  {journal} {Phys. Rev. Lett.}\ }\textbf
  {\bibinfo {volume} {123}},\ \bibinfo {pages} {246801} (\bibinfo {year}
  {2019})}\BibitemShut {NoStop}%
\bibitem [{\citenamefont {Borgnia}\ \emph {et~al.}(2020)\citenamefont
  {Borgnia}, \citenamefont {Kruchkov},\ and\ \citenamefont
  {Slager}}]{borgnia2020nonH}%
  \BibitemOpen
  \bibfield  {author} {\bibinfo {author} {\bibfnamefont {Dan~S.}\ \bibnamefont
  {Borgnia}}, \bibinfo {author} {\bibfnamefont {Alex~Jura}\ \bibnamefont
  {Kruchkov}}, \ and\ \bibinfo {author} {\bibfnamefont {Robert-Jan}\
  \bibnamefont {Slager}},\ }\bibfield  {title} {\enquote {\bibinfo {title}
  {Non-hermitian boundary modes and topology},}\ }\href {\doibase
  10.1103/PhysRevLett.124.056802} {\bibfield  {journal} {\bibinfo  {journal}
  {Phys. Rev. Lett.}\ }\textbf {\bibinfo {volume} {124}},\ \bibinfo {pages}
  {056802} (\bibinfo {year} {2020})}\BibitemShut {NoStop}%
\bibitem [{\citenamefont {Zhang}\ \emph
  {et~al.}(2019{\natexlab{a}})\citenamefont {Zhang}, \citenamefont {Yang},\
  and\ \citenamefont {Fang}}]{zhang2019correspondence}%
  \BibitemOpen
  \bibfield  {author} {\bibinfo {author} {\bibfnamefont {Kai}\ \bibnamefont
  {Zhang}}, \bibinfo {author} {\bibfnamefont {Zhesen}\ \bibnamefont {Yang}}, \
  and\ \bibinfo {author} {\bibfnamefont {Chen}\ \bibnamefont {Fang}},\
  }\bibfield  {title} {\enquote {\bibinfo {title} {Correspondence between
  winding numbers and skin modes in non-hermitian systems},}\ }\href@noop {}
  {\bibfield  {journal} {\bibinfo  {journal} {arXiv preprint arXiv:1910.01131}\
  } (\bibinfo {year} {2019}{\natexlab{a}})}\BibitemShut {NoStop}%
\bibitem [{\citenamefont {Yoshida}\ \emph {et~al.}()\citenamefont {Yoshida},
  \citenamefont {Mizoguchi},\ and\ \citenamefont
  {Hatsugai}}]{yoshida2019mirror}%
  \BibitemOpen
  \bibfield  {author} {\bibinfo {author} {\bibfnamefont {Tsuneya}\ \bibnamefont
  {Yoshida}}, \bibinfo {author} {\bibfnamefont {Tomonari}\ \bibnamefont
  {Mizoguchi}}, \ and\ \bibinfo {author} {\bibfnamefont {Yasuhiro}\
  \bibnamefont {Hatsugai}},\ }\bibfield  {title} {\enquote {\bibinfo {title}
  {Mirror skin effect and its electric circuit simulation},}\ }\href@noop {} {\
  }\Eprint {http://arxiv.org/abs/1912.12022v1} {1912.12022v1} \BibitemShut
  {NoStop}%
\bibitem [{\citenamefont {Lee}\ \emph {et~al.}({\natexlab{b}})\citenamefont
  {Lee}, \citenamefont {Li}, \citenamefont {Thomale},\ and\ \citenamefont
  {Gong}}]{lee2019unraveling}%
  \BibitemOpen
  \bibfield  {author} {\bibinfo {author} {\bibfnamefont {Ching~Hua}\
  \bibnamefont {Lee}}, \bibinfo {author} {\bibfnamefont {Linhu}\ \bibnamefont
  {Li}}, \bibinfo {author} {\bibfnamefont {Ronny}\ \bibnamefont {Thomale}}, \
  and\ \bibinfo {author} {\bibfnamefont {Jiangbin}\ \bibnamefont {Gong}},\
  }\bibfield  {title} {\enquote {\bibinfo {title} {Unraveling non-hermitian
  pumping: emergent spectral singularities and anomalous responses},}\
  }\href@noop {} {\  ({\natexlab{b}})},\ \Eprint
  {http://arxiv.org/abs/1912.06974v2} {1912.06974v2} \BibitemShut {NoStop}%
\bibitem [{\citenamefont {Longhi}(2019)}]{longhi2019topological}%
  \BibitemOpen
  \bibfield  {author} {\bibinfo {author} {\bibfnamefont {Stefano}\ \bibnamefont
  {Longhi}},\ }\bibfield  {title} {\enquote {\bibinfo {title} {Topological
  phase transition in non-hermitian quasicrystals},}\ }\href@noop {} {\bibfield
   {journal} {\bibinfo  {journal} {Physical review letters}\ }\textbf {\bibinfo
  {volume} {122}},\ \bibinfo {pages} {237601} (\bibinfo {year}
  {2019})}\BibitemShut {NoStop}%
\bibitem [{\citenamefont {Jiang}\ \emph {et~al.}(2019)\citenamefont {Jiang},
  \citenamefont {Lang}, \citenamefont {Yang}, \citenamefont {Zhu},\ and\
  \citenamefont {Chen}}]{jiang2019interplay}%
  \BibitemOpen
  \bibfield  {author} {\bibinfo {author} {\bibfnamefont {Hui}\ \bibnamefont
  {Jiang}}, \bibinfo {author} {\bibfnamefont {Li-Jun}\ \bibnamefont {Lang}},
  \bibinfo {author} {\bibfnamefont {Chao}\ \bibnamefont {Yang}}, \bibinfo
  {author} {\bibfnamefont {Shi-Liang}\ \bibnamefont {Zhu}}, \ and\ \bibinfo
  {author} {\bibfnamefont {Shu}\ \bibnamefont {Chen}},\ }\bibfield  {title}
  {\enquote {\bibinfo {title} {Interplay of non-hermitian skin effects and
  anderson localization in nonreciprocal quasiperiodic lattices},}\ }\href@noop
  {} {\bibfield  {journal} {\bibinfo  {journal} {Physical Review B}\ }\textbf
  {\bibinfo {volume} {100}},\ \bibinfo {pages} {054301} (\bibinfo {year}
  {2019})}\BibitemShut {NoStop}%
\bibitem [{\citenamefont {Zeng}\ \emph {et~al.}(2020)\citenamefont {Zeng},
  \citenamefont {Yang},\ and\ \citenamefont {Xu}}]{zeng2020topological}%
  \BibitemOpen
  \bibfield  {author} {\bibinfo {author} {\bibfnamefont {Qi-Bo}\ \bibnamefont
  {Zeng}}, \bibinfo {author} {\bibfnamefont {Yan-Bin}\ \bibnamefont {Yang}}, \
  and\ \bibinfo {author} {\bibfnamefont {Yong}\ \bibnamefont {Xu}},\ }\bibfield
   {title} {\enquote {\bibinfo {title} {Topological phases in non-hermitian
  aubry-andr{\'e}-harper models},}\ }\href@noop {} {\bibfield  {journal}
  {\bibinfo  {journal} {Physical Review B}\ }\textbf {\bibinfo {volume}
  {101}},\ \bibinfo {pages} {020201} (\bibinfo {year} {2020})}\BibitemShut
  {NoStop}%
\bibitem [{\citenamefont {Lee}\ \emph {et~al.}(2019)\citenamefont {Lee},
  \citenamefont {Li},\ and\ \citenamefont {Gong}}]{lee2019hybrid}%
  \BibitemOpen
  \bibfield  {author} {\bibinfo {author} {\bibfnamefont {Ching~Hua}\
  \bibnamefont {Lee}}, \bibinfo {author} {\bibfnamefont {Linhu}\ \bibnamefont
  {Li}}, \ and\ \bibinfo {author} {\bibfnamefont {Jiangbin}\ \bibnamefont
  {Gong}},\ }\bibfield  {title} {\enquote {\bibinfo {title} {Hybrid
  higher-order skin-topological modes in nonreciprocal systems},}\ }\href@noop
  {} {\bibfield  {journal} {\bibinfo  {journal} {Physical review letters}\
  }\textbf {\bibinfo {volume} {123}},\ \bibinfo {pages} {016805} (\bibinfo
  {year} {2019})}\BibitemShut {NoStop}%
\bibitem [{\citenamefont {Li}\ \emph {et~al.}()\citenamefont {Li},
  \citenamefont {Lee},\ and\ \citenamefont {Gong}}]{li2019topology}%
  \BibitemOpen
  \bibfield  {author} {\bibinfo {author} {\bibfnamefont {Linhu}\ \bibnamefont
  {Li}}, \bibinfo {author} {\bibfnamefont {Ching~Hua}\ \bibnamefont {Lee}}, \
  and\ \bibinfo {author} {\bibfnamefont {Jiangbin}\ \bibnamefont {Gong}},\
  }\bibfield  {title} {\enquote {\bibinfo {title} {Topology-induced spontaneous
  non-reciprocal pumping in cold-atom systems with loss},}\ }\href@noop {} {\
  }\Eprint {http://arxiv.org/abs/1910.03229v1} {1910.03229v1} \BibitemShut
  {NoStop}%
\bibitem [{\citenamefont {Mu}\ \emph {et~al.}()\citenamefont {Mu},
  \citenamefont {Lee}, \citenamefont {Li},\ and\ \citenamefont
  {Gong}}]{mu2019emergent}%
  \BibitemOpen
  \bibfield  {author} {\bibinfo {author} {\bibfnamefont {Sen}\ \bibnamefont
  {Mu}}, \bibinfo {author} {\bibfnamefont {Ching~Hua}\ \bibnamefont {Lee}},
  \bibinfo {author} {\bibfnamefont {Linhu}\ \bibnamefont {Li}}, \ and\ \bibinfo
  {author} {\bibfnamefont {Jiangbin}\ \bibnamefont {Gong}},\ }\bibfield
  {title} {\enquote {\bibinfo {title} {Emergent fermi surface in a many-body
  non-hermitian fermionic chain},}\ }\href@noop {} {\ }\Eprint
  {http://arxiv.org/abs/1911.00023v1} {1911.00023v1} \BibitemShut {NoStop}%
\bibitem [{\citenamefont {Yokomizo}\ and\ \citenamefont
  {Murakami}(2019)}]{yokomizo2019non}%
  \BibitemOpen
  \bibfield  {author} {\bibinfo {author} {\bibfnamefont {Kazuki}\ \bibnamefont
  {Yokomizo}}\ and\ \bibinfo {author} {\bibfnamefont {Shuichi}\ \bibnamefont
  {Murakami}},\ }\bibfield  {title} {\enquote {\bibinfo {title} {Non-bloch band
  theory of non-hermitian systems},}\ }\href@noop {} {\bibfield  {journal}
  {\bibinfo  {journal} {Physical review letters}\ }\textbf {\bibinfo {volume}
  {123}},\ \bibinfo {pages} {066404} (\bibinfo {year} {2019})}\BibitemShut
  {NoStop}%
\bibitem [{\citenamefont {Yang}\ \emph {et~al.}({\natexlab{a}})\citenamefont
  {Yang}, \citenamefont {Zhang}, \citenamefont {Fang},\ and\ \citenamefont
  {Hu}}]{yang2019auxiliary}%
  \BibitemOpen
  \bibfield  {author} {\bibinfo {author} {\bibfnamefont {Zhesen}\ \bibnamefont
  {Yang}}, \bibinfo {author} {\bibfnamefont {Kai}\ \bibnamefont {Zhang}},
  \bibinfo {author} {\bibfnamefont {Chen}\ \bibnamefont {Fang}}, \ and\
  \bibinfo {author} {\bibfnamefont {Jiangping}\ \bibnamefont {Hu}},\ }\bibfield
   {title} {\enquote {\bibinfo {title} {Auxiliary generalized brillouin zone
  method in non-hermitian band theory},}\ }\href@noop {} {\
  ({\natexlab{a}})},\ \Eprint {http://arxiv.org/abs/1912.05499v1}
  {1912.05499v1} \BibitemShut {NoStop}%
\bibitem [{\citenamefont {Okuma}\ \emph {et~al.}()\citenamefont {Okuma},
  \citenamefont {Kawabata}, \citenamefont {Shiozaki},\ and\ \citenamefont
  {Sato}}]{okuma2019topological}%
  \BibitemOpen
  \bibfield  {author} {\bibinfo {author} {\bibfnamefont {Nobuyuki}\
  \bibnamefont {Okuma}}, \bibinfo {author} {\bibfnamefont {Kohei}\ \bibnamefont
  {Kawabata}}, \bibinfo {author} {\bibfnamefont {Ken}\ \bibnamefont
  {Shiozaki}}, \ and\ \bibinfo {author} {\bibfnamefont {Masatoshi}\
  \bibnamefont {Sato}},\ }\bibfield  {title} {\enquote {\bibinfo {title}
  {Topological origin of non-hermitian skin effects},}\ }\href@noop {} {\
  }\Eprint {http://arxiv.org/abs/1910.02878v3} {1910.02878v3} \BibitemShut
  {NoStop}%
\bibitem [{\citenamefont {Yang}\ \emph {et~al.}({\natexlab{b}})\citenamefont
  {Yang}, \citenamefont {Wang}, \citenamefont {Jin},\ and\ \citenamefont
  {Song}}]{yang2019visualizing}%
  \BibitemOpen
  \bibfield  {author} {\bibinfo {author} {\bibfnamefont {X.~M.}\ \bibnamefont
  {Yang}}, \bibinfo {author} {\bibfnamefont {P.}~\bibnamefont {Wang}}, \bibinfo
  {author} {\bibfnamefont {L.}~\bibnamefont {Jin}}, \ and\ \bibinfo {author}
  {\bibfnamefont {Z.}~\bibnamefont {Song}},\ }\bibfield  {title} {\enquote
  {\bibinfo {title} {Visualizing topology of real-energy gapless phase arising
  from exceptional point},}\ }\href@noop {} {\  ({\natexlab{b}})},\ \Eprint
  {http://arxiv.org/abs/1905.07109v1} {1905.07109v1} \BibitemShut {NoStop}%
\bibitem [{\citenamefont {Yin}\ \emph {et~al.}(2018)\citenamefont {Yin},
  \citenamefont {Jiang}, \citenamefont {Li}, \citenamefont {L{\"u}},\ and\
  \citenamefont {Chen}}]{yin2018geometrical}%
  \BibitemOpen
  \bibfield  {author} {\bibinfo {author} {\bibfnamefont {Chuanhao}\
  \bibnamefont {Yin}}, \bibinfo {author} {\bibfnamefont {Hui}\ \bibnamefont
  {Jiang}}, \bibinfo {author} {\bibfnamefont {Linhu}\ \bibnamefont {Li}},
  \bibinfo {author} {\bibfnamefont {Rong}\ \bibnamefont {L{\"u}}}, \ and\
  \bibinfo {author} {\bibfnamefont {Shu}\ \bibnamefont {Chen}},\ }\bibfield
  {title} {\enquote {\bibinfo {title} {Geometrical meaning of winding number
  and its characterization of topological phases in one-dimensional chiral
  non-hermitian systems},}\ }\href@noop {} {\bibfield  {journal} {\bibinfo
  {journal} {Physical Review A}\ }\textbf {\bibinfo {volume} {97}},\ \bibinfo
  {pages} {052115} (\bibinfo {year} {2018})}\BibitemShut {NoStop}%
\bibitem [{\citenamefont {Jiang}\ \emph {et~al.}(2018)\citenamefont {Jiang},
  \citenamefont {Yang},\ and\ \citenamefont {Chen}}]{jiang2018topological}%
  \BibitemOpen
  \bibfield  {author} {\bibinfo {author} {\bibfnamefont {Hui}\ \bibnamefont
  {Jiang}}, \bibinfo {author} {\bibfnamefont {Chao}\ \bibnamefont {Yang}}, \
  and\ \bibinfo {author} {\bibfnamefont {Shu}\ \bibnamefont {Chen}},\
  }\bibfield  {title} {\enquote {\bibinfo {title} {Topological invariants and
  phase diagrams for one-dimensional two-band non-hermitian systems without
  chiral symmetry},}\ }\href@noop {} {\bibfield  {journal} {\bibinfo  {journal}
  {Physical Review A}\ }\textbf {\bibinfo {volume} {98}},\ \bibinfo {pages}
  {052116} (\bibinfo {year} {2018})}\BibitemShut {NoStop}%
\bibitem [{\citenamefont {Jiang}\ \emph {et~al.}()\citenamefont {Jiang},
  \citenamefont {L\"u},\ and\ \citenamefont {Chen}}]{jiang2019topological}%
  \BibitemOpen
  \bibfield  {author} {\bibinfo {author} {\bibfnamefont {Hui}\ \bibnamefont
  {Jiang}}, \bibinfo {author} {\bibfnamefont {Rong}\ \bibnamefont {L\"u}}, \
  and\ \bibinfo {author} {\bibfnamefont {Shu}\ \bibnamefont {Chen}},\
  }\bibfield  {title} {\enquote {\bibinfo {title} {Topological invariants, zero
  mode edge states and finite size effect for a generalized non-reciprocal
  su-schrieffer-heeger model},}\ }\href@noop {} {\ }\Eprint
  {http://arxiv.org/abs/1906.04700v1} {1906.04700v1} \BibitemShut {NoStop}%
\bibitem [{\citenamefont {Majorana}(1932)}]{majorana_1932}%
  \BibitemOpen
  \bibfield  {author} {\bibinfo {author} {\bibfnamefont {Ettore}\ \bibnamefont
  {Majorana}},\ }\bibfield  {title} {\enquote {\bibinfo {title} {Atomi
  orientati in campo magnetico variabile},}\ }\href {\doibase
  10.1007/bf02960953} {\bibfield  {journal} {\bibinfo  {journal} {Il Nuovo
  Cimento}\ }\textbf {\bibinfo {volume} {9}},\ \bibinfo {pages} {43–50}
  (\bibinfo {year} {1932})}\BibitemShut {NoStop}%
\bibitem [{\citenamefont {Bloch}\ and\ \citenamefont
  {Rabi}(1945)}]{bloch_rabi_1945}%
  \BibitemOpen
  \bibfield  {author} {\bibinfo {author} {\bibfnamefont {F.}~\bibnamefont
  {Bloch}}\ and\ \bibinfo {author} {\bibfnamefont {I.~I.}\ \bibnamefont
  {Rabi}},\ }\bibfield  {title} {\enquote {\bibinfo {title} {Atoms in variable
  magnetic fields},}\ }\href {\doibase 10.1103/revmodphys.17.237} {\bibfield
  {journal} {\bibinfo  {journal} {Reviews of Modern Physics}\ }\textbf
  {\bibinfo {volume} {17}},\ \bibinfo {pages} {237–244} (\bibinfo {year}
  {1945})}\BibitemShut {NoStop}%
\bibitem [{\citenamefont {Biedenharn}\ and\ \citenamefont
  {Dam}(1965)}]{biedenharn_dam_1965}%
  \BibitemOpen
  \bibfield  {author} {\bibinfo {author} {\bibfnamefont {Lawrence~C.}\
  \bibnamefont {Biedenharn}}\ and\ \bibinfo {author} {\bibfnamefont
  {Hendrik~Van}\ \bibnamefont {Dam}},\ }\href@noop {} {\emph {\bibinfo {title}
  {Quantum theory of angular momentum: a collection of reprints and original
  papers}}}\ (\bibinfo  {publisher} {Academic Press},\ \bibinfo {year}
  {1965})\BibitemShut {NoStop}%
\bibitem [{\citenamefont {Hannay}(1998)}]{hannay_1998}%
  \BibitemOpen
  \bibfield  {author} {\bibinfo {author} {\bibfnamefont {J~H}\ \bibnamefont
  {Hannay}},\ }\bibfield  {title} {\enquote {\bibinfo {title} {The berry phase
  for spin in the majorana representation},}\ }\href {\doibase
  10.1088/0305-4470/31/2/002} {\bibfield  {journal} {\bibinfo  {journal}
  {Journal of Physics A: Mathematical and General}\ }\textbf {\bibinfo {volume}
  {31}} (\bibinfo {year} {1998}),\ 10.1088/0305-4470/31/2/002}\BibitemShut
  {NoStop}%
\bibitem [{\citenamefont {Bruno}(2012)}]{bruno_2012}%
  \BibitemOpen
  \bibfield  {author} {\bibinfo {author} {\bibfnamefont {Patrick}\ \bibnamefont
  {Bruno}},\ }\bibfield  {title} {\enquote {\bibinfo {title} {Quantum geometric
  phase in majorana’s stellar representation: Mapping onto a many-body
  aharonov-bohm phase},}\ }\href {\doibase 10.1103/physrevlett.108.240402}
  {\bibfield  {journal} {\bibinfo  {journal} {Physical Review Letters}\
  }\textbf {\bibinfo {volume} {108}} (\bibinfo {year} {2012}),\
  10.1103/physrevlett.108.240402}\BibitemShut {NoStop}%
\bibitem [{\citenamefont {Liu}\ and\ \citenamefont {Fu}(2014)}]{liu_fu_2014}%
  \BibitemOpen
  \bibfield  {author} {\bibinfo {author} {\bibfnamefont {H.D.}\ \bibnamefont
  {Liu}}\ and\ \bibinfo {author} {\bibfnamefont {L.B.}\ \bibnamefont {Fu}},\
  }\bibfield  {title} {\enquote {\bibinfo {title} {Representation of berry
  phase by the trajectories of majorana stars},}\ }\href {\doibase
  10.1103/physrevlett.113.240403} {\bibfield  {journal} {\bibinfo  {journal}
  {Physical Review Letters}\ }\textbf {\bibinfo {volume} {113}} (\bibinfo
  {year} {2014}),\ 10.1103/physrevlett.113.240403}\BibitemShut {NoStop}%
\bibitem [{\citenamefont {Yang}\ \emph {et~al.}(2015)\citenamefont {Yang},
  \citenamefont {Guo}, \citenamefont {Fu},\ and\ \citenamefont
  {Chen}}]{yang_guo_fu_chen_2015}%
  \BibitemOpen
  \bibfield  {author} {\bibinfo {author} {\bibfnamefont {Chao}\ \bibnamefont
  {Yang}}, \bibinfo {author} {\bibfnamefont {Huaiming}\ \bibnamefont {Guo}},
  \bibinfo {author} {\bibfnamefont {Li-Bin}\ \bibnamefont {Fu}}, \ and\
  \bibinfo {author} {\bibfnamefont {Shu}\ \bibnamefont {Chen}},\ }\bibfield
  {title} {\enquote {\bibinfo {title} {Characterization of symmetry-protected
  topological phases in polymerized models by trajectories of majorana
  stars},}\ }\href {\doibase 10.1103/physrevb.91.125132} {\bibfield  {journal}
  {\bibinfo  {journal} {Physical Review B}\ }\textbf {\bibinfo {volume} {91}}
  (\bibinfo {year} {2015}),\ 10.1103/physrevb.91.125132}\BibitemShut {NoStop}%
\bibitem [{\citenamefont {Liu}\ and\ \citenamefont
  {Fu}(2016)}]{PhysRevA.94.022123}%
  \BibitemOpen
  \bibfield  {author} {\bibinfo {author} {\bibfnamefont {H.~D.}\ \bibnamefont
  {Liu}}\ and\ \bibinfo {author} {\bibfnamefont {L.~B.}\ \bibnamefont {Fu}},\
  }\bibfield  {title} {\enquote {\bibinfo {title} {Berry phase and quantum
  entanglement in majorana's stellar representation},}\ }\href {\doibase
  10.1103/PhysRevA.94.022123} {\bibfield  {journal} {\bibinfo  {journal} {Phys.
  Rev. A}\ }\textbf {\bibinfo {volume} {94}},\ \bibinfo {pages} {022123}
  (\bibinfo {year} {2016})}\BibitemShut {NoStop}%
\bibitem [{\citenamefont {Zak}(1989)}]{zak1989berry}%
  \BibitemOpen
  \bibfield  {author} {\bibinfo {author} {\bibfnamefont {J}~\bibnamefont
  {Zak}},\ }\bibfield  {title} {\enquote {\bibinfo {title} {Berry’s phase for
  energy bands in solids},}\ }\href@noop {} {\bibfield  {journal} {\bibinfo
  {journal} {Physical review letters}\ }\textbf {\bibinfo {volume} {62}},\
  \bibinfo {pages} {2747} (\bibinfo {year} {1989})}\BibitemShut {NoStop}%
\bibitem [{\citenamefont {Berry}(1984)}]{berry1984quantal}%
  \BibitemOpen
  \bibfield  {author} {\bibinfo {author} {\bibfnamefont {Michael~Victor}\
  \bibnamefont {Berry}},\ }\bibfield  {title} {\enquote {\bibinfo {title}
  {Quantal phase factors accompanying adiabatic changes},}\ }\href@noop {}
  {\bibfield  {journal} {\bibinfo  {journal} {Proceedings of the Royal Society
  of London. A. Mathematical and Physical Sciences}\ }\textbf {\bibinfo
  {volume} {392}},\ \bibinfo {pages} {45--57} (\bibinfo {year}
  {1984})}\BibitemShut {NoStop}%
\bibitem [{\citenamefont {Brody}(2013)}]{brody2013biorthogonal}%
  \BibitemOpen
  \bibfield  {author} {\bibinfo {author} {\bibfnamefont {Dorje~C}\ \bibnamefont
  {Brody}},\ }\bibfield  {title} {\enquote {\bibinfo {title} {Biorthogonal
  quantum mechanics},}\ }\href@noop {} {\bibfield  {journal} {\bibinfo
  {journal} {Journal of Physics A: Mathematical and Theoretical}\ }\textbf
  {\bibinfo {volume} {47}},\ \bibinfo {pages} {035305} (\bibinfo {year}
  {2013})}\BibitemShut {NoStop}%
\bibitem [{\citenamefont {Zhang}\ \emph
  {et~al.}(2019{\natexlab{b}})\citenamefont {Zhang}, \citenamefont {Wang},\
  and\ \citenamefont {Gong}}]{zhang_wang_gong_2019_2}%
  \BibitemOpen
  \bibfield  {author} {\bibinfo {author} {\bibfnamefont {Da-Jian}\ \bibnamefont
  {Zhang}}, \bibinfo {author} {\bibfnamefont {Qing-Hai}\ \bibnamefont {Wang}},
  \ and\ \bibinfo {author} {\bibfnamefont {Jiangbin}\ \bibnamefont {Gong}},\
  }\bibfield  {title} {\enquote {\bibinfo {title} {Quantum geometric tensor in
  pt-symmetric quantum mechanics},}\ }\href {\doibase
  10.1103/physreva.99.042104} {\bibfield  {journal} {\bibinfo  {journal}
  {Physical Review A}\ }\textbf {\bibinfo {volume} {99}} (\bibinfo {year}
  {2019}{\natexlab{b}}),\ 10.1103/physreva.99.042104}\BibitemShut {NoStop}%
\bibitem [{\citenamefont {Zhang}\ \emph
  {et~al.}(2019{\natexlab{c}})\citenamefont {Zhang}, \citenamefont {Wang},\
  and\ \citenamefont {Gong}}]{zhang_wang_gong_2019}%
  \BibitemOpen
  \bibfield  {author} {\bibinfo {author} {\bibfnamefont {Da-Jian}\ \bibnamefont
  {Zhang}}, \bibinfo {author} {\bibfnamefont {Qing-Hai}\ \bibnamefont {Wang}},
  \ and\ \bibinfo {author} {\bibfnamefont {Jiangbin}\ \bibnamefont {Gong}},\
  }\bibfield  {title} {\enquote {\bibinfo {title} {Time-dependent pt-symmetric
  quantum mechanics in generic non-hermitian systems},}\ }\href {\doibase
  10.1103/physreva.100.062121} {\bibfield  {journal} {\bibinfo  {journal}
  {Physical Review A}\ }\textbf {\bibinfo {volume} {100}} (\bibinfo {year}
  {2019}{\natexlab{c}}),\ 10.1103/physreva.100.062121}\BibitemShut {NoStop}%
\bibitem [{\citenamefont {Mong}\ and\ \citenamefont
  {Shivamoggi}(2011)}]{mong2011edge}%
  \BibitemOpen
  \bibfield  {author} {\bibinfo {author} {\bibfnamefont {Roger~SK}\
  \bibnamefont {Mong}}\ and\ \bibinfo {author} {\bibfnamefont {Vasudha}\
  \bibnamefont {Shivamoggi}},\ }\bibfield  {title} {\enquote {\bibinfo {title}
  {Edge states and the bulk-boundary correspondence in dirac hamiltonians},}\
  }\href@noop {} {\bibfield  {journal} {\bibinfo  {journal} {Physical Review
  B}\ }\textbf {\bibinfo {volume} {83}},\ \bibinfo {pages} {125109} (\bibinfo
  {year} {2011})}\BibitemShut {NoStop}%
\bibitem [{\citenamefont {K{\'a}roly}(2016)}]{karoly2016short}%
  \BibitemOpen
  \bibfield  {author} {\bibinfo {author} {\bibfnamefont {Asb{\'o}th~J{\'a}nos}\
  \bibnamefont {K{\'a}roly}},\ }\href@noop {} {\emph {\bibinfo {title} {A Short
  Course on Topological Insulators: Band-structure Topology and Edge States in
  One and Two Dimensions}}}\ (\bibinfo  {publisher} {Springer},\ \bibinfo
  {year} {2016})\BibitemShut {NoStop}%
\bibitem [{\citenamefont {Rhim}\ \emph {et~al.}(2017)\citenamefont {Rhim},
  \citenamefont {Behrends},\ and\ \citenamefont
  {Bardarson}}]{PhysRevB.95.035421}%
  \BibitemOpen
  \bibfield  {author} {\bibinfo {author} {\bibfnamefont {Jun-Won}\ \bibnamefont
  {Rhim}}, \bibinfo {author} {\bibfnamefont {Jan}\ \bibnamefont {Behrends}}, \
  and\ \bibinfo {author} {\bibfnamefont {Jens~H.}\ \bibnamefont {Bardarson}},\
  }\bibfield  {title} {\enquote {\bibinfo {title} {Bulk-boundary correspondence
  from the intercellular zak phase},}\ }\href {\doibase
  10.1103/PhysRevB.95.035421} {\bibfield  {journal} {\bibinfo  {journal} {Phys.
  Rev. B}\ }\textbf {\bibinfo {volume} {95}},\ \bibinfo {pages} {035421}
  (\bibinfo {year} {2017})}\BibitemShut {NoStop}%
\bibitem [{\citenamefont {Chen}\ \emph {et~al.}(2019)\citenamefont {Chen},
  \citenamefont {Chang},\ and\ \citenamefont {Kao}}]{chen2019zak}%
  \BibitemOpen
  \bibfield  {author} {\bibinfo {author} {\bibfnamefont {Han-Ting}\
  \bibnamefont {Chen}}, \bibinfo {author} {\bibfnamefont {Chia-Hsun}\
  \bibnamefont {Chang}}, \ and\ \bibinfo {author} {\bibfnamefont {Hsien-chung}\
  \bibnamefont {Kao}},\ }\bibfield  {title} {\enquote {\bibinfo {title} {The
  zak phase and winding number},}\ }\href@noop {} {\bibfield  {journal}
  {\bibinfo  {journal} {arXiv preprint arXiv:1908.06700}\ } (\bibinfo {year}
  {2019})}\BibitemShut {NoStop}%
\bibitem [{\citenamefont {Motohiko}\ \emph {et~al.}()\citenamefont {Motohiko},
  \citenamefont {Yukio}, \ and\ \citenamefont
  {Naoto}}]{ezawa_tanaka_nagaosa_2013}%
  \BibitemOpen
  \bibfield  {author} {\bibinfo {author} {\bibfnamefont {Ezawa}\ \bibnamefont
  {Motohiko}}, \bibinfo {author} {\bibfnamefont {Tanaka}\ \bibnamefont {Yukio}}, \ and\
  \bibinfo {author} {\bibfnamefont {Nagaosa}\ \bibnamefont {Naoto}},\
  }\bibfield  {title} {\enquote {\bibinfo {title} {Topological Phase Transition without Gap Closing},}\ }\href@noop {} {\bibfield  {journal} {\bibinfo  {journal} {Scientific Reports}\ }\textbf {\bibinfo {volume} {3}},\ \bibinfo {pages} {2790} (\bibinfo
  {year} {2013})} \BibitemShut {NoStop}%
\bibitem [{\citenamefont {Mailybaev}\ \emph {et~al.}(2005)\citenamefont
  {Mailybaev}, \citenamefont {Kirillov},\ and\ \citenamefont
  {Seyranian}}]{mailybaev2005geometric}%
  \BibitemOpen
  \bibfield  {author} {\bibinfo {author} {\bibfnamefont {Alexei~A}\
  \bibnamefont {Mailybaev}}, \bibinfo {author} {\bibfnamefont {Oleg~N}\
  \bibnamefont {Kirillov}}, \ and\ \bibinfo {author} {\bibfnamefont
  {Alexander~P}\ \bibnamefont {Seyranian}},\ }\bibfield  {title} {\enquote
  {\bibinfo {title} {Geometric phase around exceptional points},}\ }\href@noop
  {} {\bibfield  {journal} {\bibinfo  {journal} {Physical Review A}\ }\textbf
  {\bibinfo {volume} {72}},\ \bibinfo {pages} {014104} (\bibinfo {year}
  {2005})}\BibitemShut {NoStop}%
\bibitem [{\citenamefont {Xu}\ \emph {et~al.}()\citenamefont {Xu},
  \citenamefont {Liu}, \citenamefont {Zhang},\ and\ \citenamefont
  {Liang}}]{xu2020ms}%
  \BibitemOpen
  \bibfield  {author} {\bibinfo {author} {\bibfnamefont {Xingran}\ \bibnamefont
  {Xu}}, \bibinfo {author} {\bibfnamefont {Haodi}\ \bibnamefont {Liu}},
  \bibinfo {author} {\bibfnamefont {Zhidong}\ \bibnamefont {Zhang}}, \ and\
  \bibinfo {author} {\bibfnamefont {Zhaoxin}\ \bibnamefont {Liang}},\
  }\bibfield  {title} {\enquote {\bibinfo {title} {The non-hermitian
  geometrical property of 1d lieb lattice under majorana's stellar
  representation},}\ }\href@noop {} {\ }\Eprint
  {http://arxiv.org/abs/2002.01344v1} {2002.01344v1} \BibitemShut {NoStop}%
\end{thebibliography}
%

\end{document}